\documentclass[]{input/pasj02}
\usepackage[switch,mathlines]{lineno} 
\jyear{2025}
\Received{}
\Accepted{}

\begin{document} 
\title{The Galactic-Centre Arms inferred from ACES (ALMA CMZ Exploration Survey)}
 
\def\afmark{\altaffilmark}
\def\aftext{\altaffiltext} 
\author{
Yoshiaki \textsc{Sofue}\afmark{1}\orcid{0000-0002-4268-6499}\email{sofue@ioa.s.u-tokyo.ac.jp}\aftext{1}{Institute of Astronomy, The University of Tokyo, Mitaka, Tokyo 181-0015, Japan},
 Tomoharu \textsc{Oka}\afmark{2}\orcid{0000-0002-5566-0634}\aftext{2}{Department of Physics, Faculty of Science and Technology, Keio University, 3-14-1 Hiyoshi, Yokohama, Kanagawa 223-8522, Japan}, 
\orcid{0000-0001-6353-0170}{Steven N. \textsc{Longmore}}\afmark{3,4},\aftext{3}{Astrophysics Research Institute, Liverpool John Moores University, IC2, Liverpool Science Park, 146 Brownlow Hill, Liverpool L3 5RF, UK}\aftext{4}{Cosmic Origins Of Life (COOL) Research DAO, https://coolresearch.io}
\orcid{0000-0001-7330-8856}{Daniel \textsc{Walker}}\afmark{5}\aftext{5}{UK ALMA Regional Centre Node, Jodrell Bank Centre for Astrophysics, Oxford Road, The University of Manchester, Manchester M13 9PL, United Kingdom},
\orcid{0000-0001-6431-9633}{Adam \textsc{Ginsburg}}\afmark{6},\aftext{6}{Department of Astronomy, University of Florida, P.O. Box 112055, Gainesville, FL 32611}
\orcid{0000-0001-9656-7682}{Jonathan D. \textsc{Henshaw}}\afmark{3,7},\aftext{7}{Max Planck Institute for Astronomy, K\"{o}nigstuhl 17, D-69117 Heidelberg, Germany}
\orcid{0000-0001-8135-6612}{John \textsc{Bally}}\afmark{8},\aftext{8}{Center for Astrophysics and Space Astronomy; Department of Astrophysical and Planetary Sciences; University of Colorado, Boulder, CO 80389, USA}
\orcid{0000-0003-0410-4504}{Ashley T. \textsc{Barnes}}\afmark{9},\aftext{9}{European Southern Observatory (ESO), Karl-Schwarzschild-Strasse 2, 85748 Garching, Germany}
\orcid{0000-0002-6073-9320}{Cara \textsc{Battersby}}\afmark{10},\aftext{10}{Department of Physics, University of Connecticut, 196A Auditorium Road, Unit 3046, Storrs, CT 06269, USA}
\orcid{0000-0001-8064-6394}{Laura \textsc{Colzi}}\afmark{11},\aftext{11}{Centro de Astrobiología (CAB), CSIC-INTA, Carretera de Ajalvir km 4, Torrejón de Ardoz, 28850 Madrid, Spain}
Paul \textsc{Ho}\afmark{12},\aftext{12}{AS/NTU Astronomy-Mathematics Building,  Roosevelt Rd, Taipei 10617, Taiwan}\orcid{0000-0002-3412-4306}
Izaskun \textsc{Jimenez}-\textsc{Serra}\afmark{11}
\orcid{0000-0003-4493-8714},
\orcid{0000-0002-8804-0212}J.~M.~Diederik \textsc{Kruijssen}\afmark{13,4},\aftext{13}{Technical University of Munich, School of Engineering and Design, Department of Aerospace and Geodesy,  Arcisstr. 21, 80333 Munich, Germany}
Elizabeth \textsc{Mills}\afmark{14},\aftext{14}{ Department of Physics and Astronomy, University of Kansas, 1251 Wescoe Hall Drive, Lawrence, KS 66045, USA}\orcid{0000-0001-8782-1992}
\orcid{0000-0002-6362-8159}Maya A. \textsc{Petkova}\afmark{15},\aftext{15}{Space, Earth and Environment Department, Chalmers University of Technology, SE-412 96 Gothenburg, Sweden}
\orcid{0000-0001-6113-6241}{Mattia C. \textsc{Sormani}}\afmark{16,17},\aftext{16}{Universit{\`a} dell’Insubria, via Valleggio 11, 22100 Como, Italy}\aftext{17}{Department of Physics, University of Surrey, Guildford GU2 7XH, UK}
\orcid{0000-0000-0000-0000}Jen \textsc{Wallace}\afmark{10},
\orcid{0000-0003-3341-6144}{Jairo  \textsc{Armijos-Abenda\~no}}\afmark{18},\aftext{18}{Observatorio Astron\'omico de Quito, Observatorio Nacional, Escuela Polit\'ecnica Nacional, Interior del Parque La Alameda, 170136, Quito, Ecuador}
\orcid{0000-0003-0980-6871}{Katarzyna M. \textsc{Dutkowska}}\afmark{19},\aftext{19}{Leiden Observatory, Leiden University, P.O. Box 9513, 2300 RA Leiden, The Netherlands}
\orcid{0000-0001-0000-0000}{Rei \textsc{Enokiya}}\afmark{20},\aftext{20}{Kyuushu Sangyo University, Fukuoka, Japan}
\orcid{0000-0002-8966-9856}{Yasuo \textsc{Fukui}}\afmark{21},\aftext{21}{Department of Physics, Nagoya University, Chikusa-ku, Nagoya 464-8602, Japan}
Pablo \textsc{Garc\'ia}\afmark{22,23},\aftext{22}{Chinese Academy of Sciences South America Center for Astronomy, National Astronomical Observatories, CAS, Beijing 100101, China}\aftext{23}{Instituto de Astronom\'ia, Universidad Cat\'olica del Norte, Av. Angamos 0610, Antofagasta, Chile}
Andres \textsc{Guzman}\afmark{24},\aftext{24}{Joint ALMA Observatory, Alonso de Cordova 3107, Vitacura 763-0355, Santiago de Chile, Chile}
\orcid{0000-0002-7495-4005}{Christian \textsc{Henkel}}\afmark{25},\aftext{25}{MPIfR, Auf dem H\"ugel 69, Bonn, Germany}
\orcid{0000-0001-9155-3978}{Pei-Ying \textsc{Hsieh}}\afmark{26},\aftext{26}{National Astronomical Observatory of Japan, 2-21-1 Osawa, Mitaka, Tokyo 181-8588, Japan}
Yue \textsc{Hu}\afmark{27},\aftext{27}{Institute for Advanced Study, 1 Einstein Drive, Princeton, NJ 08540, USA}\orcid{0000-0002-8455-0805}
\orcid{0000-0003-4140-5138}{Katharina \textsc{Immer}}\afmark{9},
\orcid{0000-0003-0416-4830}{Desmond \textsc{Jeff}}\afmark{6,28},\aftext{28}{National Radio Astronomy Observatory, 520 Edgemont Road, Charlottesville, VA 22903, USA}
Ralf S.\ \textsc{Klessen}\afmark{29,30,31,32},\aftext{29}{Universit\"{a}t Heidelberg, Zentrum f\"{u}r Astronomie, Institut f\"{u}r Theoretische Astrophysik, Albert-Ueberle-Str.\ 2, 69120 Heidelberg, Germany}\aftext{30}{Universit\"{a}t Heidelberg, Interdisziplin\"{a}res Zentrum f\"{u}r Wissenschaftliches Rechnen, Im Neuenheimer Feld 225, 69120 Heidelberg, Germany}\aftext{31}{Center for Astrophysics $\vert$ Harvard \& Smithsonian, 60 Garden Street, Cambridge, MA, 02138, USA}\aftext{32}{Elizabeth S. and Richard M. Cashin Fellow at the Radcliffe Institute for Advanced Studies at Harvard University, 10 Garden Street, Cambridge, MA 02138, USA}\orcid{0000-0002-0560-3172}
Kotaro \textsc{Kohno}\afmark{1},\orcid{0000-0002-4052-2394}
Mark R. \textsc{Krumholz}\afmark{33},\aftext{33}{Research School of Astronomy and Astrophysics, Australian National University, Cotter Road, Weston ACT 2611, Australia}\orcid{0000-0003-3893-854X}
\orcid{0000-0002-5776-9473}{Dani \textsc{Lipman}}\afmark{10}, 
Mark R. \textsc{Morris}\afmark{34},\aftext{34}{Department of Physics and Astronomy, University of California, Los Angeles, CA 90095, USA} \orcid{0000-0002-6753-2066} 
\orcid{0000-0002-6379-7593}{Francisco \textsc{Nogueras-Lara}}\afmark{9}, 
\orcid{0000-0000-0000-0000}{Mairi \textsc{Nonhebel}\afmark{9,35},\aftext{35}{SUPA,School of Physics and Astronomy, University of St. Andrews, North Haugh, St. Andrews KY16 9SS, UK}
\orcid{0000-0001-8224-1956}{J\"urgen \textsc{Ott}}\afmark{36},\aftext{36}{National Radio Astronomy Observatory, P.O. Box O, 1011 Lopezville Road, Socorro, NM 87801, USA} 
\orcid{0000-0002-3972-1978}{Jaime E. \textsc{Pineda}}\afmark{37},\aftext{37}{MPI for Extraterrestrial Physics, Giessenbachstr. 1, D-85748, Garching by Muenchen, Germany}
\orcid{0000-0001-9281-2919}{Sergio \textsc{Mart\'in}}\afmark{38,39},\aftext{38}{European Southern Observatory, Alonso de C\'ordova, 3107, Vitacura, Santiago 763-0355, Chile}\aftext{39}{Joint ALMA Observatory, Alonso de C\'ordova, 3107, Vitacura, Santiago 763-0355, Chile}
\orcid{0009-0009-5346-7329}{Miguel Angel \textsc{Requena-Torres}}\afmark{40},\aftext{40}{Department of Physics, Astronomy, and Geosciences, Towson University, Towson, MD 21252, USA} 
\orcid{0000-0002-2887-5859}{Víctor M. \textsc{Rivilla}}\afmark{11}, 
\orcid{0000-0001-5389-0535}{Denise \textsc{Riquelme-V\'asquez}}\afmark{41},\aftext{41}{Departamento de Astronom\'ia, Universidad de La Serena, Ra\'ul Bitr\'an 1305, La Serena, Chile}
\orcid{0000-0002-3078-9482}{\'Alvaro S\'anchez-Monge}\afmark{42,43},\aftext{42}{Institut de Ci\`encies de l'Espai (ICE), CSIC, Campus UAB, Carrer de Can Magrans s/n, E-08193, Bellaterra (Barcelona), Spain}\aftext{43}{Institut d'Estudis Espacials de Catalunya (IEEC), E-08860, Castelldefels (Barcelona), Spain}
\orcid{0000-0002-3941-0360}
{Miriam \textsc{G. Santa-Maria}}\afmark{6}, 
Howard A.~\textsc{Smith}\afmark{31}, 
Tabassum S \textsc{Tanvir}\afmark{44},\aftext{44}{Department of Physics and Astronomy, Iowa State University, 2323 Osborn Drive, Ames, IA 50010, USA} \orcid{0000-0002-0862-0701}
\orcid{0000-0003-1841-2241}Volker \textsc{Tolls}\afmark{31},
and
\orcid{0000-0002-9279-4041}{Q. Daniel \textsc{Wang}}\afmark{45}\aftext{45}{Department of Astronomy, University of Massachusetts, Amherst, MA 01003, USA}
%
%
}}  

\maketitle
 
\def\be{\begin{equation}}\def\ee{\end{equation}}\def\vlsr{v_{\rm LSR}} \def\Vlsr{\vlsr}\def\Msun{M_\odot} \def\vr{v_{\rm r}}\def\deg{^\circ}  \def\d{^\circ}
\def\vrot{V_{\rm rot}} \def\Vrot{\vrot}\def\co{$^{12}$CO } \def\coth{$^{13}$CO $(J=1-0)$} \def\Xco{X_{\rm {^{12}}CO}}   \def\Tb{T_{\rm B}} \def\Tp{T_{\rm p}}\def\Htwo{H$_2$} \def\htwo{H$_2$}  \def\Kkms{K km s$^{-1}}  \def\Hcc{{\rm H \cm^{-3}}} \def\kms{km s$^{-1}$} \def\Ico{I_{\rm CO}}  \def\Kkms{K \kms }\def\mH{m_{\rm H}}  \def\Ico{I_{\rm ^{12}CO}} \def\Icoth{I_{\rm ^{13}CO}} 
 \def\htwo{H$_2$} \def\Tb{T_{\rm B}}   \def\mH{m_{\rm H}} \def\ekms{{\rm \ km \ s^{-1}}}  \def\epc{{\rm \ pc} }  \def\Hii{HII} \def\apj{ApJ} \def\aap{AA} \def\mnras{MNRAS} \def\pasj{PASJ} \def\aj{AJ} \def\xcounit{H$_2$ cm $^{-2}$ [K km s$^{-1}]^{-1}$}  \def\log{{\rm log}}  \def\tc{t_{\rm C}} \def\fc{f_{\rm C}} \def\SFR{{\rm SFR}} \def\sfr{{\rm SFR}}\def\tc{t_{\rm c}}\def\lc{l_{\rm c}}\def\vc{v_{\rm c}}\def\tp{t_{\rm p}}\def\rc{r_{\rm c}}\def\nc{n_{\rm c}}\def\pcc{p_{\rm c}}  
\def\Msun{M_{\odot \hskip-5.2pt \bullet}}    \def\kms{km s$^{-1}$}  \def\deg{^\circ}   \def\Htwo{H$_2$\ }  \def\fmol{f_{\rm mol}} \def\Fmol{ $f_{\rm mol}$ }  \def\sfu{\Msun~{\rm y^{-1}~kpc^{-2}}} \def\sfuvol{\Msun~{\rm y^{-1}~kpc^{-3}}}\def\log{{\rm log}}
\def\hcc{{\rm H~cm^{-3}}} \def\Hcc{ $\hcc$ }\def\Htot{ H$_{\rm tot}$ } \def\ssfr{\Sigma_{\rm SFR}} \def\vsfr{\rho_{\rm SFR}} \def\sfr{{\rm SFR}}\def\H{{\rm H}}\def\cm{{\rm cm}}\def\kpc{{\rm kpc}} \def\bc{\begin{center}}\def\ec{\end{center}} 
\def\xcounit{\Htwo cm$^{-2}$ [K \kms]} \def\pc{{\rm pc}}\def\My{{\rm My}} \def\kpc{{\rm kpc}}\def\rc{r_{\rm c}} \def\vc{v_{\rm c}}
\def\urho{\Msun \pc^{-3}}\def\urhohtwo{{\rm H_2} \cm^{-3}} \def\nc{n_{\rm c}} 
\def\vexpa{v_{\rm expa}} \def\rbub{r_{\rm b}}\def\x{\times}\def\xfour{\times 10^4}\def\xthree{\times 10^3}\def\xfive{\times 10^5}\def\xfifty{\times 10^{50}}\def\xmtwe{\times 10^{-20}} \def\sigv{\sigma_v} \def\Rbow{R_{\rm bow}} \def\Rzero{R_0} \def\Lcone{L_{\rm cone}} \def\Lsun{L_\odot}\def\Rhii{R_{\rm HII}} \def\nuv{n_{\rm UV}} \def\ni{n_{\rm i}} \def\ne{n_{\rm e}}\def\ar{a_{\rm r}}
\def\Te{T_{\rm e}}\def\Tn{T_{\rm n}}\def\cosh{{\rm cosh}}\def\({\left(} \def\){\right)}\def\[{\left[} \def\]{\right]}\def\Hcc{H cm$^{-3}} \def\Hsqcm{H cm$^{-2}$} \def\L{\mathcal{L}}\def\Rc{R_{\rm c}} \def\rhom{\rho_{\rm m}} \def\rhoc{\rho_{\rm c}} \def\V{V_{\rm rot}} \def\Vp{V_{\rm pat}} \def\vpat{V_{\rm pattern}}\def\red{\textcolor{red}} \def\blue{\textcolor{blue}}
\def\ss{\subsection}\def\sss{\subsubsection}
\def\hcn{H$^{13}$CN $(J=1-0)$}\def\cs{CS ($J=2-1$)}
\def\hcnaces{H$^{13}$CN $(J=1-0)$}\def\csaces{CS ($J=2-1$)}
\def\hcnaste{HCN ($J=4-3$)}
\def\dvdl{d\vlsr/dl}
\def\sgrastar{Sgr A$^*$} 
\def\vex{V_{\rm ex}}
\def\Jybeam{Jy beam$^{-1}$}
\def\cc{cm$^{-3}$}
\def\asec{''\!\!}
\def\degd{^\circ\!\!}
\def\red{\textcolor{red}}
\def\kmsperd{\kms deg$^{-1}$}
\def\htwocc{H$_2$ cm$^{-3}$}
   

\begin{abstract}
Analyzing longitude-velocity diagrams (LVDs) in the \csaces\ and \hcnaces\ molecular lines from the internal release data of the ALMA Central-Molecular-Zone Exploration Survey (ACES) and in the \coth\ line from the Nobeyama Galactic-Centre (GC) survey, we identify six GC Arms as prominent straight LV ridges. In addition to the currently known Arms I to IV, we identify a new inner arm, Arm V, and further highlight the circum-nuclear disc (CND) as Arm VI. Integrated intensity maps of the Arms on the sky suggest that most of the Arms compose ring-like structures inclined from the Galactic plane. We determine the radii (curvatures) of the Arms using the velocity-gradient ($dv/dl$) method, assuming that the arms are rotating on circular orbits at a constant velocity of $\sim 150$ \kms. We show that Arms I and II compose the main ring structure of the CMZ with radii $\sim 100$--120 pc; Arm III is a dense arm 42 pc from the GC; Arm IV is a clear and narrow arm 20 pc from the GC; and Arm V is a faint, long arm of 8.2 pc radius. We show that the circum-nuclear disc (CND) composes the sixth arm, Arm VI, of radius $\sim 2.3$ pc associated with bifurcated spiral fins. We also discuss the association of the 20- and 50-\kms clouds with these Arms. The radii of the arms fall on an empirical relation $R\sim 630 (2/5)^N$ for $N=1$ (Arm I) to 6 (VI), suggesting either discrete rings or a logarithmic spiral with pitch angle $\sim 22\deg$. The vertical full extent of the arm increases with radius and is represented by $z\sim 0.7 (R/1 \epc)^{0.7}$ pc. The tilt angle of the arms from the Galactic plane, or the warping, increases rapidly toward the GC. 
\end{abstract}
 

\section{Introduction}
\label{intro}  

Because we see the Galactic Central Molecular Zone (CMZ) edge-on, its true 3D structure is challenging to decipher and remains substantially uncertain
\citep{morris+1996,henshaw+2016,sof2022,henshaw+2023}.
Kinematic analysis of longitude-velocity diagrams (LVDs), assuming Galactic rotation, offers one way to help to resolve line-of-sight degeneracy 
\citep{bally+87,bally+88,sof1995,sof2022,oka+1998,1999ApJS..120....1T,kruijssen+2015,henshaw+2016,henshaw+2023}. 
Here, we exploit the special behavior of a rotating arm or a ring in the longitude-velocity diagram (LVD), which makes the LV ridge to appear sharpest near its intersection with the rotation axis at $l\sim 0\deg$.
The absorption of line emission against the background continuum helps to distinguish the far and near sides of clouds relative to Sgr A$^*$ \citep{sawada+2004,2017MNRAS.471.2523Y,sof2022}.

Coherent ridges on the LVD suggest that the CMZ is structured into multiple arms. The densest and most prominent LV ridge seen in the \coth\ line is called Galactic-Centre (GC) Arm I, the second is Arm II, and further arms (III and IV) have been proposed \citep{sof1995}. 
There seems to be consensus that Arms I and II compose a ring structure of radius $\sim 100$-120 pc (the "120-pc ring") \citep{sof1995,oka+1998,tok+2019,henshaw+2016,henshaw+2023},
which is understood as due to a large-scale accretion of gas from the outer Galactic disc \citep{2011ApJ...735L..33M,2012ApJ...751..124K,kruijssen+2015,2015MNRAS.453..739K,2017MNRAS.466.1213K,2017MNRAS.469.2251R,sorma+2019,sorma+2020,tress+2020}.

However, the more internal structure of molecular gas within the CMZ inside 120 pc ring appears to have not yet been fully explored, but is believed to consist of a continuous disc, arm/rings or a hole (empty space), or a combination of these. 
Among these the arm/ring structure can be most easily recognized using high-resolution molecular-line mapping data, while the other structures may be obtained as the residual.
For the purpose to map arms/rings of molecular gas, we analyze the data cubes observed with the Nobeyama 45-m telescope in the \coth line, ASTE (Atacama Submm Telescope Experiments) 10-m telescope in \hcnaste, and ALMA (Atacama Large Millimeter/submillimeter Array) in the course of the large project ACES (ALMA CMZ Exploration Survey) in \csaces\ and \hcnaces\ lines 
(ALMA Program, 2021.1.00172.L. PI: S. Longmore et al. in preparation).  
Thanks to the high spatial resolution offered by ALMA, we can much improve arm-identification at the very center inside $|l|\lesssim 0\deg.2 (\sim 30\ \epc)$.
 
The region around Sgr A$^*$ contains many well-known molecular clouds including 
the 50-\kms\ cloud (hereafter 50kmC) and 20-\kms\ cloud (20kmC) \citep{1990ApJ...356..160G,2009PASJ...61...29T,Takekawa17a}, high-velocity compact clouds (HVCCs) \citep{oka+1999,2023ApJ...950...25I}, the circum-nuclear disk (CND) \citep{2001ApJ...551..254W,2018PASJ...70...85T}, and the mini-spirals around \sgrastar\ \citep{2017ApJ...842...94T}. 
We try to understand these innermost structures under a unified view of a molecular disc with arms/rings rapidly rotating in the deep gravitational potential which reaches the specific kinetic energies of $\sim \vrot^2/2 \sim 10^{14}$ erg g$^{-1}$ at $\vrot \sim 100$--150 \kms.

In this study, we combine \coth\ and \hcnaste\ single-dish data with the ACES interferometric mosaics in the \csaces and \hcnaces\ at $|l|\lesssim 0\deg.2$ to determine the internal kinematical structure of the CMZ by constraining the radii and vertical extents of the arms/rings present within $\sim 100$ pc. This will help us to visualize the gaseous structure in the circum-nuclear region, which is essential to carry out the 3D modeling of the CMZ. 

We adopt a Solar galactocentric distance $R_0=8.2$ kpc, close to the recent measurement 
\citep{gravity+2019}, for convenience to compare with the other works. 
The coordinates of \sgrastar\ is taken to be
$(l,b)=(359\deg.944227, -0\deg.046157)=(359\deg 56'39\asec.2, -00\deg 02'46\asec.2)$, 
and the LSR (Local Standard of Rest) velocity is assumed to be $\vlsr=0$ \kms.

\section{Data and analysis}
\label{sec_data}

\ss{Single-dish data}
We used the archival data cube of the \coth\ line emission at 110.27 GHz taken from the CMZ survey obtained using the Nobeyama 45-m telescope \citep{tok+2019}.
The data cube had sampling grids of $(7\asec.5 \times 7\asec.5 \times 2 \ekms)$ with an effective resolution of $16.7''$ which yielded an rms noise of $\sim 0.15$ K in brightness temperature, $\Tb$.
We also used an archival data cube of the HCN ($J=4-3$) line emission at 354.5 GHz from the GC survey with the ASTE 10-m telescope \citep{2018ApJS..236...40T}, which had ($8\asec.5 \times 8\asec.5\times 2$ \kms) grids with an effective angular resolution of $24''$ and rms noise of 0.14 K in $\Tb$.   
  
\ss{ACES} 

The molecular-line cubes from ALMA used in this work were taken from the internal release version of the 12m+7m+TP (total power)-mode data from the ALMA cycle 8 Large Program "ALMA Central Molecular Zone Exploration Survey" (ACES, 2021.1.00172.L; Longmore et al. in preparation). ACES observed the CMZ in ALMA Band 3, covering a frequency range of $\sim$ 86 - 101~GHz across six spectral windows of varying spectral resolution and bandwidth.

The ALMA pipeline calibrated measurement sets were produced using CASA 6.4.1.12, and all of the 12m and 7m data were re-imaged using CASA 6.4.3-2. In general the imaging parameters were the same as those used by the pipeline to produce the delivered data, but there were instances where parameters were changed, in particular to fix divergent channels, and to undo size mitigation performed by the default pipeline parameters. 

We also found that the ALMA pipeline often did not perform optimally when identifying the line-free channels in the data, resulting in residual continuum emission after performing the continuum subtraction. The pipeline also used a polynomial fit order of 1, which often resulted in poor baselines after continuum subtraction, particularly in the narrow spectral windows which are often filled with broad line emission. To fix these issues, we first re-ran the continuum subtraction using a fit order of 0, and then additionally used \texttt{statcont} \citep{2018A&A...609A.101S} to remove the residual continuum emission. After re-imaging and subtracting the continuum, we combined the 12m, 7m, and TP data using the \texttt{feather} task in CASA. We first combined the 7m and TP cubes, then combine this 7m+TP cube with the 12m data.

The ACES coverage is split into 45 individual sub-mosaics, each with approximately 150 pointings. For each line/SPW, we used the \texttt{radio-beam} and \texttt{reproject} Python packages to convolve all sub-mosaics to a common beam, and then project on to the full ACES footprint. The resulting cubes provide a contiguous mosaic of the CMZ.

For this work, we used the cubes in the \cs\ line at a frequency 97.9810 GHz with FWHM angular resolution $2\asec.21$ of the synthesized beam and rms noise of 0.0038 \Jybeam\ (0.10 K)
with velocity channel increment of 1.45 \kms, and \hcn\ line at 86.3399 GHz with resolution $2\asec.72$ and rms noise 0.0046 \Jybeam\ (0.10 K) with a velocity channel increment 0.88 \kms.
The intensity scales are \Jybeam\ (1 \Jybeam$=26.1$ and 22.2 K in brightness temperature at 98 and 86 GHz, respectively).

The cubes cover the CMZ at $-0\degd.6 \lesssim l \lesssim +0\degd.9$ and $-0\degd.3 \lesssim b \lesssim +0\degd.1$, and velocity ranges were $-220 \le \vlsr \le +220$ \kms~and $-150 \le \vlsr \le +150$ \kms, respectively, with spatial and velocity grids of ($0\asec.5 \times 0\asec.5\times 0.15$ \kms).
We further cut out a more interior region at  $-0\degd.25 \le l \le +0\degd.15$ and $-0\degd.1 \le b \le +0\deg$ for a detailed analysis of the circum-nuclear region centered on Sgr A$^*$.
 
\ss{Definition and identification of the Arms}

The goal of this paper is to identify spiral arms, rings, and/or segments thereof (hereafter "arms") in the CMZ that rotate within the Galactic gravitational potential, and to quantify their galacto-central radii or curvatures.
An "arm" is here defined by a tilted ridge on the LVD that extends straightly for $\sim \pm 100$ \kms, indicating a coherent ring-like structure rotating at $\sim 100$--150 \kms.
 
We used the \coth\ line from the single dish in order to trace the arms and extended structure in the entire CMZ.
The lines \csaces\ and \hcnaces\ from the ALMA cubes were chosen to trace the innermost arms, which are supposed to consist of relatively denser molecular gas \citep{2015PASP..127..299S}, for a complementary analysis to the \coth\ line from the single dish.

To identify the arms, we exploit a special property of the LVD that, due to the degeneracy of the radial velocity, arms appear clearest and brightest near their intersection with the axis of rotation at $l\sim 0\deg$.
Although some arms overlap at $l\sim 0\deg$, they can be distinguished from one another because their ridges run at different tilt angles \citep{sof2006}.
Moreover, fortunately, non-circular motions overlap in most spiral arms, which yields displacements of $\vlsr$ from zero near $l\sim 0\deg$, so in most cases the degeneracy is resolved, or the LV ridges do not overlap at the intersection point.

Note that there are four ways to display LVD as follows.\\
Type 1: channel LVD at every latitude as one channel of the cube;\\
Type 2: total average over all latitudes;\\
Type 3: average over selected latitudes; and  \\
Type 4 LVD: maximum intensity (peak $\Tb$) across latitudes. \\
They are all used in this work depending on the purpose of each figure. 

Nevertheless, especially in the central region, the LV arms are often buried in bright extended features.
In order to abstract such buried arms by subtracting more extended structures, we apply the IMSHIFT relieving technique described in Appendix \ref{aprelief}.
This method is particularly useful for the single-dish data with lower angular resolution.

By tracing the tilted LVD ridge (LVR), we measure the longitudinal velocity gradient $dv/dl$, which is related to the radius or the curvature of the arm as explained in detail in section \ref{dvdl}.
For identifying a spiral arm or a ring, this method is more sensitive and accurate than measuring the terminal velocity ends at the farthest longitudes, because the inherent width and velocity dispersion make it difficult to define the exact terminal longitude and velocity.

\ss{LVD "Arms"} 

\begin{figure}   
\begin{center}  
Arm I: LVD (Type 1): \coth: 45m\\
\hskip 5mm\includegraphics[width=8.5cm]{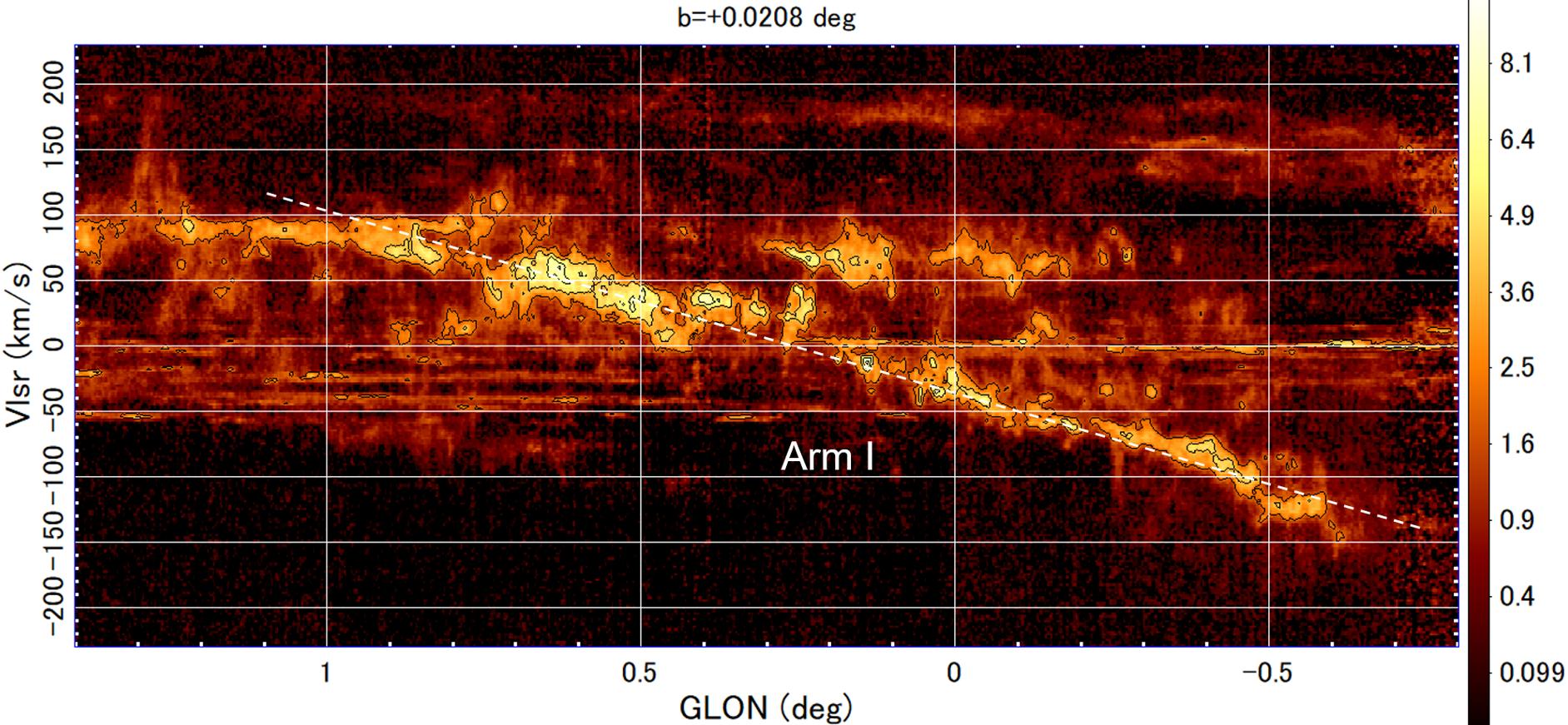} 
\end{center}
\caption{Longitude-velocity diagram (LVD) showing the Galactic-Centre (GC) Arm I  in 
\coth\ from the 45-m telescope at a representative latitude (Type-1 LVD). Color bars indicate the brightness temperature in K. {Alt text: LVD (Longitude-velocity diagram) of Arm I by 45m telescope.}}
\label{ArmI}	 

\begin{center}   
Arm II; LVD (Type 1); \coth; 45m\\
 \includegraphics[width=8.5cm]{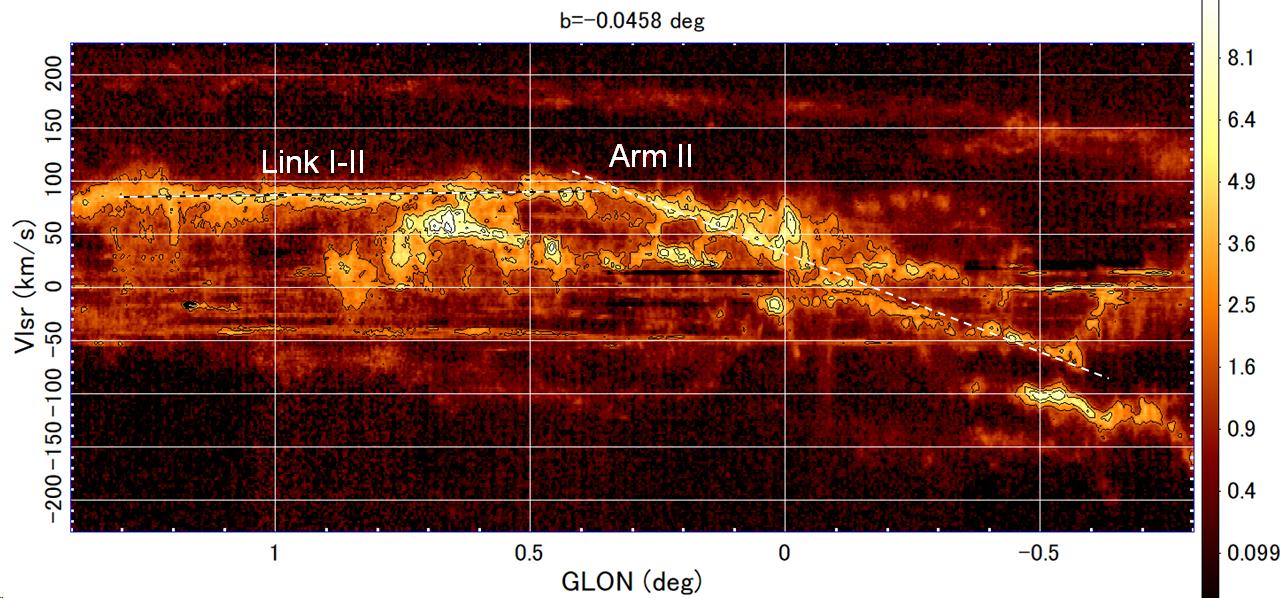}  
\end{center}
\caption{Arm II on the LVD in \coth\ from the 45-m telescope at a representative latitude that is indicated in each panel. The horizontal line named as Link I-II indicates possible connection of Arm II with the Arm I as well as to the outer disc. Color bars indicate the brightness temperature in K.
 {Alt text: LVD of Arm II by 45m.}}
\label{ArmII}	

\begin{center}   
Arm III; LVD (Type 1); \coth; 45m\\
\includegraphics[width=8.5cm]{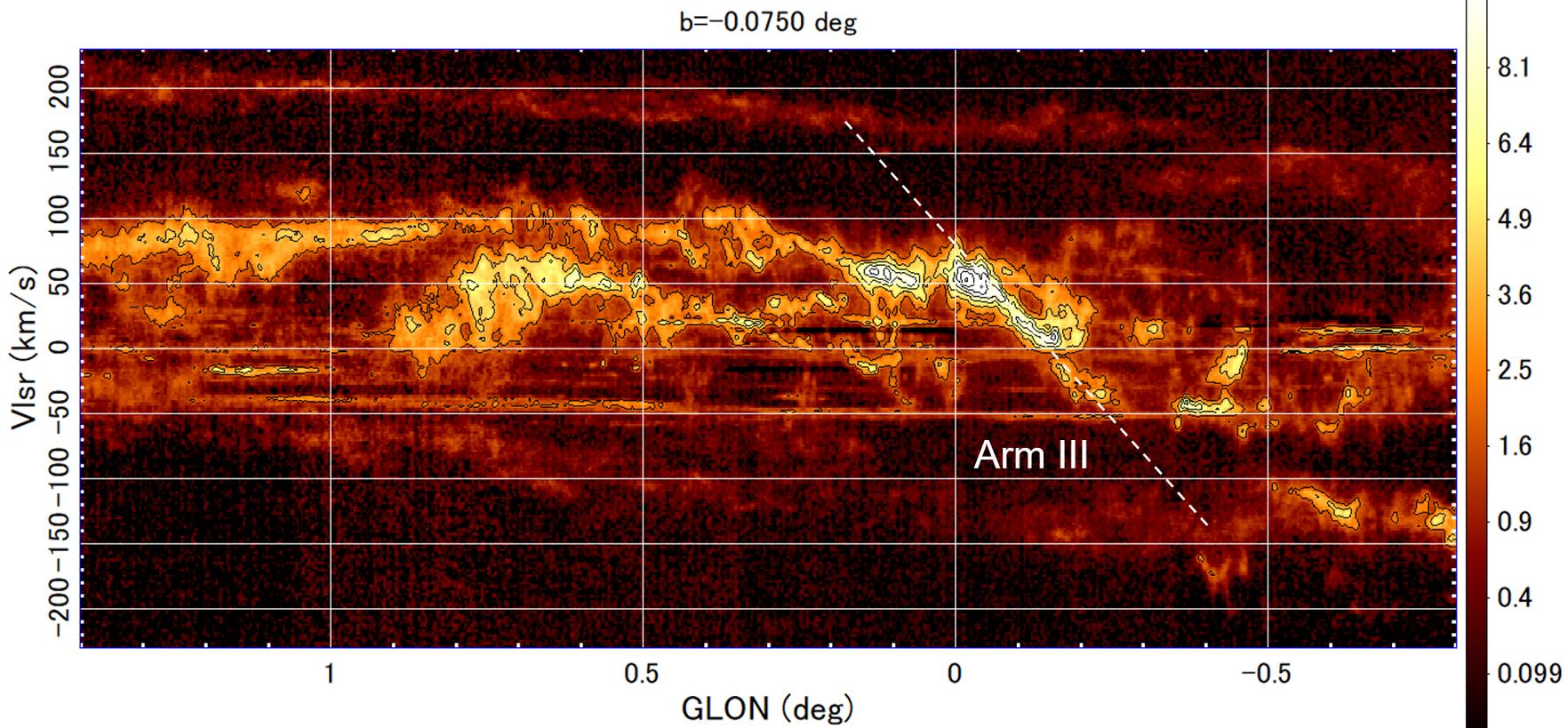}\\    
45m, Relief LVD (Type 1), \coth\\
\includegraphics[width=8.5cm]{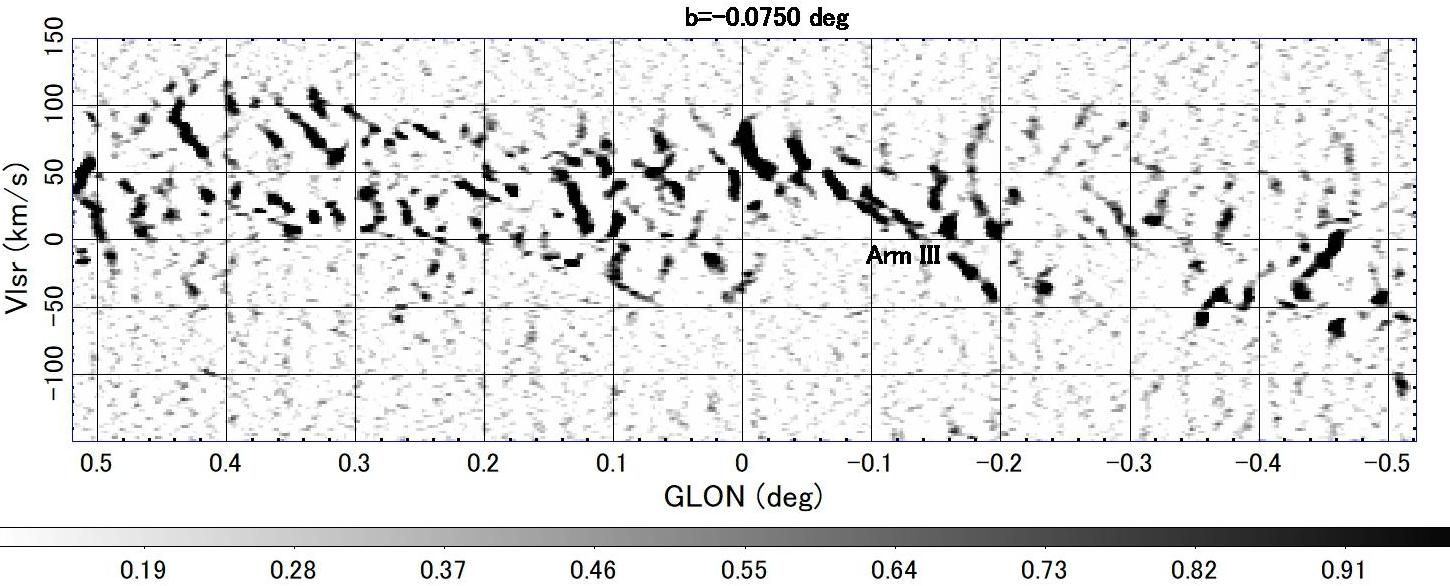}  
\end{center}
\caption{[Top] Arm III LVD in \coth\ from the 45-m telescope at representative latitude $b=-0\degd.075$. 
[Bottom] Same, but relieved LVD (see Appendix \ref{aprelief} for the relieving method) for the central region. 
Color bars indicate the brightness temperature in K.  {Alt text: LVD of Arm III by 45m.}}
\label{ArmIII}	 
\end{figure}

\begin{figure}   
\begin{center}   
Arm IV; LVD (Type 1); \coth; 45m\\
\includegraphics[width=8.5cm]{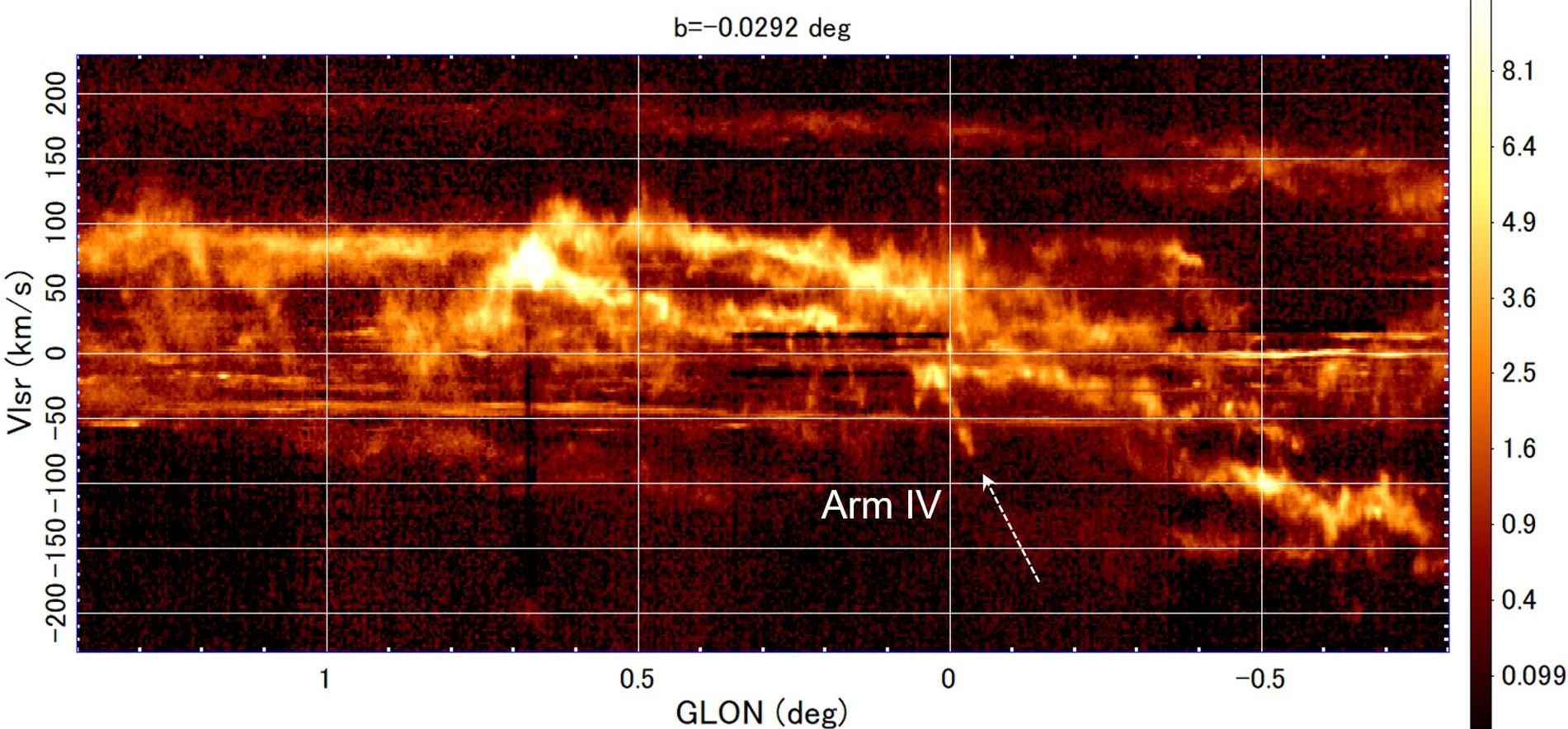}\\    
Arm IV; \coth; LVD (Type 3) Avr. $b=-0\degd.11$ to $-0\degd.01$; 45m\\
\includegraphics[width=8.5cm]{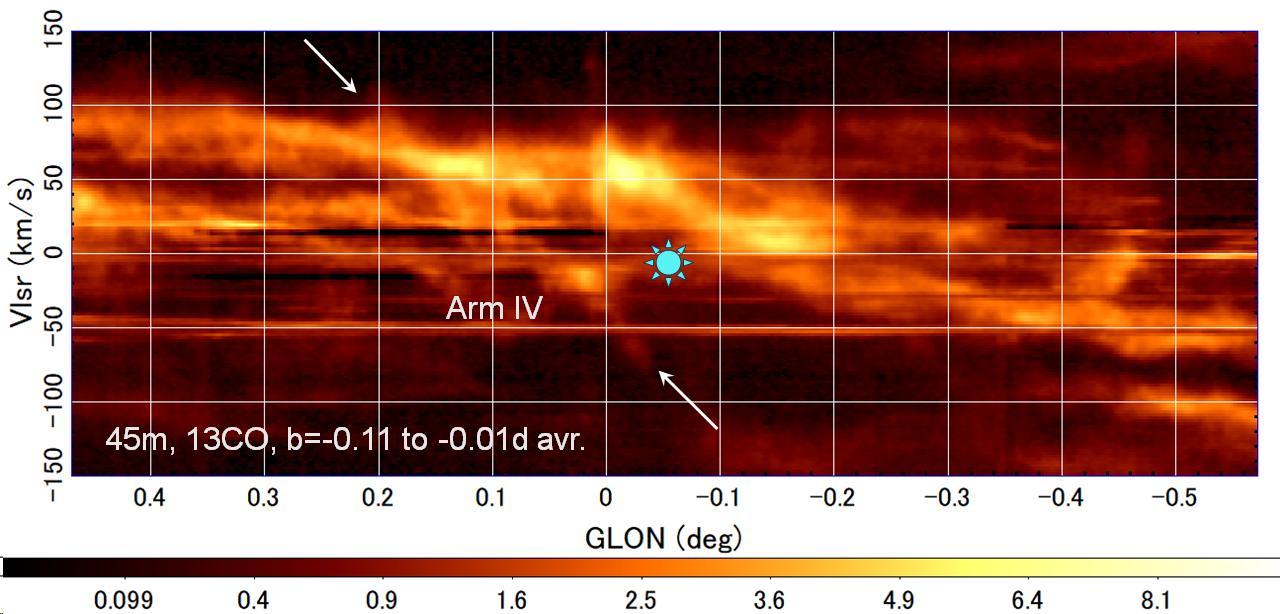} \\
Arm IV; \coth; 45m Relief LVD (Type 1)\\
\includegraphics[width=8.5cm]{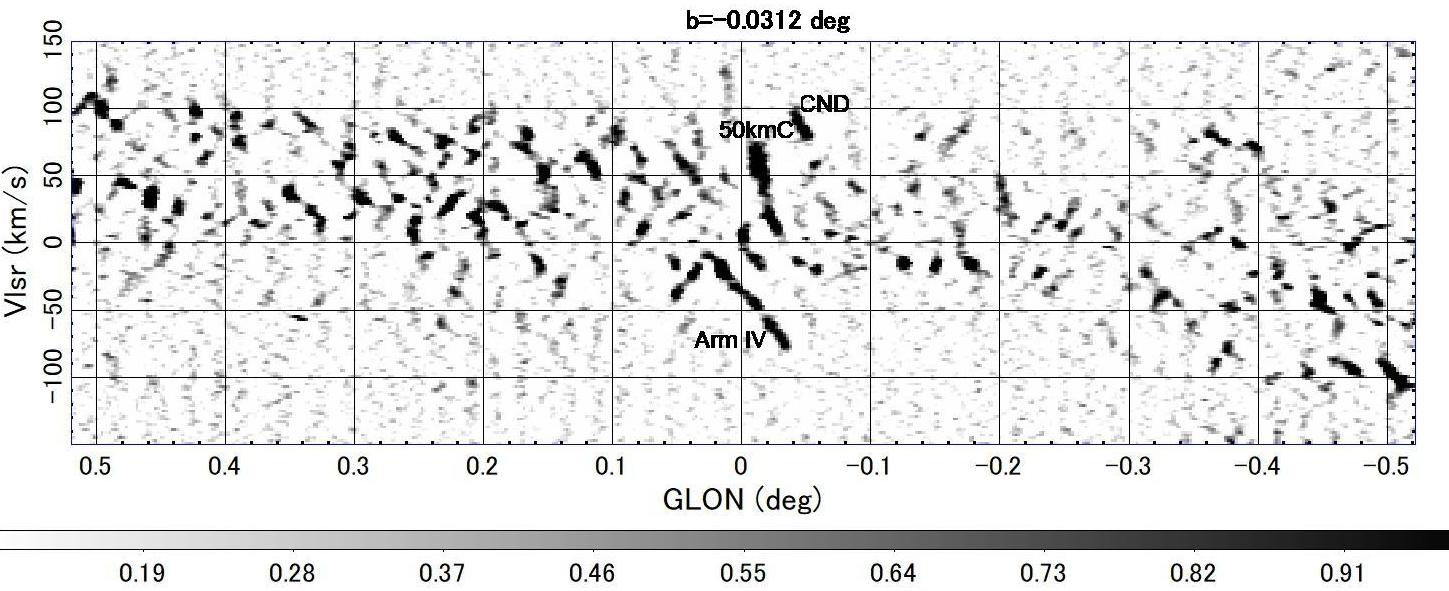}  
\end{center}
\caption{Arm IV: [Top] LVD (Type 1) in \coth\ from the 45-m telescope at representative latitude, $b=-0\degd.03$.
[Middle] LVD averaged between $b=0\degd.11$ and $-0\degd.01$ (Type-3 LVD), showing the entire arm including positive-velocity extensions.
[Bottom] Relieved LVD (Type 1) at $b=-0\degd.03$, showing the negative velocity ridge. Color bars indicate the brightness temperature in K.
The blue symbol is \sgrastar. 
 {Alt text: LVD of Arm IV by 45m.} }
\label{ArmIV}	 
\end{figure}

\begin{figure}   
\begin{center}     
Arm V; LVD (Type 1) at $b=-0\degd.056$; \coth; 45m Relief\\ 
\includegraphics[width=8.5cm]{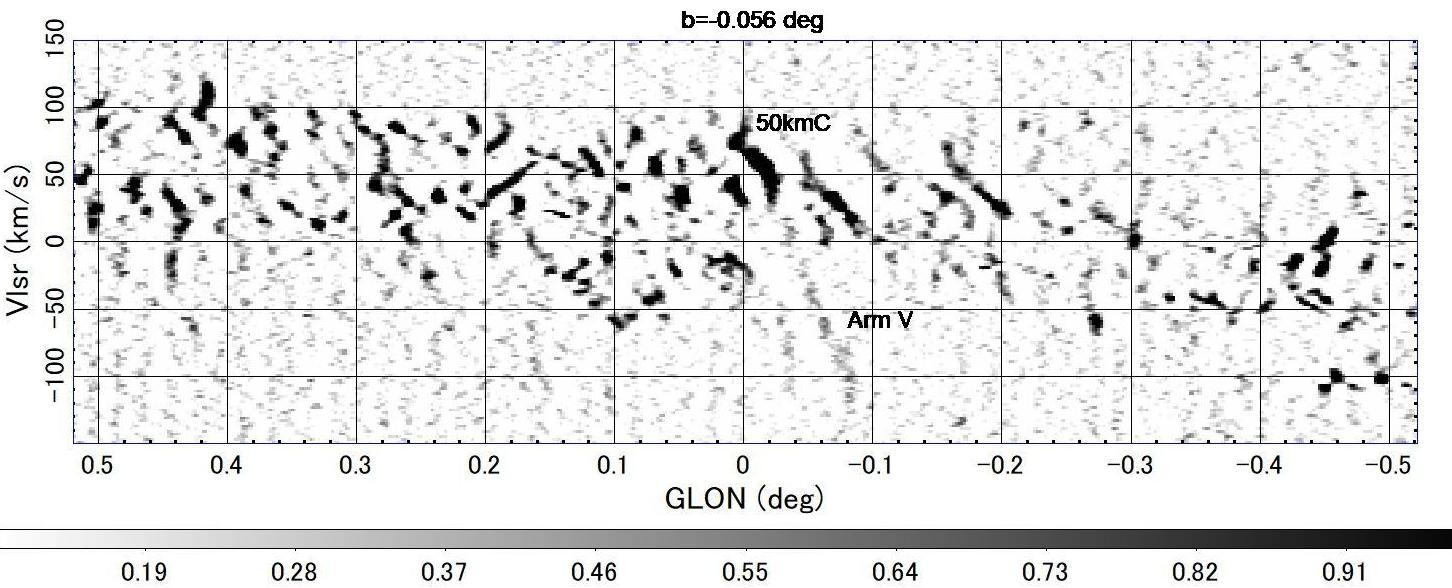}  \\
Arm IV \& V; LVD (Type 2) avr. $-0\degd.1$ to $0\deg$; \coth; 45m Relief\\
 \includegraphics[width=8.5cm]{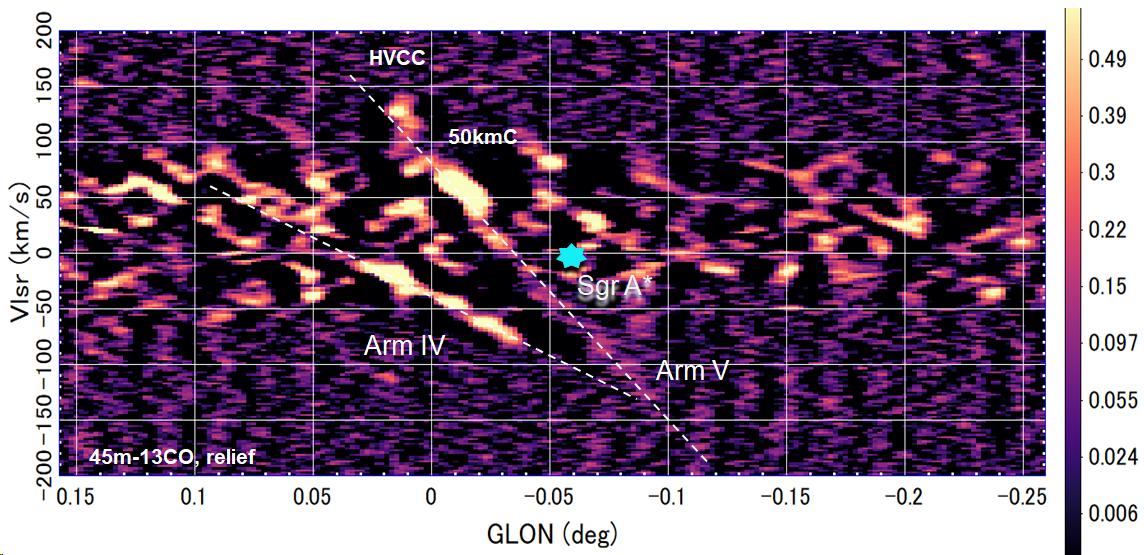} \\ 
Arm V; LVD (Type 3) avr. $b=-0\degd.048$ $\pm 5$;  \csaces; ACES \\
\includegraphics[width=8.5cm]{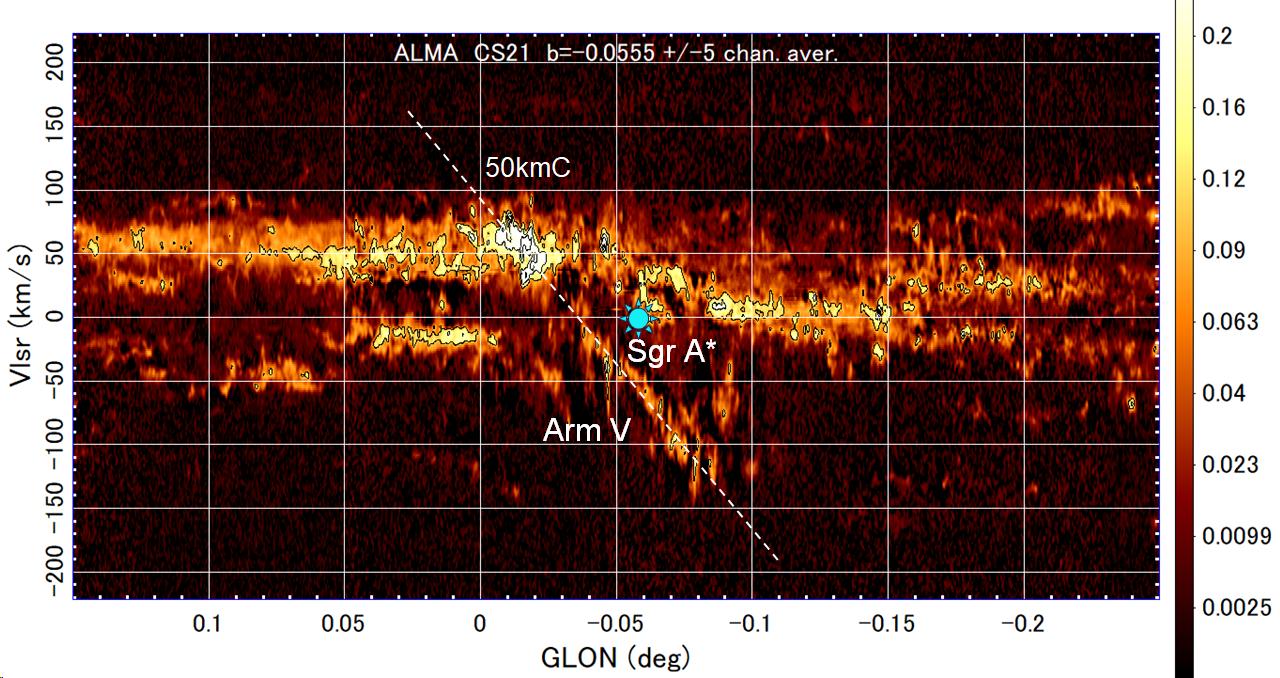}   
\end{center}
\caption{Arm V: [Top] Releaved LVD (Type 1) in \coth\ from the 45-m telescope at representative latitudes $b=-0\degd.056$.
[Middle] Relieved LVD (Type 3) in \coth\ averaged between $-0\degd.1$ to $0\deg$.
[Bottom] LVD (Type 3) in \csaces\ averaged over $\pm 5$ ($\pm 4$ \kms) channels around $b=-0\degd.0555$ by ALMA.
 {Alt text: LVD of Arm V by 45m.}}
\label{ArmV}	 
\end{figure}

\begin{figure}   
\begin{center}    \vskip -5mm
Arm VI = CND\\   
LVD (Type 1) Relief $b=-0\degd.048$; \coth\ Nobeyama 45m \\
 \includegraphics[width=8.5cm]{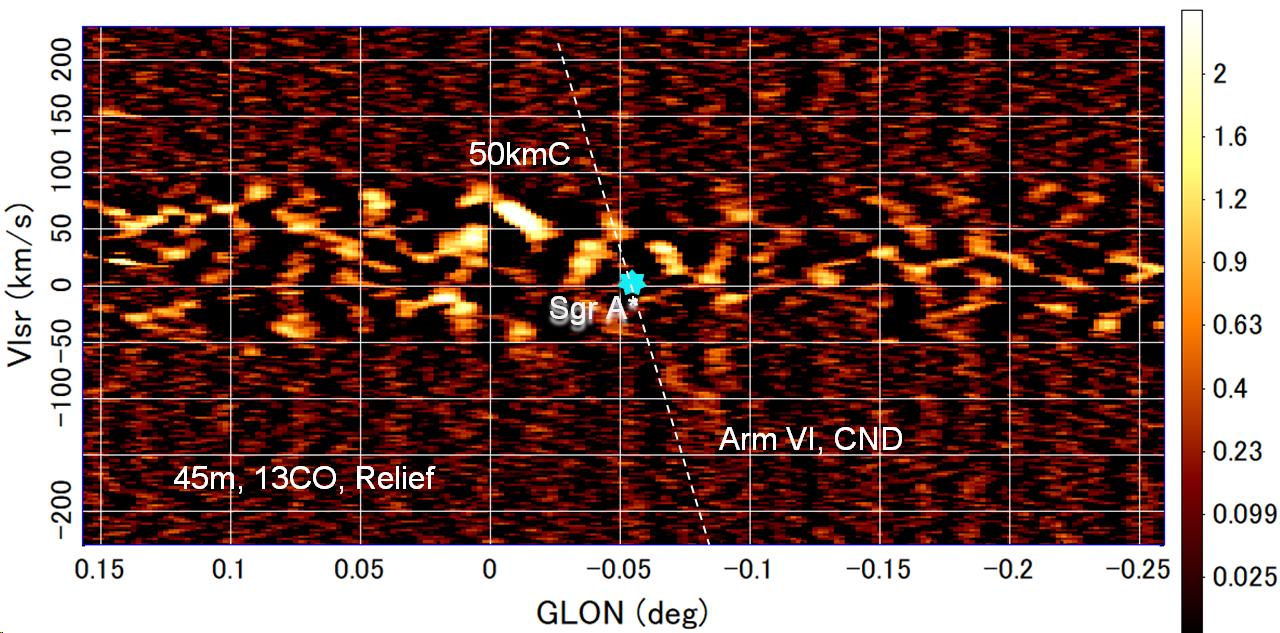}  \\ 
LVD (Type 1) $b=-0\degd.048$; \csaces\ ACES \\ 
 \includegraphics[width=8.5cm]{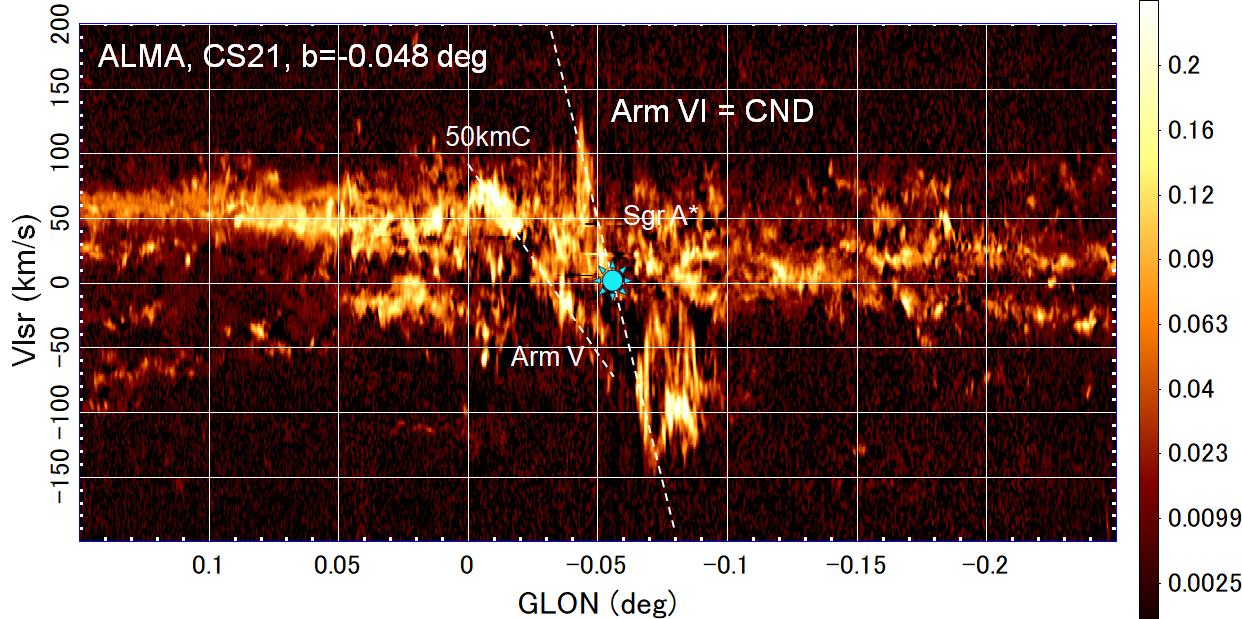}\\
LVD (Type 3) Avr. $b=-0\degd.048\ \pm 25$ channels; \hcnaces\ ACES \\ 
 \includegraphics[width=8.5cm]{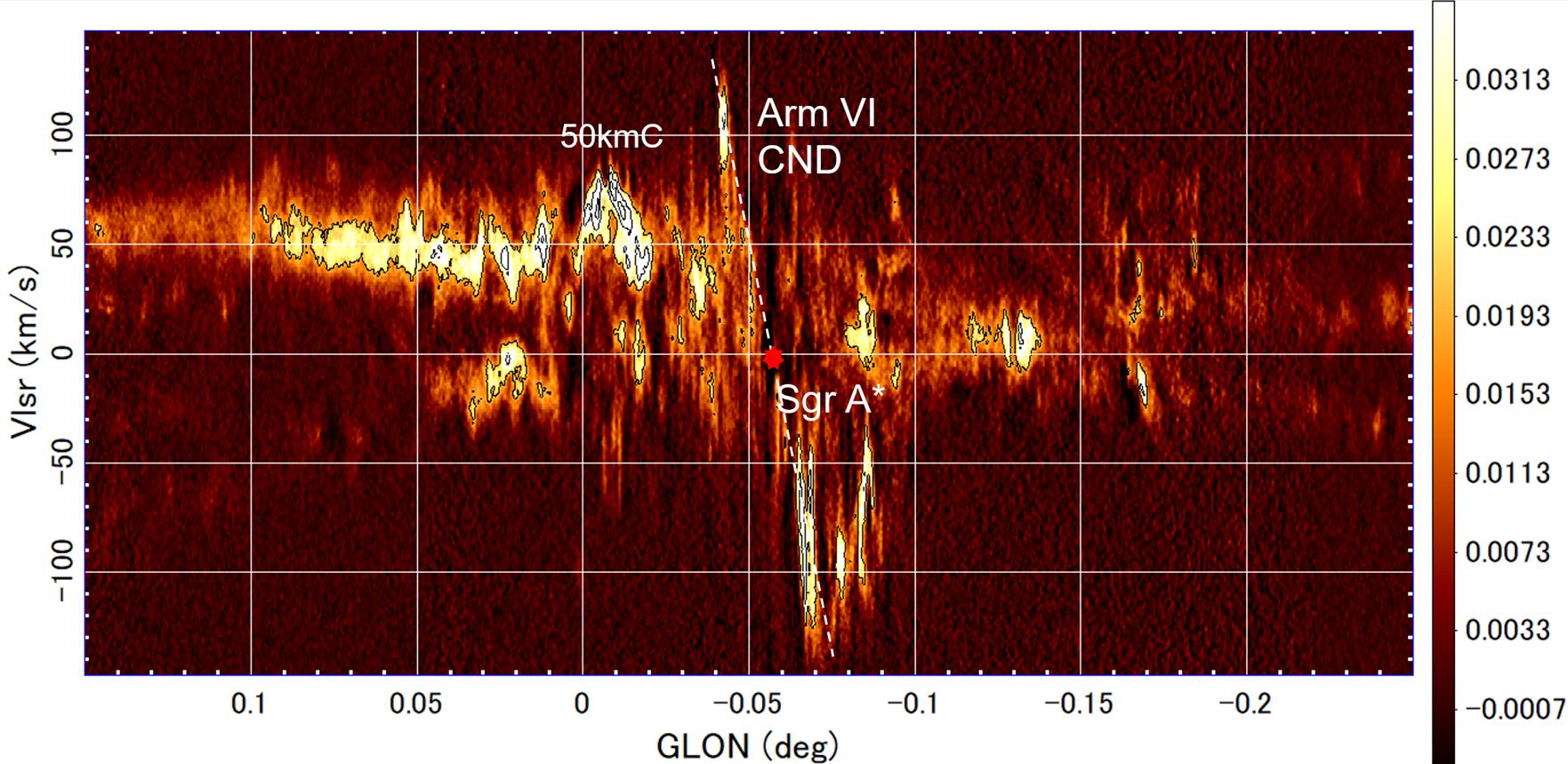}   \\
LVD (Type 2) Avr. all; Non relief; \hcnaste\ ASTE 10m \\   
\includegraphics[width=8.5cm]{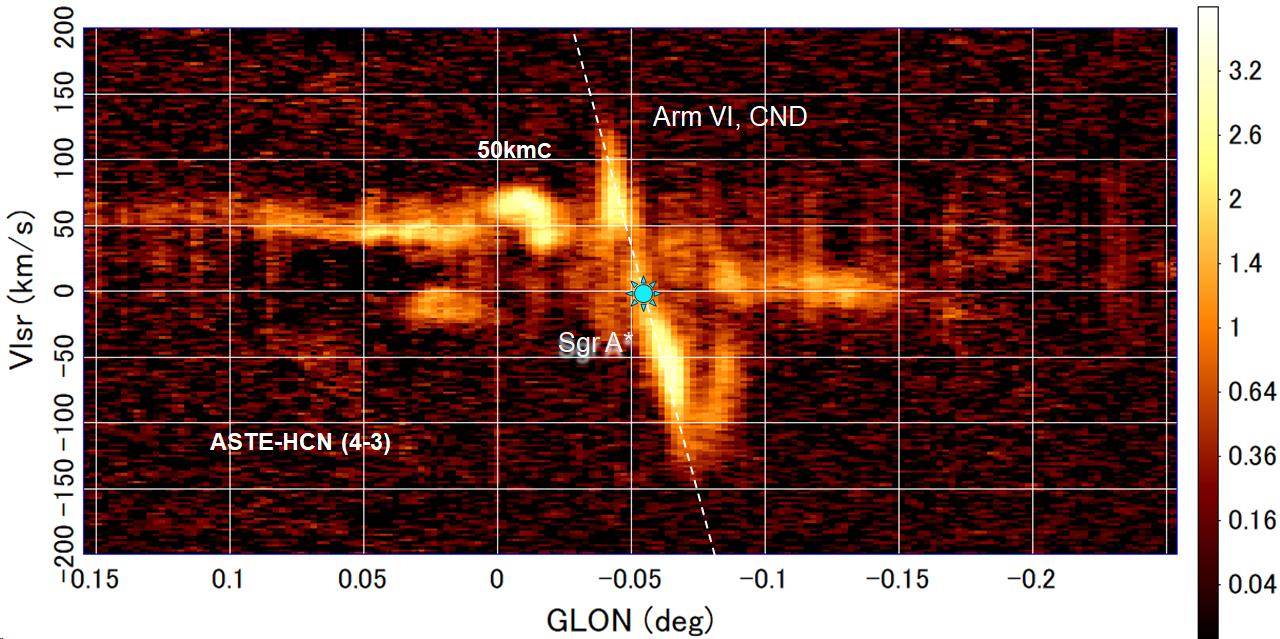}\\
\end{center}
\caption{Arm VI:
[Top] 45m LVD (Type 1, relief) in \coth\ at $b=-0\degd.048$. 
[2nd] ALMA \csaces\ LVD (Type 1) from ACES internal release data.
[3rd] ALMA \hcnaces\ LVD (Type 3).
[Bottom] ASTE \hcnaces\ LVD (Type 2) Averaged over all latitudes.   
Color bars indicate the brightness temperature in K for top panel (45m) , and in Jy beam$^{-1}$ for others (ACES).
 {Alt text: LVD of Arm VI by 45m.}}
\label{ArmVI}	 
\end{figure}

We search for arms in all channels of the cubes as LV ridges with extents over +/- 100 km/s.  
In figures \ref{ArmI} to \ref{ArmVI} we show the LVDs at representative latitudes of the thus recognized GC Arms in the \coth\ line from Nobeyama 45 m, \hcnaste\ from ASTE 10 m, and \csaces\ and \hcnaces\ from ALMA. 
Details of the data are shown in individual panels of the figures and captions.
The identified arms are marked by the white dashed lines or arrows with corresponding names.

The arms are often too thin and faint to be recognized on a single or an averaged LVD.
In order to convincingly trace such arms in more detail, we also use channel LVDs as presented in
Appendix \ref{apchannel}. 

\begin{figure}    
\begin{center} 
All Arms I to VI; LVDs ACES ($b\sim -0\degd.3$ to $+0\degd.2$)\\ 
\csaces\, LVD (Type 2) Mean \\
\includegraphics[width=8.5cm]{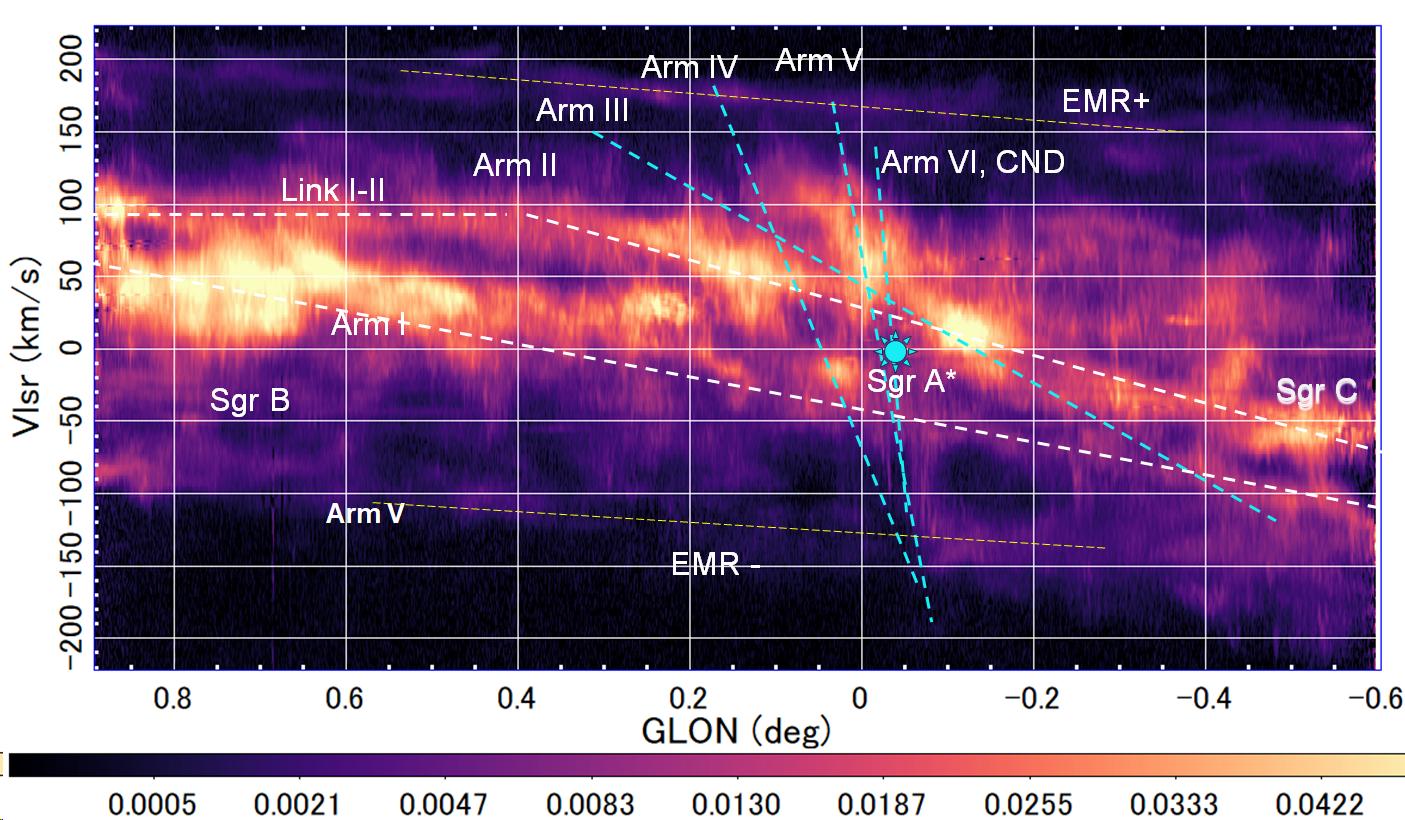}   \\
\csaces\, LVD (Type 4) Max in $b\sim -0\degd.3$ to $+0\degd.2$ \\
\includegraphics[width=8.5cm]{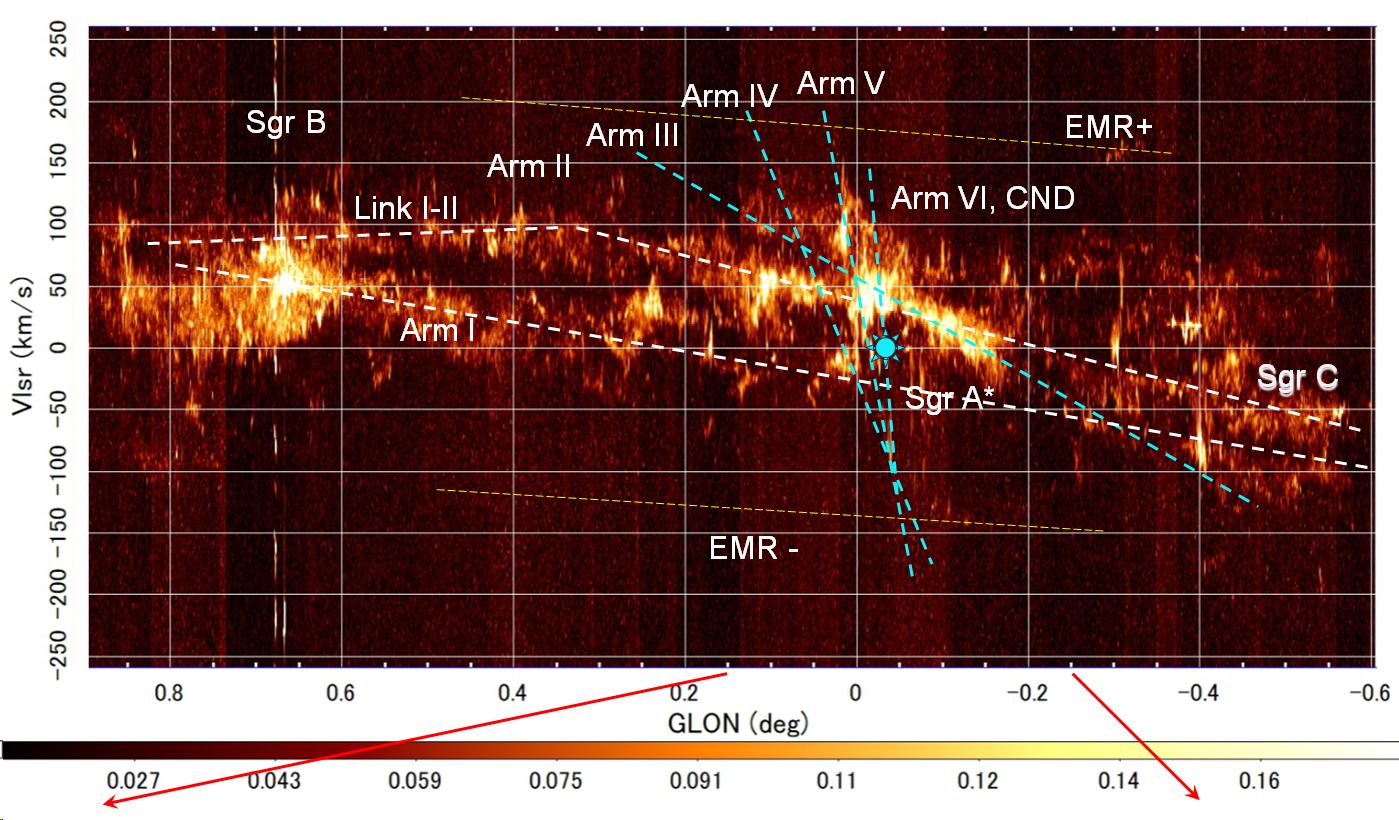}   \\
\csaces, LVD (Type 4) Max\\
\includegraphics[width=8.5cm]{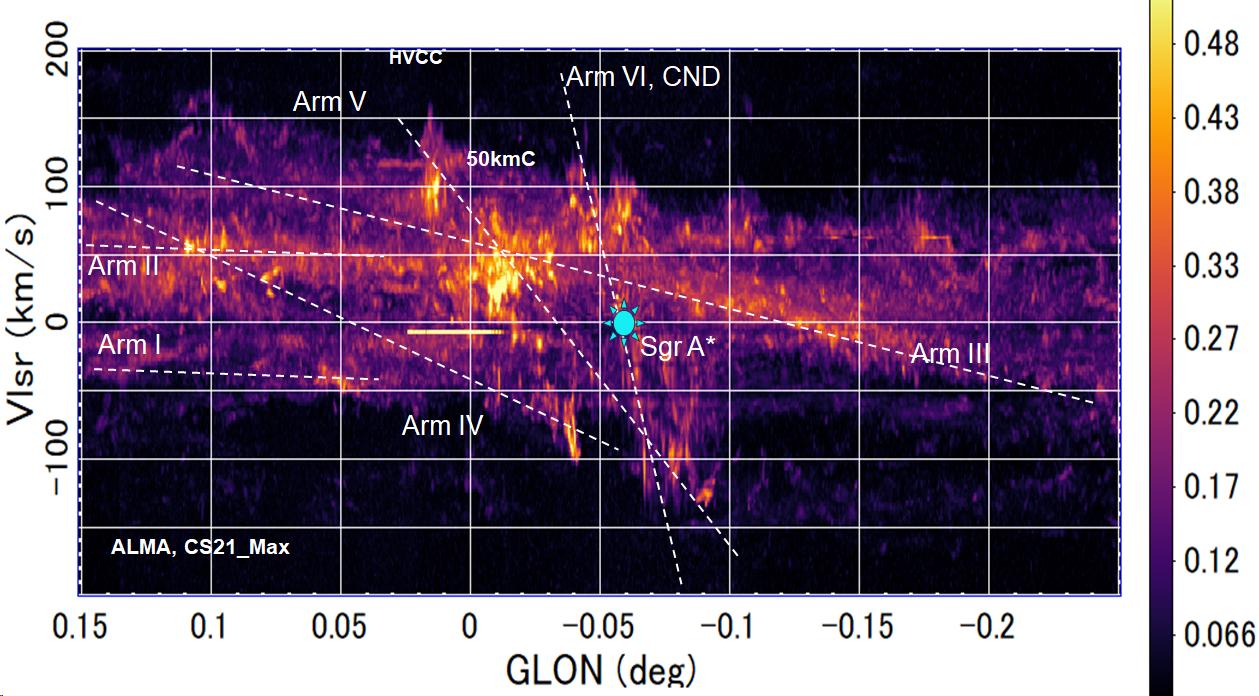} \\ 
\hcnaces, LVD (Type 4) Max \\
\includegraphics[width=8.5cm]{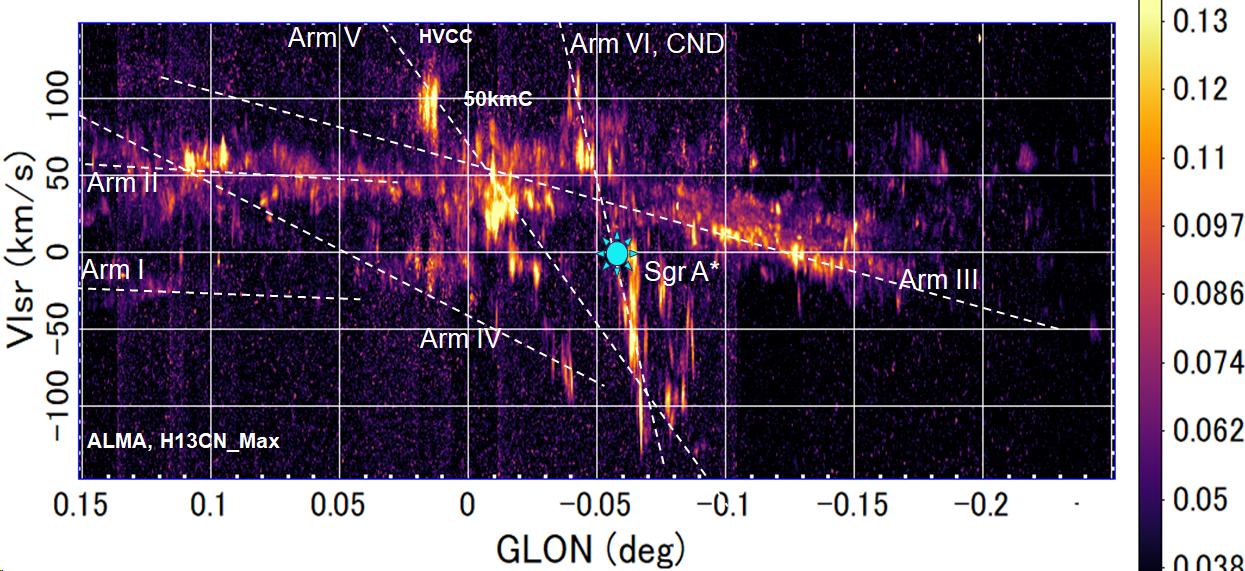}  
\end{center}
\caption{[Top] LVD of latitudinal mean \cs\ intensity. 
The identified Arms are indicated by the dashed lines.
EMR stands for expanding molecular ring (+/- for positive and negative $\vlsr$, respectively).  
[2nd] LVD of maximum (peak) intensity along each latitude in \csaces.
Arm III shows up clearly in this map.
[3rd] Same, but the central region.
[Bottom] Same, but in \hcnaces\. 
Color bars indicate the intensity scale in Jy beam$^{-1}$.
 {Alt text: LVDs of Arm I to VI by ACES.}} 
\label{aces_lv_max}	
\end{figure}

In figure \ref{aces_lv_max} we summarize all the identified arms by dashed lines superposed on the maximum-intensity LVD from ALMA in the whole mapped area by ACES in \csaces\ from $b=-0\degd.3$ to $+0\degd.2$.
The middle and bottom panels show the same in \csaces\ and \hcnaces, but for the central region 
from $l=-0\degd.25$ to $0\degd.15$, and $b=-0\degd.1$ to $0\deg$. These LVDs were made by creating maximum intensity projections along the latitude axis of the cubes using the \texttt{spectral-cube} Python package.

\ss{Vertical profiles of the Arms}
\label{bprofile}

\begin{figure*}   
\begin{center}   
\includegraphics[height=4.5cm]{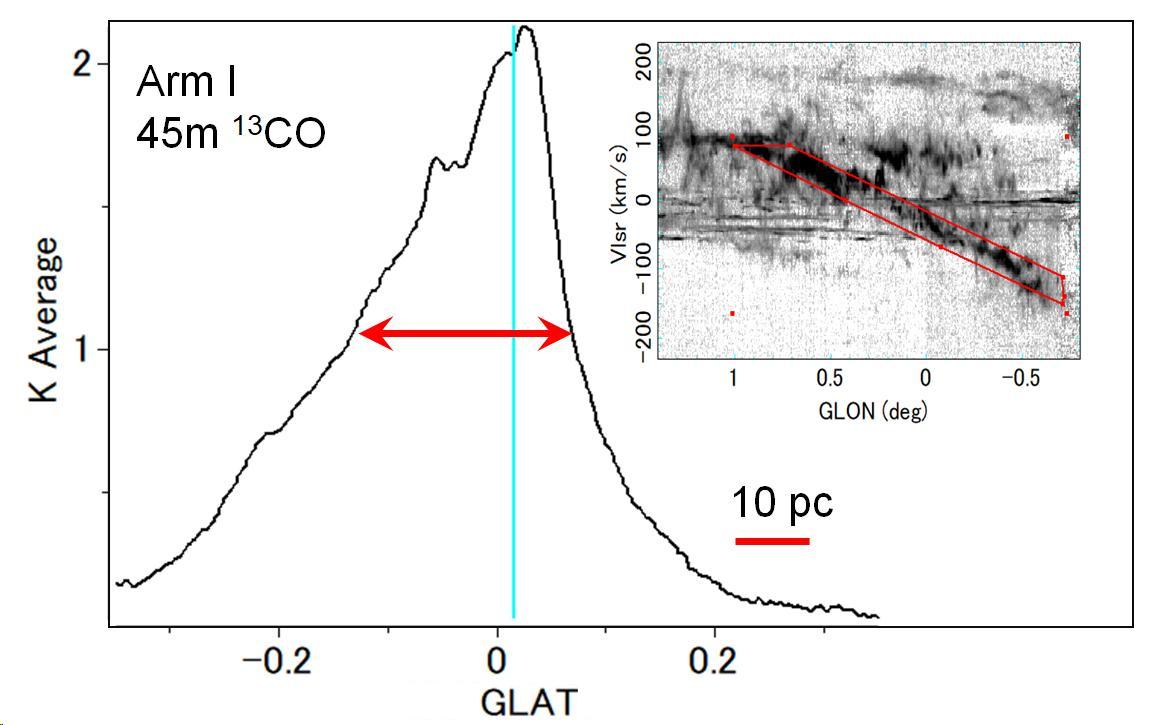} 
\includegraphics[height=4.5cm]{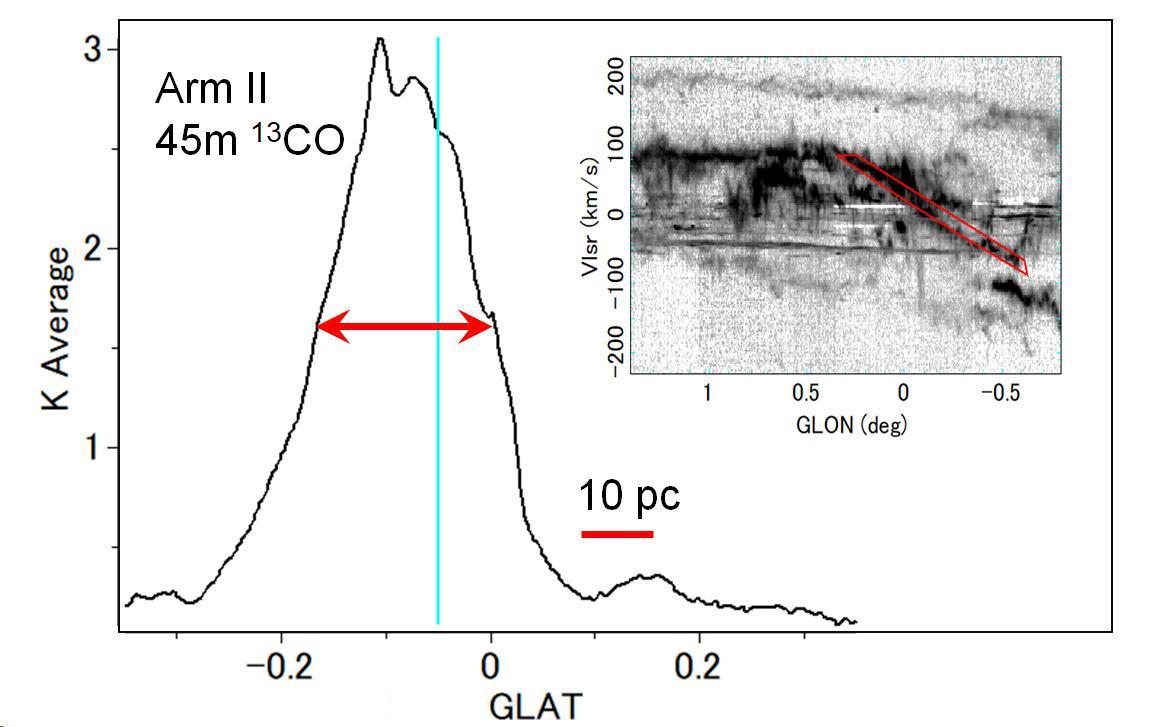} \\
\vskip 3mm
\includegraphics[height=4.5cm]{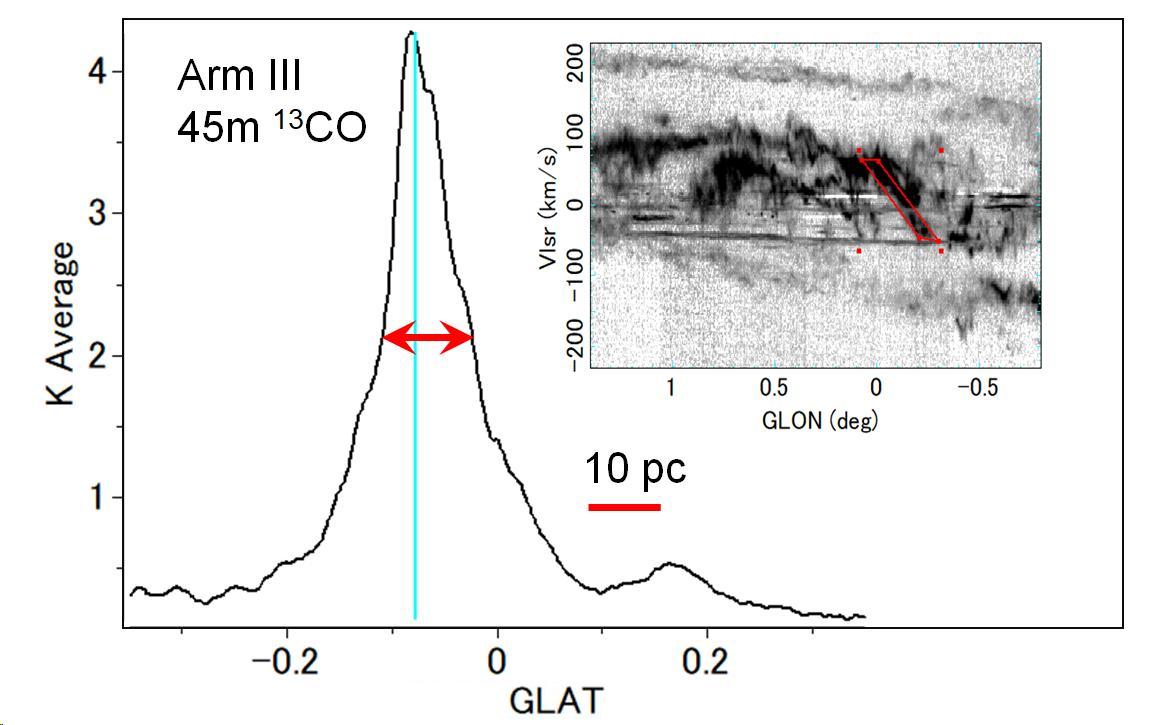} 
\includegraphics[height=4.5cm]{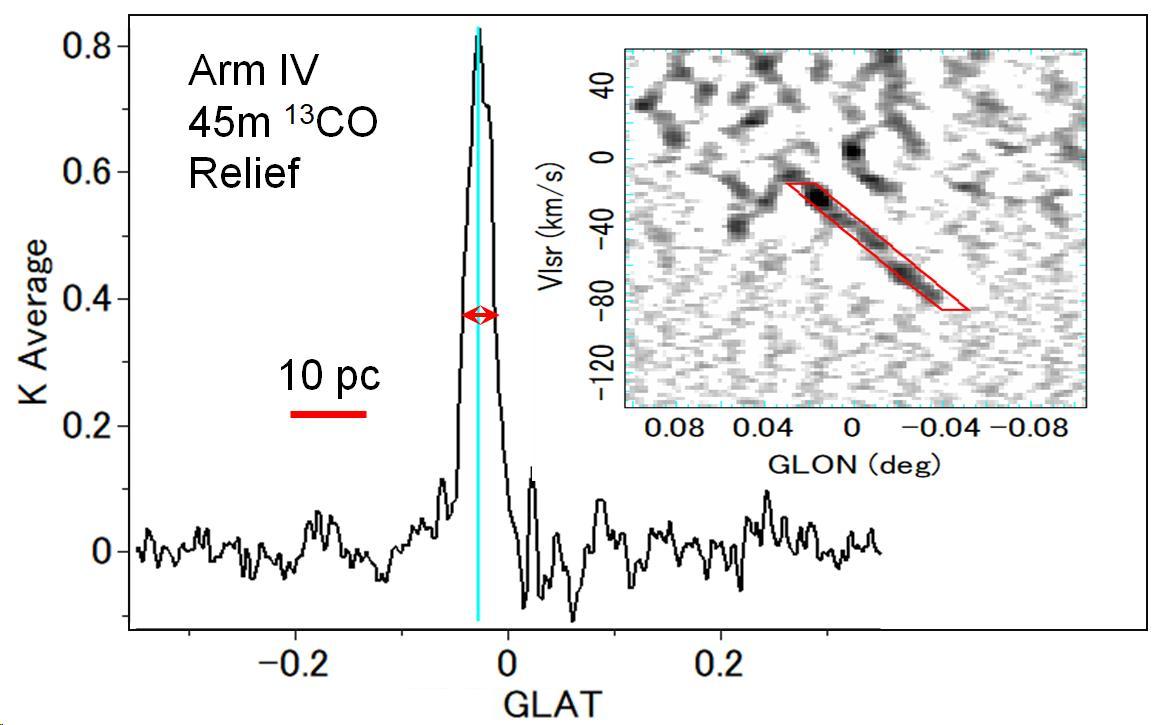} \\
\vskip 3mm
\includegraphics[height=4.5cm]{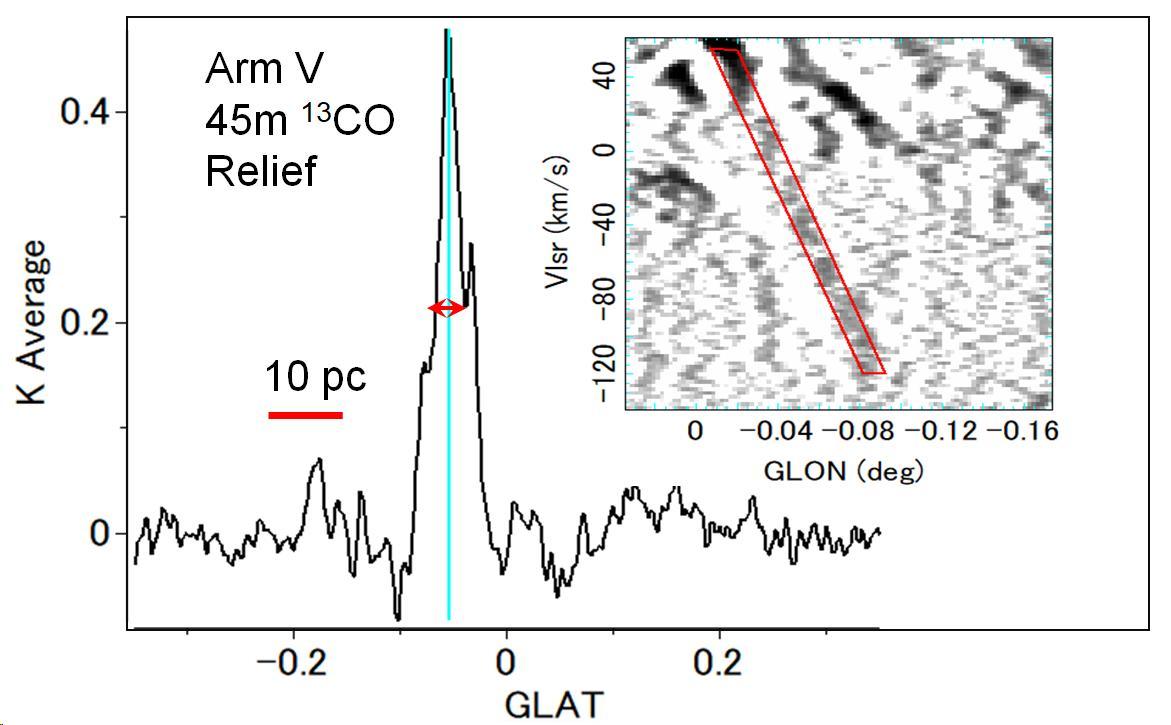}  
\includegraphics[height=4.5cm]{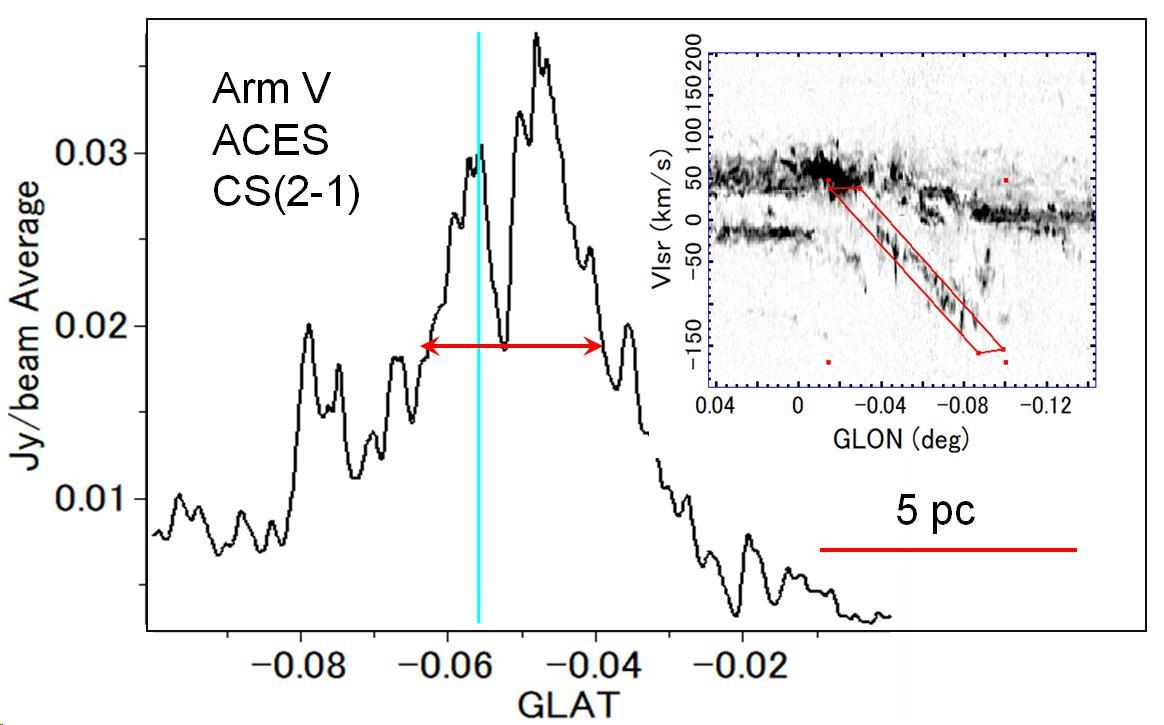}   \\
\vskip 3mm
\includegraphics[height=4.7cm]{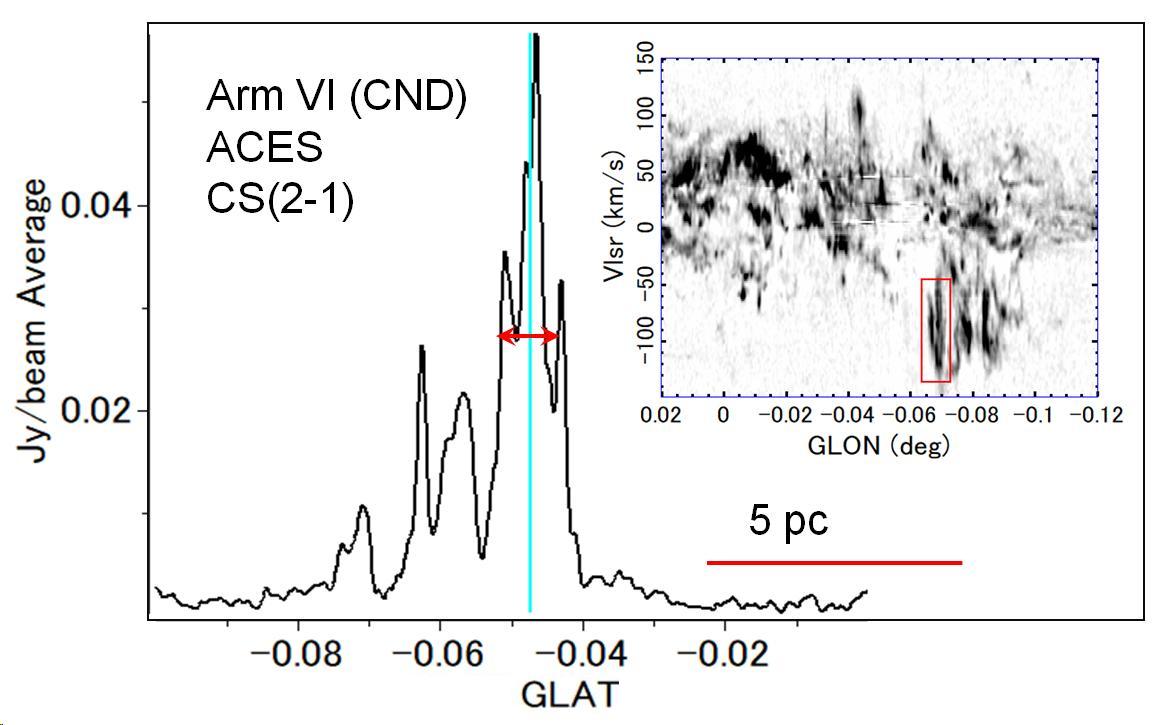}     
\hskip 10mm
\includegraphics[height=4.2cm]{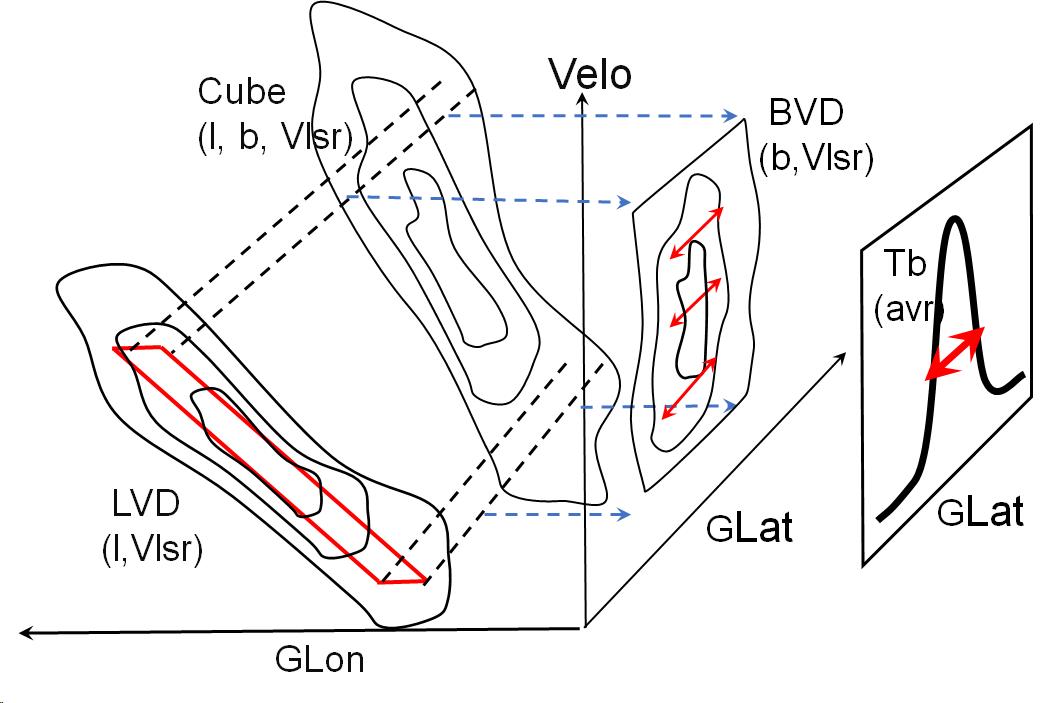}   
\end{center}
\caption{{Latitudinal profiles in the \coth\ and \csaces\ line emissions of Arm I to VI averaged in the regions along the LV ridges as marked by the red boxes in the insertions. The latitudinal full width at half maximum (FWHM), $z$, was measured as the red arrows.  
{The bottom-right insertion explains the procedure to obtain the latitudinal profile using SAOimage (ds9).}}
 {Alt text: Latitude profiles of Arm I to VI.}}
\label{arm_b_prof}	 
\end{figure*}

Figure \ref{arm_b_prof} shows latitudinal intensity profiles of the Arms 
{averaged along the LV ridges of the individual arms inside the red boxes as shown in the inserted LVD in each panel. 
We then measure the full width at half maximum (FWHM) of the profile, as indicated by the red arrow, and define the width as the "full vertical extent $z\sim 2h_z$" of the arm, where $h_z$ is the scale height of the disc.
The bottom-right insertion explains the procedure to obtain the latitudinal profile of an arm.}

{The full-width vertical extents of the arms are thus obtained to be $z\sim 27.2$ pc for Arm I,
$z\sim 22.9$ pc for Arm II;
$\sim 12.9$ to 5.7 pc for Arm III in \coth\ and \csaces\, respectively; 
$\sim 4.3$ pc for Arm IV; 
$\sim 0 3.1$ to 3.3 pc for Arm V in \coth\ and \csaces;
and 
 $\sim \sim 1.3$ pc for Arm VI (CND).}
We find that the vertical extent of the arms decreases from Arm I to VI, which will be discussed in section \ref{dvdl}.
The derived quantities are listed in table \ref{tab1}.

\ss{Arms on the sky (moment 0 maps along LV ridges): The LV-masking method}
\label{LVmasking}

\def\vrid{v_{\rm ridge}}

We then produce integrated intensity (moment 0) maps around the LV ridges of Arm I to VI (CND) indicated by the dashed lines in the LVDs in figures \ref{ArmI} to \ref{aces_lv_max}. 
We produce these maps by using a "masked cube" created by convolving the original cube with a cube of the same size (masking cube) representing a Gaussian function, or the "LV-masking function",
 \be
f(l,\vlsr)={\rm exp} \left[ -\left({\vlsr-\vrid (l)}\over {\delta v}\right)^2 \right].
\ee
Here, $\vrid (l)$ represents $\vlsr$ of the LV ridge at longitude $l$, and is expressed by a linear or a bent-linear (curved) function of $l$ to represent the dashed line of each arm in figures \ref{ArmI} to \ref{aces_lv_max}.
So, in most cases
\be
\vrid(l)=A l+B,
\ee
where $A=d\vlsr/dl$ and $B=v_{l=0\deg}$ and  are taken to be constants, and were measured along each of the dashed lines in figure \ref{ArmI} to \ref{aces_lv_max}.
The velocity half width was estimated to be $\delta v=10$ and 7.5 \kms for \coth\ and \csaces, respectively, using LVDs around the clearest parts of Arms I to III (figures \ref{ArmI} to \ref{ArmIII}), and Arms IV and V (figures \ref{ArmIV} and \ref{ArmV}). 

\begin{figure*}   
\begin{center} 
(T) Moment 0, total intensity integrated from $\vlsr=-230$ to 230 \kms, \coth, 45m\\  
\includegraphics[width=10cm]{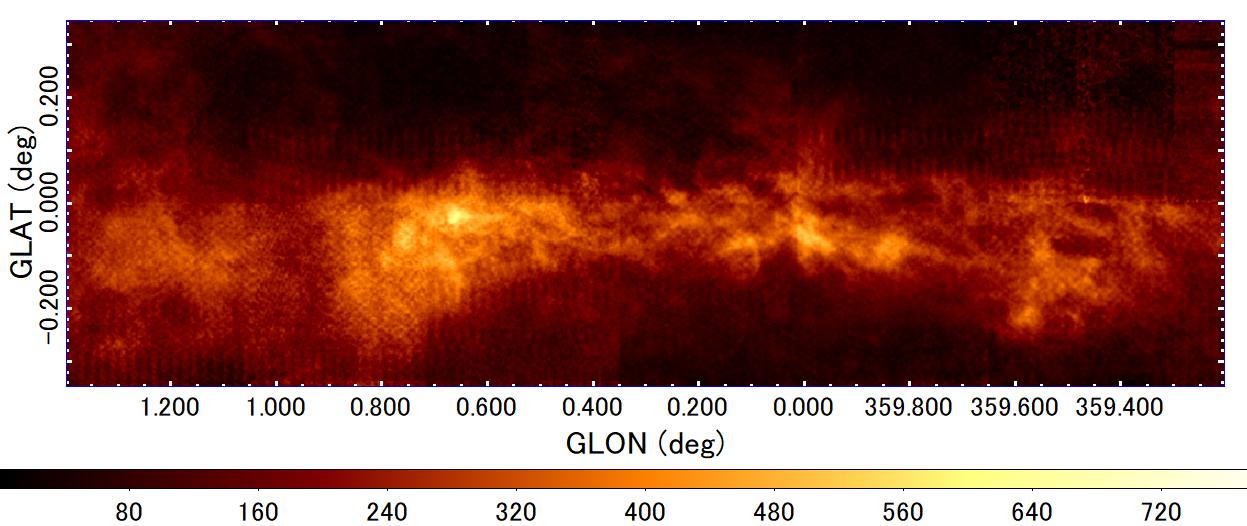}   \\
LV-masked ($\pm 10$ \kms) moment 0 ; \coth; 45m\\ 
(A) Arm I+II+Link \hskip 7cm (B) Arm I\\ \includegraphics[width=8.5cm]{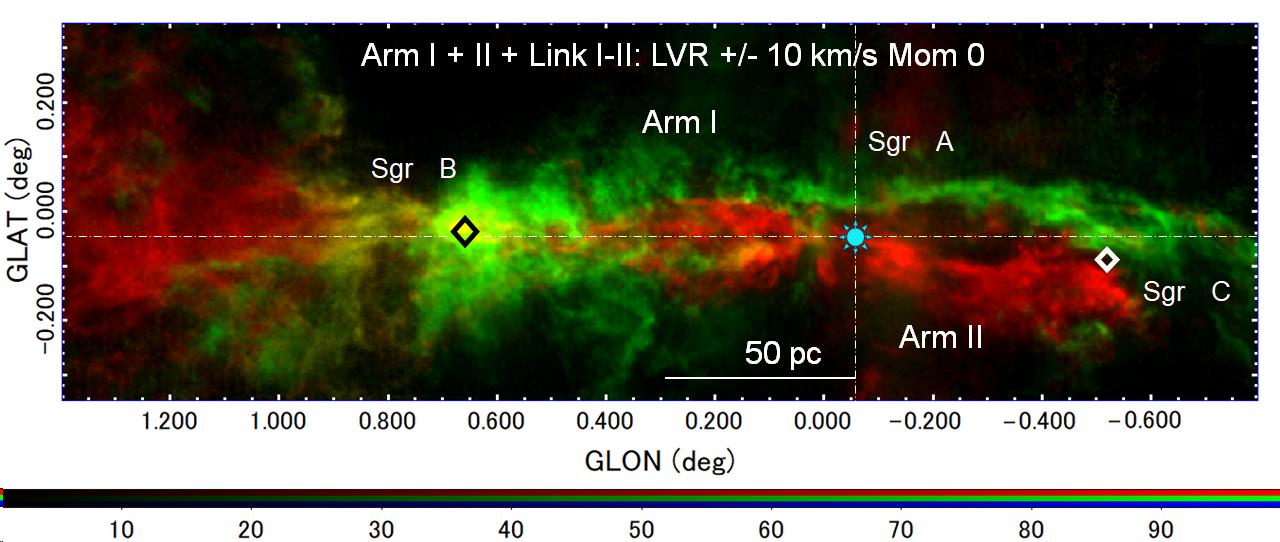}   
\includegraphics[width=8.5cm]{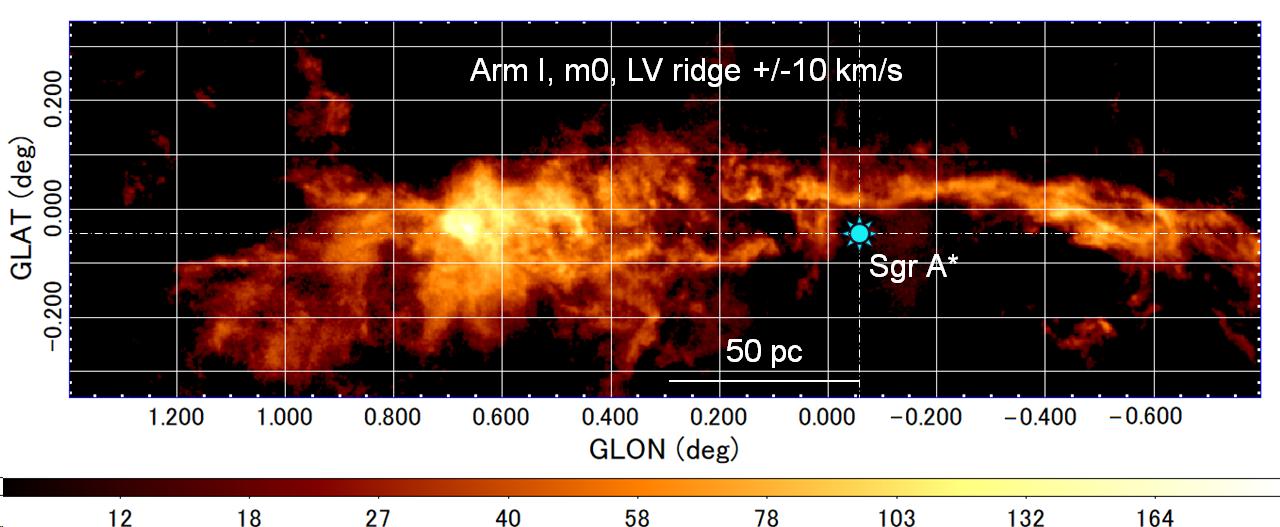} \\
(C) Arm II+Link I-II\hskip 7cm (D) Arm II \\\includegraphics[width=8.5cm]{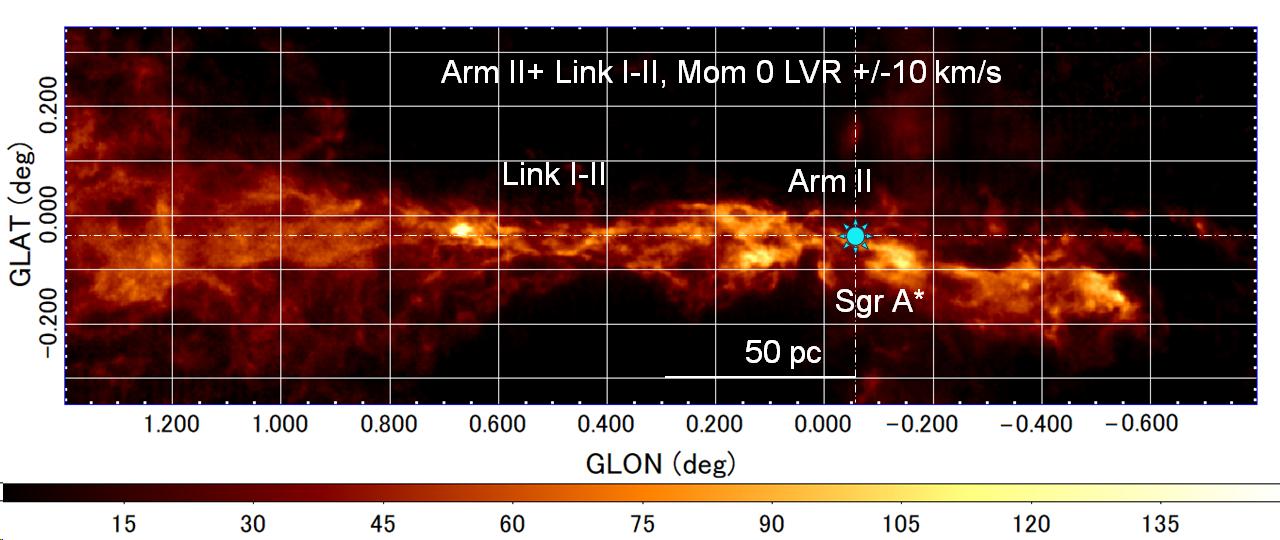}  
\includegraphics[width=8.5cm]{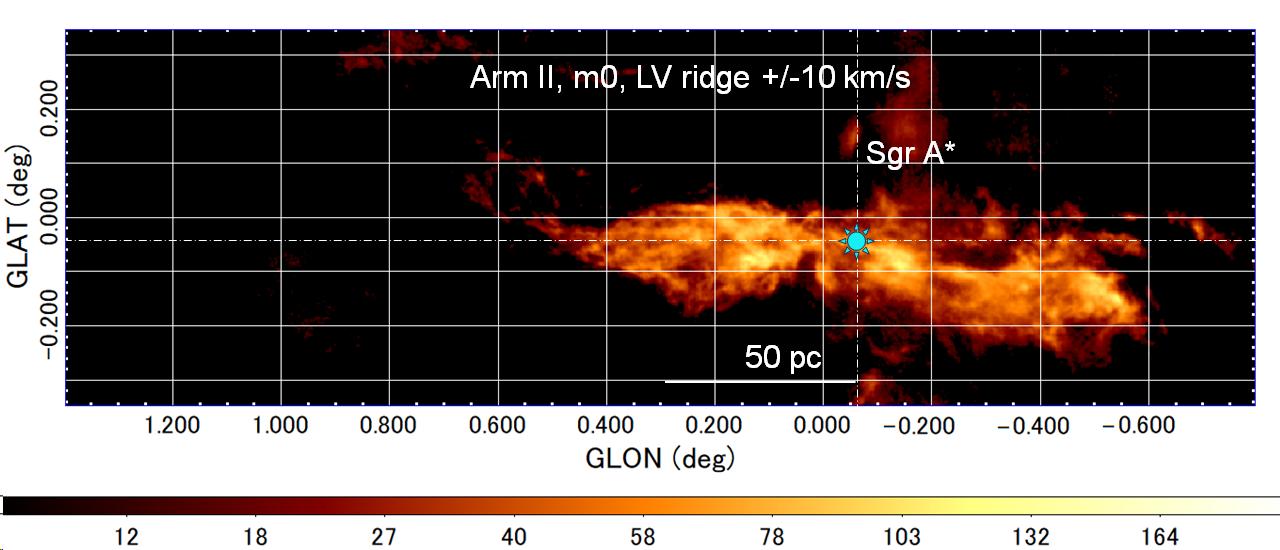} \\ 
(E) Arm III \hskip 7cm (F) Arm IV\\
\includegraphics[width=8.5cm]{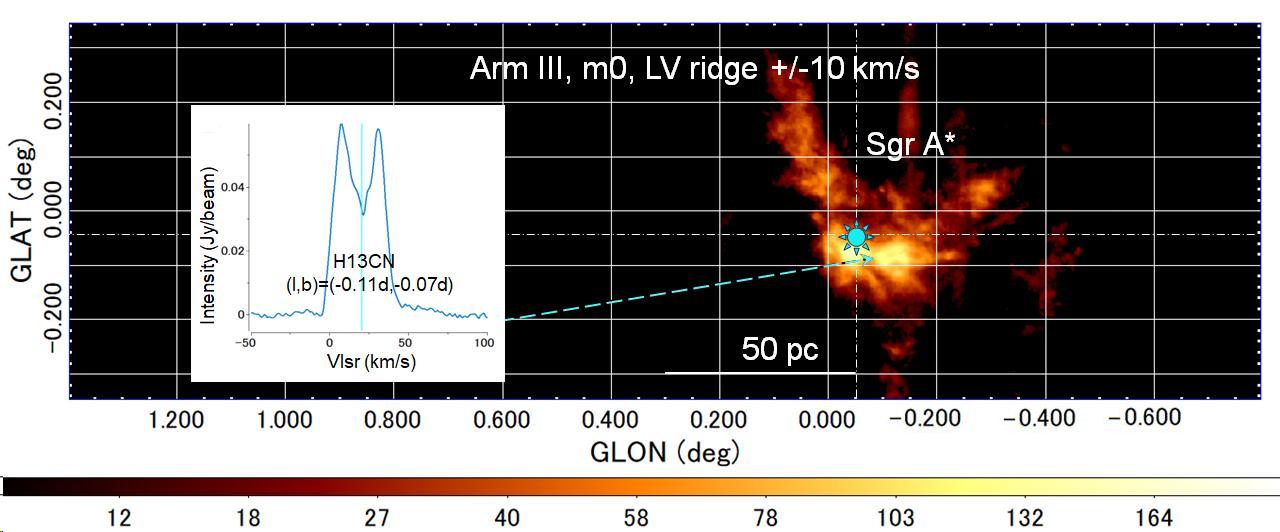} 
\includegraphics[width=8.5cm]{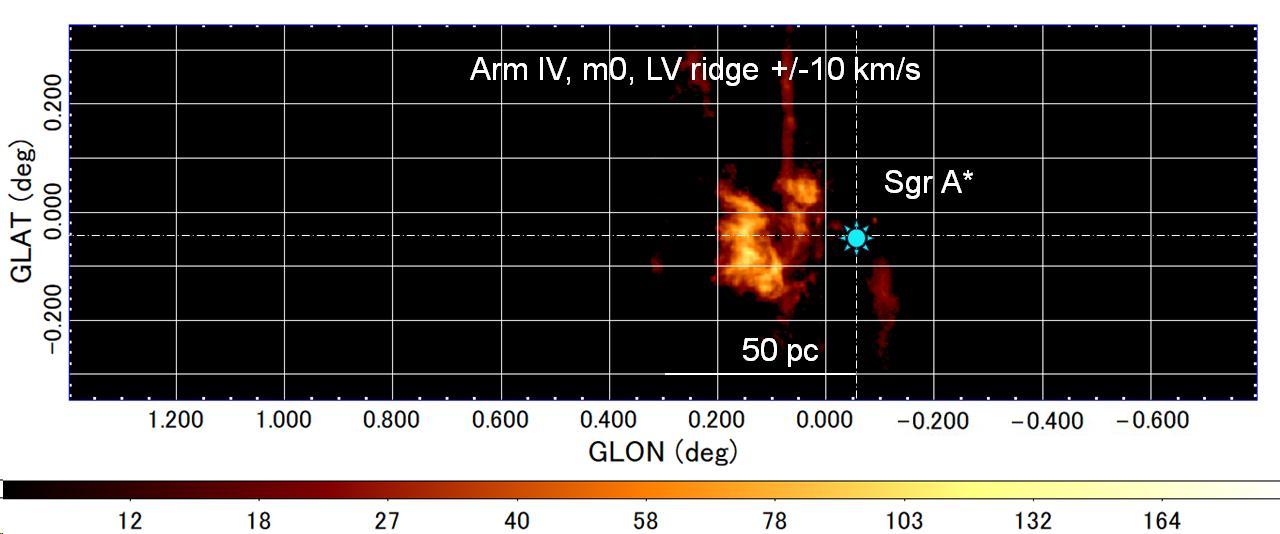} 
\end{center}
\caption{Integrated intensity (moment 0, [K \kms]) maps in \coth\ of the GC Arms from Nobeyama 45 m at a resolution of $16.7''$.
(T) Whole area total integrated intensity (moment 0) map in the \coth\ line by 45-m telescope.
(A) Arm I (green) + II + Link I-II (red) integrated intensity within $\pm 10$ \kms from the LV ridge (same below);
(B) Arm I; 
(C) Arm II + Link I-II along the bent LVR in figure \ref{ArmII}; 
(D) Arm II (straight LVR alone); 
(E) Arm III; and
(F) Arm IV. 
The color bars indicate the integrated intensity (moment 0) in K \kms.
A \hcnaces\-line spectrum of Arm III at $(l,b)=(-0\deg.11,-0\deg.07)$ is inserted in panel (E). 
 {Alt text: Moment 0 maps using masked LVDs for Arms I to III by 45m.}}
\label{moment_45m}	 
\end{figure*}

\begin{figure}   
\begin{center}    
(T) Moment 0, Total integ from -222 to 222 \kms, \csaces; ACES\\  
\includegraphics[width=9cm]{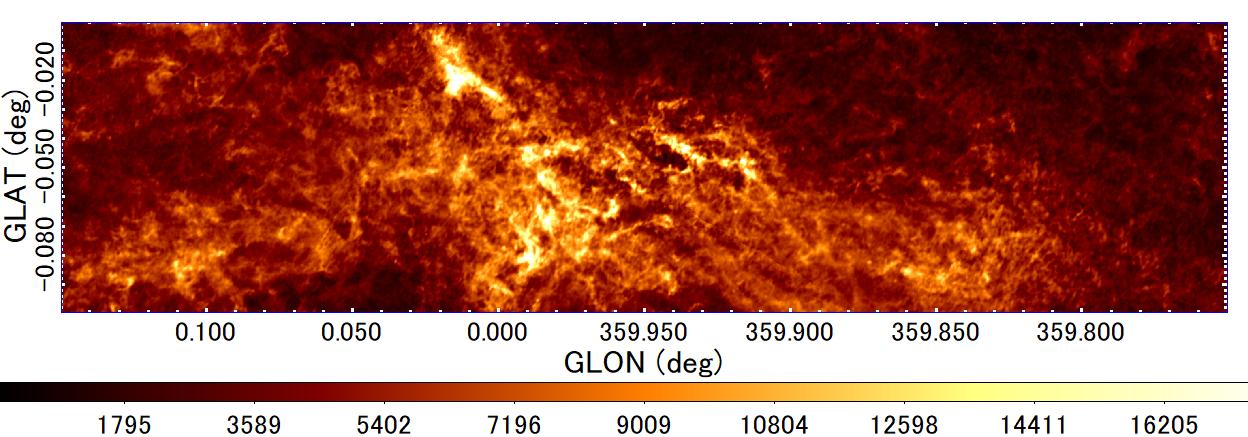} \\  
Moment 0; LV-masking ($\pm 7$ \kms)\\ 
(A) Arm III\\
\includegraphics[width=9cm]{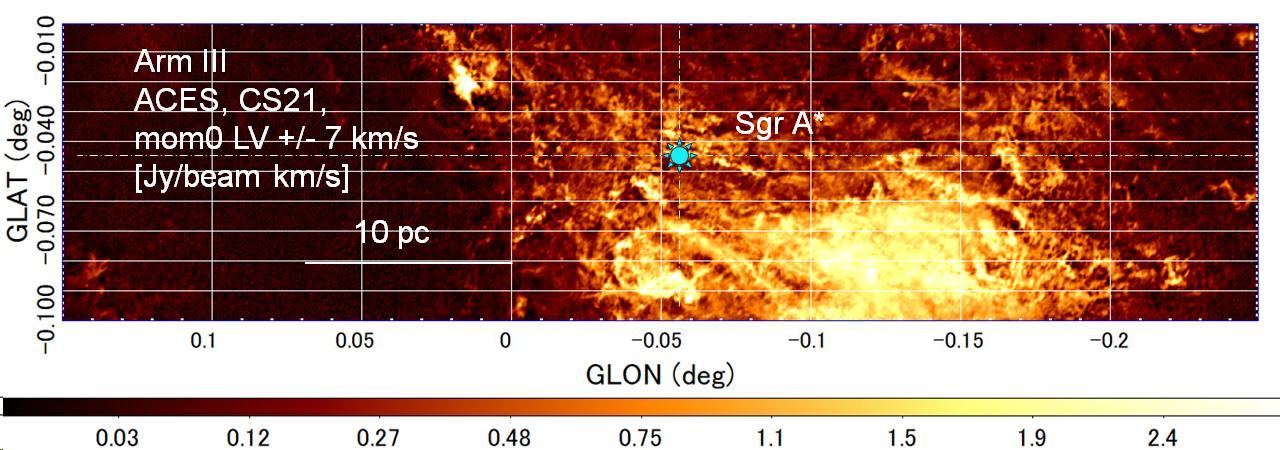} \\  
(B) Arm IV\\
\includegraphics[width=9cm]{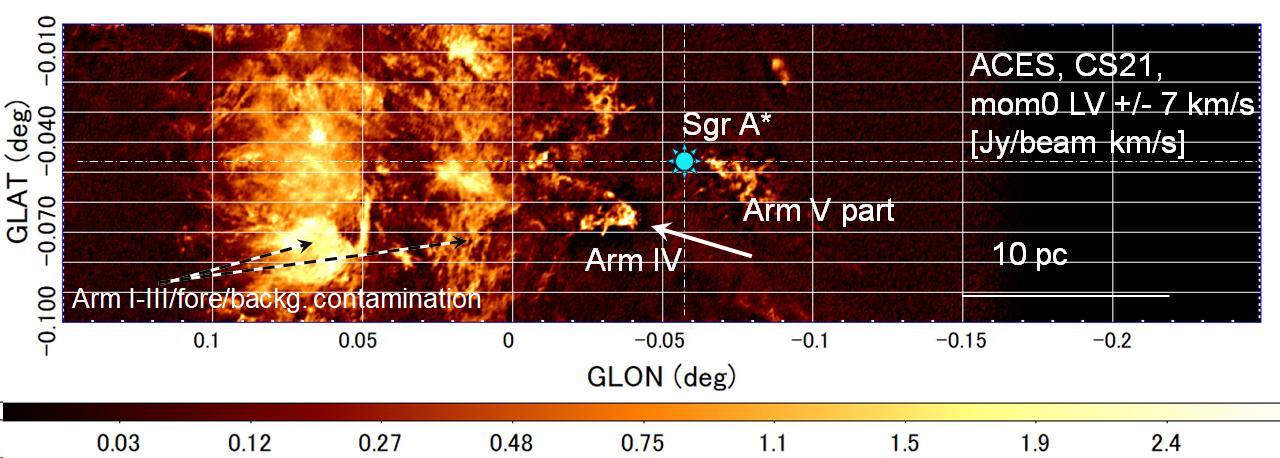} \\ 
(C) Arm V\\
\includegraphics[width=9cm]{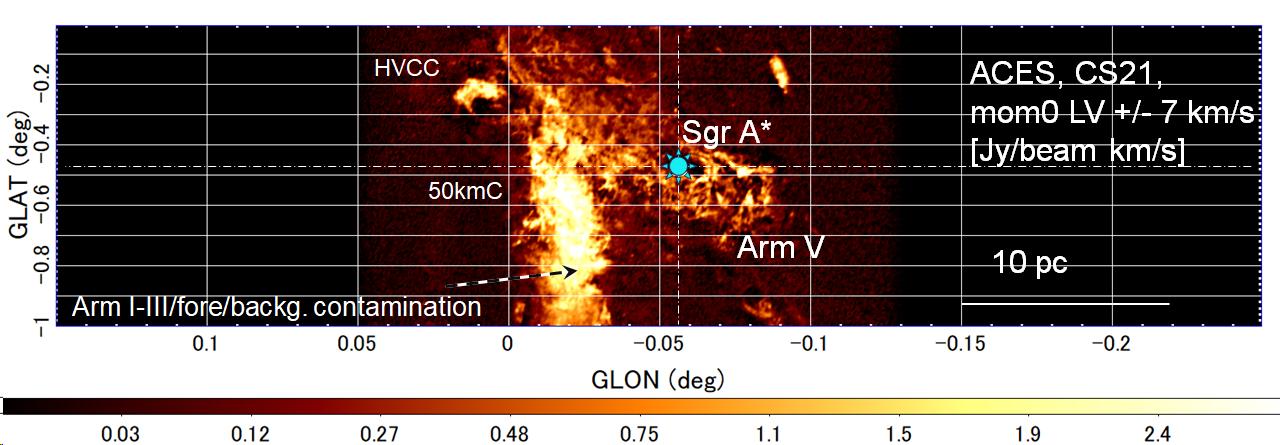} \\  
(D) Arm VI = CND\\
\includegraphics[width=9cm]{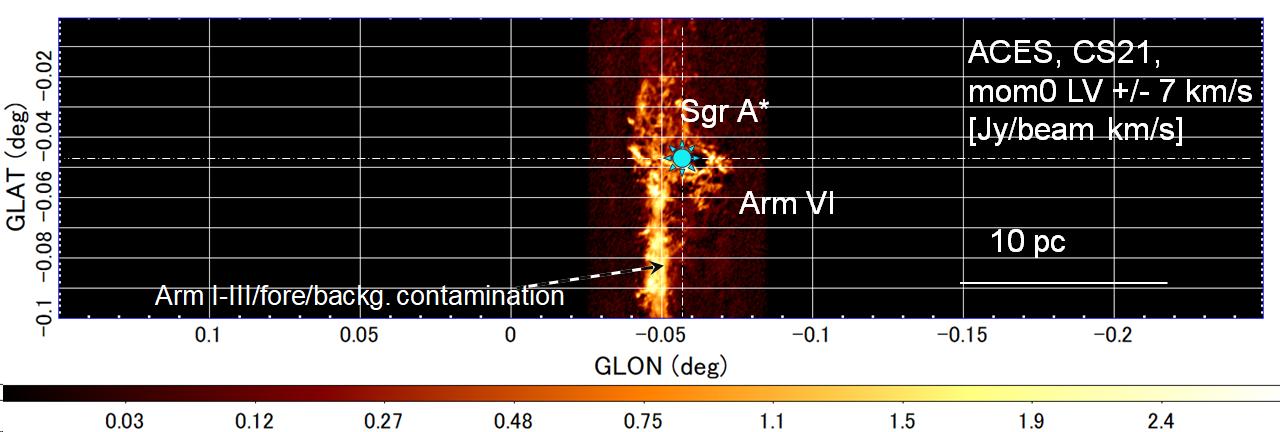}  
\end{center}
\caption{(T) ACES \cs\ integrated intensity (Moment 0; [Jy beam$^{-1}$\kms]) over entire velocity range from $-220$ to 220 \kms, and (A)-(D) along LV ridges with $\vlsr$ within $\delta v =\pm 7.5$ \kms for Arms III to VI (CND) at a resolution of $2.2''$. 
Vertical broad bands are contamination of the extended emission of CMZ and Arms I and II.
Color bars indicate the integrated intensity (moment 0) in Jy beam$^{-1}$ \kms. 
{Alt text: Moment 0 maps using masked LVDs for Arms IV to VI by ACES.}}
\label{moment_aces}	  
\end{figure}

The obtained moment 0 maps along LVR from the 45-m telescope  in \coth\ and from ALMA in \csaces\ and \hcnaces\ are shown in  figures \ref{moment_45m} and \ref{moment_aces}, respectively.
The vertical broad and bright bands in the maps of Arms IV to VI from ALMA are contaminations of the local disc and  the "fore-/background CMZ" including a part of Arms I and II.

\section{Galactic-Centre Arms}
\label{sec_GCA}

The so far identified Arms I to VI are summarized in the LVD of figure \ref{aces_lv_max} by the dashed lines. 
In this section we describe the individual arms based on the LVDs as well as the LV-masked moment 0 maps.
We highlight the inner Arms III and IV, and report the exixtence of a new arm, which we name Arm V. 
We further identify an even inner arm, naming it Arm VI, which is an alternate view of the circum-nuclear disk (CND) \citep{2001ApJ...551..254W,2018PASJ...70...85T}.
We also discuss the mini-spirals around \sgrastar\ \citep{2017ApJ...842...94T}, which we consider to compose the innermost family of the arms, and thus name Arm VII.

\ss{Arm I}
\label{ssArmI}

Arm I, or the Sgr-B Arm, appears as the most prominent LV ridge of the CMZ, and is considered to be a spiral arm or a ring on the nearer side of \sgrastar, the Milky Way's nucleus.
Moment 0 maps of this arm shown in panels (A) and (B) of figure \ref{moment_45m} reveal a long and sharp arm tailing from the Sgr B cloud complex.
Active star forming regions Sgr B1 and B2 are located on this arm \citep{sof1995,sof2022,oka+1998,tok+2019,henshaw+2016,henshaw+2023}.

\ss{Arm II}
\label{ssArmII}

Arm II, or the Sgr-C Arm, is the second-brightest arm associated with Sgr C, and is rotating on the far side of Sgr A.
Arms I and II are linked by a horizontal LV belt (Link I-II) as indicated by the dashed line in figure \ref{aces_lv_max}.

Link I-II appears to be connected to the more outer disc at velocities at $\vlsr \sim 100$ \kms, which may suggest a gaseous arm connecting the CMZ and the Galactic disc.
However, we do not discuss this feature in this paper, because it is far outside the ACES field. 

In panel (C) of figure \ref{moment_45m} we show a moment 0 map integrated along the bent LVR Arm II and Link I-II (figure \ref{ArmII}).
This map indicates that Arm II is tailing from Sgr C and extends nearly symmetrically to Arm I from Sgr B, and extends further to the west along the horizontal LVR Link I-II beyond the edge of Arm II.
Panel (D) presents a part of Arm II in the moment 0 map traced by the tilted straight LVR alone.

The molecular gas mass of the CMZ is shared mostly by these two main Arms I and II as discussed below in section (\ref{volumeratio}). 
The two arms have been proposed to compose the main 120-pc ring \citep{sof1995} of the CMZ, and have been extensively studied in order to derive the 3D structure of the CMZ \citep{2011ApJ...735L..33M,kruijssen+2015,henshaw+2016,tok+2019,sof2022}. 
It is further suggested that the Arms are related to the outer star-formation region Sgr E ($l\sim -1\deg$) and supernova remnants Sgr D ($\sim +1\degd.2$), drawing a double infinity ($\infty \hskip -1.5mm \infty$) on the sky \citep{sof2022}.
However, the degenerate Arms I and II are resolved in the LV space, and the masked moment 0 map (figure \ref{moment_45m}) shows a simple tilted ring. 

\ss{Arm III}
\label{ssArmIII}

Arm III has not been studied in detail so far in spite of its high brightness.
It is visible in \coth\ in the LVD in figure \ref{ArmIII} and in the moment 0 maps in figures \ref{moment_45m} and \ref{moment_aces}. 
The LV ridge of this arm is composed of two parallel stripes in \hcnaces\ and  \hcnaste\ lines. 
The line spectrum shows a clear center-velocity absorption along this arm as shown by an insertion in panel E of figure \ref{moment_45m}.
The absorption belt along Arm III will be discussed in some detail later in subsection \ref{ss20vs50}. 

This arm seems to consist of the GMC M-0.13-0.08 (20kmC) (see \citet{Takekawa17a}).
The moment 0 map in figure \ref{ArmIII} shows a rather short arm on the sky, being led by a bright clump of 20kmC.
If 20kmC is physically associated with Arm III, its 3D position can be determined kinematically, as will be done in section \ref{dvdl}.
Thereby, we assume that Arm III is in front of \sgrastar, following the face-on geometry proposed by \citet{Takekawa17a}.

However, it has also been suggested that there is a physical contact between the +20-\kms\ cloud and the CND \citep{Takekawa17a}.
If this is the case, a different view is required, and Arm III may be re-defined as a long bright ridge only in negative $\vlsr$, extending to $(l,\vlsr)\sim (-0\degd.2,-50 \ekms)$.
Another concern is its possible relation to the 50kmC: 
As shown in the relieved LVD in figure \ref{ArmIII}, the 50kmC is located on the increasing-longitude extension of Arm III.
In this paper, we examine another possibility that 50kmC is related to Arm V not only for the LV position but also for the large $dv/dl$ value close to that of Arm V and large velocity width (section \ref{ss20vs50}. 
 
\ss{Arm IV}
\label{ssArmIV}

Arm IV is the most clearly visible arm in the relieved \coth\ LVD at negative $\vlsr$ as shown in the bottom panel of figure \ref{ArmIV}.
Its positive-$\vlsr$ extension is visible in the original LVD as indicated in the middle panel of this figure, but is strongly disturbed by the contamination from Arms I and II as well as the extended CMZ emission.
  
This arm can be clearly traced on the LVD in figure \ref{ArmIV}, and is also visible in the moment 0 map in \csaces\ from ACES. 
The arm runs westward from $l\sim 0\deg.02$ and stops at $l\sim -0\deg.04$.

A part of this Arm has been identified as "C1" clump at 
$(l, b, \vlsr) \simeq (-0\deg.03, -0\deg.06, -70 \ekms)$, 
which shows 
intense  CS emission \citep{Oka+11}. 

\ss{Arm V}
\label{ssArmV}

Arm V is a straight and long LV ridge composed of a low-brightness stripe extending from $(l,\vlsr)=(-0\deg.02,+30 \ekms)$ to $(-0\deg.1,-130 \ekms)$, as shown in figure \ref{ArmV}. 
In the moment 0 map in \csaces\ (figures \ref{moment_aces} (C)), Arm V runs at position angle ${\rm PA}\sim 25\deg$, and is approximately fitted by an ellipse of axial ratio $b/a\sim 0.25$, or inclination angle of $i\sim 76\deg$ (tilt angle $\hat{i}^sim 14\deg$).  
The 50kmC is located on the positive-velocity extension of Arm V in the LVD, and the velocity gradient is about equal to that of Arm V (see section \ref{ss20vs50}).  

\begin{figure}   
\begin{center}     
Arm VI (CND); Mom. 0; LV-masking $\pm 7.5$  \kms; \\ 
\csaces; ACES\\
\includegraphics[width=7.5cm]{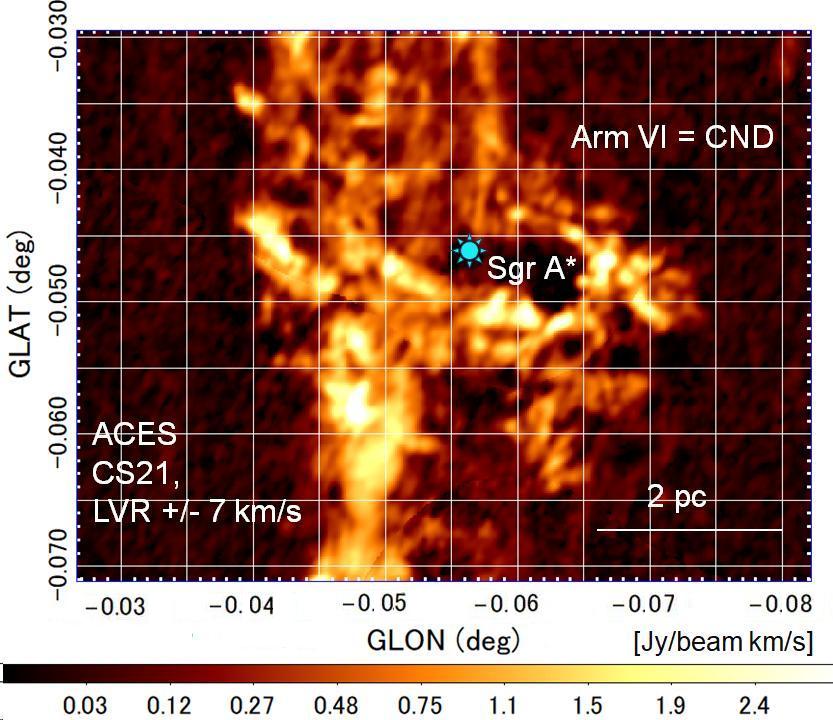} \\
\includegraphics[width=7.5cm]{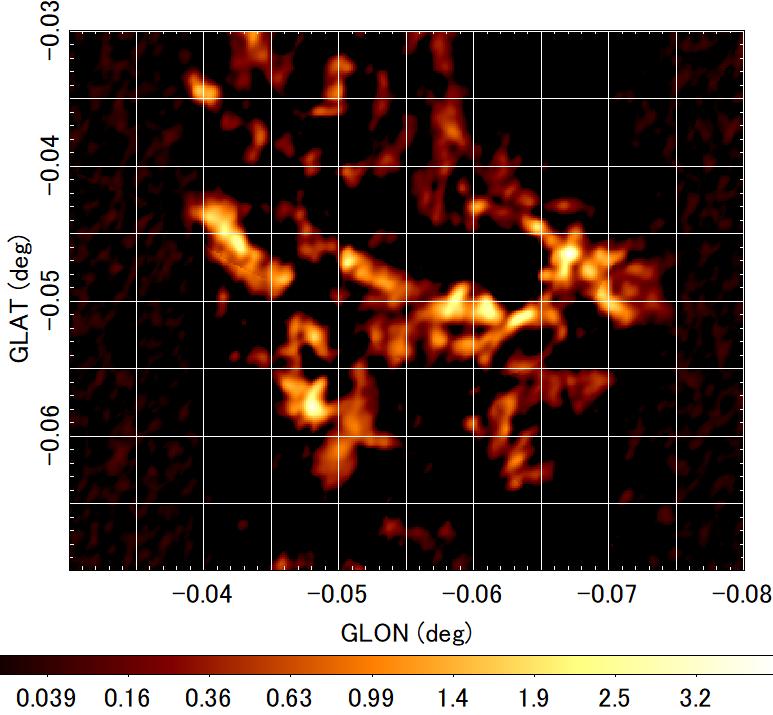} \\
\hcnaces; ACES \\
\includegraphics[width=7.5cm]{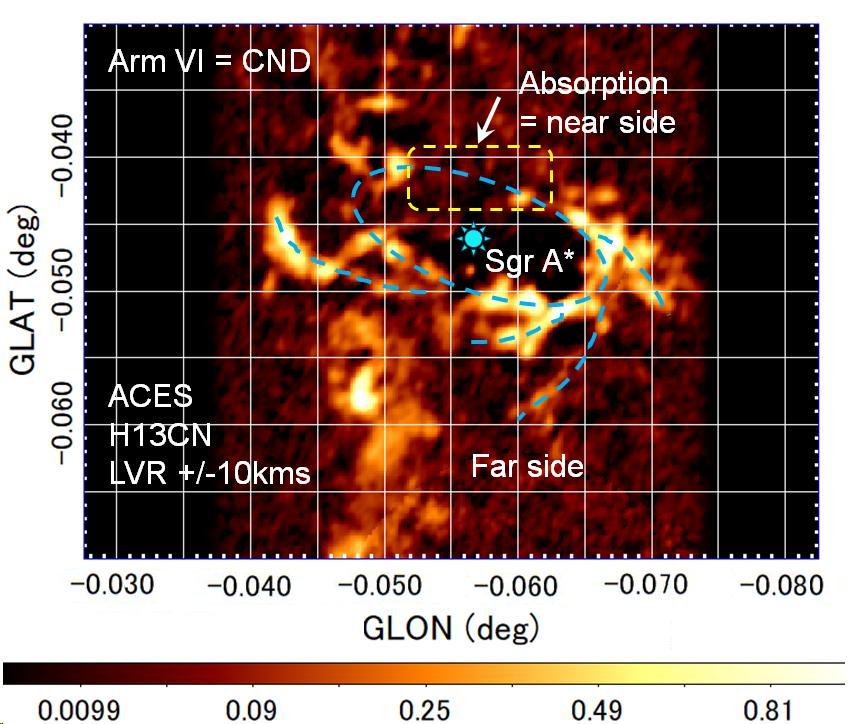} 
\end{center}
\caption{[Top] Arm VI (CND) in \csaces\, same as figure \ref{moment_aces}, but close up.
[Middle] Same, but the vertically extended components due to fore/background emission have been removed.
[Bottom] Same, but in \hcnaces.  
An ellipse is seen associated with spiral fins, indicating a ring with inclination angle $i\sim 65\deg$. 
 {Alt text: Masked moment 0 maps of Arm VI in \csaces\ and \hcnaces\ by ACES.}}
\label{ArmVI_up}	 
 \end{figure}

\ss{Arm VI -- CND --}
\label{ssArmVI}

The CND \citep{2001ApJ...551..254W,2009ApJ...695.1477M,2012A&A...539A..29M,2014A&A...570A...2F,2017ApJ...850..192M,2018PASJ...70...85T,2021ApJ...913...94H}
is recognized in the \csaces\ and \hcn-lines in the LVDs in figure \ref{ArmVI} as a highly-tilted bright ridge extending from $(-0\degd.04,+120\ \ekms)$ to $(-0\degd.07,-130\ \ekms)$ indicating a rotating ring.
We call this ring Arm VI, which is equivalent to CND.

We stress that the arm is hardly visible in the \coth\ line, becoming clearer in the LVDs of \csaces, \hcnaces, and \hcnaste.  
This demonstrates that the molecular gas in the CND (Arm VI) is much denser and warmer compared to 
the general CMZ clouds such as in Arms I and II.
The LV ridge exhibits a double-peaked rotating ring structure whose central part is missing. 
A curved spur is extending from the negative-velocity end of Arm VI, indicating that the arm is associated with a high-velocity non-circular flow. 

Figure \ref{ArmVI_up} shows moment 0 maps of Arm VI in \csaces\ and \hcnaces\ as integrated within  $\delta \vlsr=\pm 7$ and $\pm 10$ \kms\ from the LV ridge (LVR), respectively, which are in good agreement with the CS $(J=7-6)$ map using ALMA by \citet{2018PASJ...70...85T}.  
The figure exhibits an elongated ellipse centered on \sgrastar\ as marked by the dashed line in the bottom panel of figure \ref{ArmVI_up}, which has an axial ratio of 
$b/a\sim 0\degd.0085/0\degd.023=0.405$ at position angle of the major axis of PA$\sim 70\deg$. 
If it is a circular ring, the inclination angle is $i\simeq 66\deg$ (tilt angle $\hat{i}\sim 24\deg$).
Several spiral arms and fins are bifurcating from the ring, trailing in the sense of counter-clockwise rotation in the figure (on the sky). 

{The positive-latitude side of the ring (northern wing) is missing (figure \ref{ArmVI_up}, bottom panel), although it is visible in the warm H$_2$ (2.4-$\mu$m) emission \citep{2014A&A...570A...2F,2017ApJ...850..192M}.
The difference of the feature in mm and $\mu$m emissions may be attributed to absorption of the Sgr A's continuum emission by the CS and H$^{13}$CN molecules. 
On the contrary, the negative-latitude side (southern wing) draws a nearly perfect ellipse, indicating no absorption. 
These suggests that the northern wing of the ring is in front of Sgr A and the southern half is beyond it.}
This is consistent with the counter-clockwise rotation of the disc seen from below the Galactic plane.
There is a hole in the center of the ring coinciding with \sgrastar, which is due to the absence of molecular gas around the minispirals at high temperature as well as due to absorption of Sgr A's continuum emission.

\ss{{Arm VII -- Minispirals --}}
\label{ssArmVII} 

The minispirals are composed of ionized gas \citep{2017ApJ...842...94T} and are not detected in the present data.
They rotate around \sgrastar\ in elliptical orbits of radius $\sim 1.2$ pc and inclination $i\sim 126\deg$ ($\hat{i}\sim 36\deg$) \citep{2009ApJ...699..186Z}, and therefore have a vertical extent $z\sim 1.4$ pc.
Interestingly, they are on a natural extension of the radius-vertical extent and radius-inclination relationships among the Arms as discussed in subsection \ref{faceon}.
Although they are out of our direct analysis based on the molecular line observations, we here suggest to interpret the minispirals as the innermost family of the arms of the CMZ and call them Arm VII.

\ss{Fine LV stripes}

Besides the large-scale, grand-designed arm structures that trace the tilted LV ridges spanning $\sim \pm 100$ \kms\ like Arms I to VI, there are numerous fine stripes composed of shorter ($\sim \pm 10$--30 \kms) vertical LV ridges seen in the relieved LVDs from the 45-m telescope (Appendix \ref{aprelief}, figure \ref{45m13lvx5}) and those in the ACES LVDs at higher resolution with $\sim \pm 10$ \kms (figure \ref{aces_lv_max}).
They are mostly individual molecular clouds not resolved in the longitude direction, indicating that the CMZ contains numerous clouds with sizes less than the beam width, $\sim 2''\sim 0.1$ pc and velocity dispersions of $\sim 10$--30 \kms). 

We point out that some of such LV ridges (individual clouds) are inclined in the same sense as that caused by the Galactic rotation, but generally with tilt angles steeper than those of the main Arms I and II.
This suggests that some of the LV ridges, except for the innermost LV ridges discussed in the next section, represent individual clouds locally rotating more rapidly than the disc's rotation due to self contraction with the cloud's angular momentum being conserved.
 
\section{Arm radius and vertical extent}
\label{sec_armradius}

\ss{Radii of GC Arms using $dv/dl$ method}
\label{dvdl}

Here we introduce three observable quantities to describe the identified arms on the LVD:
(1) Velocity gradient (slope) $dv/dl$ of the LV ridge, 
(2) velocity intersection $\vlsr^*$ of the ridge at the longitude of Sgr A$^*$ at $l=-0\degd.056$, and 
(3) peri-/apocentric longitude offset $\Delta l_0$ from \sgrastar, at which the motion of gas becomes perpendicular to the line of sight so that $\vlsr=0$ \kms.
Table \ref{tab1} lists the values of $\dvdl,\ \vlsr^*$ and $\Delta l$ measured by eye-fitting to the corresponding LV ridges in the LV plane.
 
If we assume that an extended object is rotating around a certain center in an edge-on disc, the curvature of the flow line, $R$, is related to the velocity gradient by 
\be
R\simeq R_0 \vrot \left( \frac{d\vlsr}{dl}\right)^{-1},
\label{eq_dvdl}
\ee 
where $\vrot\sim 150$ \kms~is the rotation velocity. 
If we assume that the flow is circular around \sgrastar, the radius is equal to the galactocentric distance $R$. 

The error in the curvature propagates from that in the $dv/dl$ measurement, which is about $\pm 5$\%, and the uncertainty of the rotation velocity, which causes an error of a factor of 1.5 (100--150 \kms). 
Therefore, the error of the curvatures/radii of the Arms determined in this paper is a factor of $\sim 1.5$ mostly attributed to the accuracy of assumed rotation velocity.
 
Note that this method measures the local curvature of the streamlines for a given flow velocity. 
Even if the flow is not circular, e.g., elliptical, hyperbolic, etc., it gives the local streamline curvature.
It works even if the flow is overlapped by radial expansion or contraction, because $dv_{\rm expa}/dl \simeq {\rm O}(\theta^2)\simeq 0$ near the rotation axis with $v_{\rm expa}$ and $\theta$ being the expanding/contracting velocity of the disc (arm) and Galacto-centric longitude, respectively.
This method measures the curvature of the streamline, but not the curvature of the density distribution such as a filament or arm that may be inclined at an angle to the streamline in such a case of galactic shock waves.  

In the case of Arm IV, for example, we measure the velocity gradient to be $d \vlsr/dl \simeq$ 200 \kms per $0\degd.02$ in the LVDs, yielding $R\simeq 21$ pc, assuming $i\sim 90\deg$ and $j\sim 0\deg$.  
The rotation period is then $P=2\pi R/\vrot \sim 0.86$ Myr.

\begin{figure}   
\begin{center}   
\includegraphics[width=7cm]{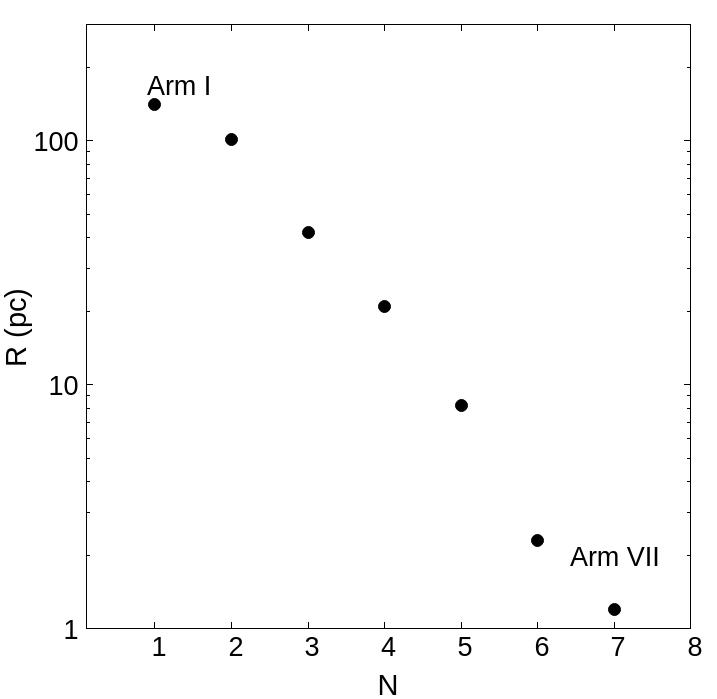}    
\includegraphics[width=7cm]{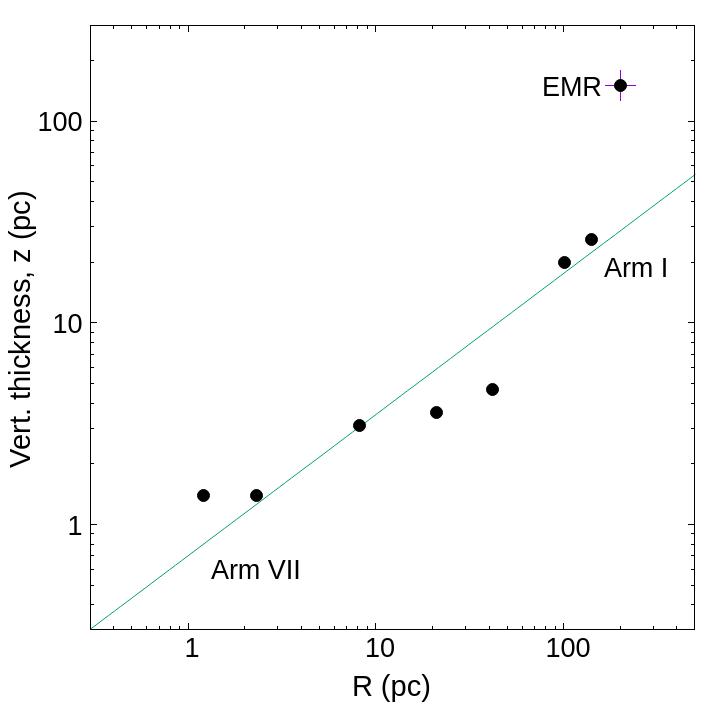}
\includegraphics[width=7cm]{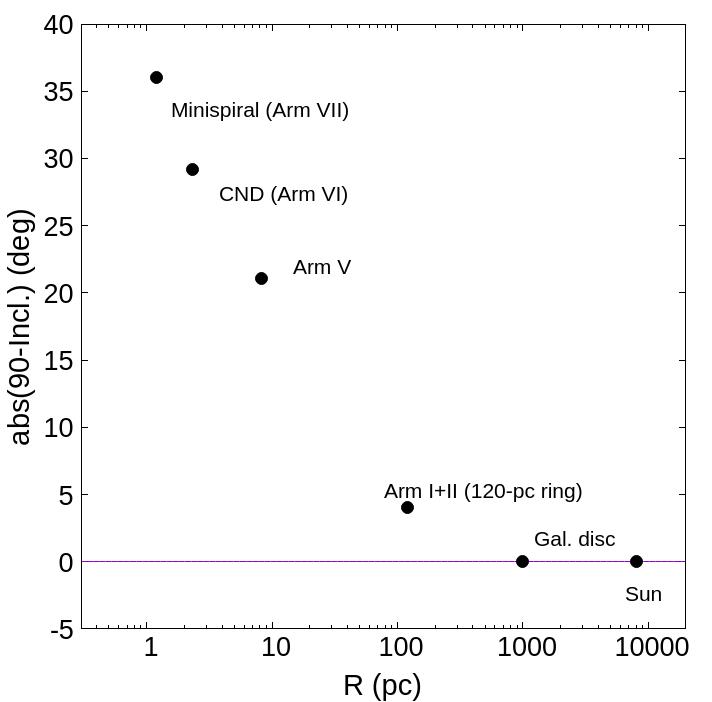}      
\end{center}    
\caption{ 
[Top] Arm radius-to-number relation, which is expressed by $R\sim 630\times (2/5)^N$ pc with an error of factor $\sim 2$. 
[Middle] Vertical {extent}, $z$, of Arm VI to Arm VII (minispials) plotted against radius $R$. 
The straight line represents 
$z=0.7 (R/{\rm 1 \ pc})^{0.7}$ pc.
The big cross indicates EMR.
[Bottom] Absolute values of $90\deg-$inclination angle of the arms/rings plotted against radius.   
 {Alt text: Plots of radius vs Arm number, vertical extent vs radius, and tilt angle vs radius.}} 
\label{arm-plot}	 
\end{figure}

Table \ref{tab1} lists the estimated values of $R$ and $P$ for the identified arms obtained for $R_0=8.2$ kpc and $\vrot=150$ \kms. 
In the top panel of figure \ref{arm-plot} we show the "arm-radius relation", or a plot of radii against arm number I to VII, where Arm VII represents  the minispirals as discussed in later section.
The plot is approximately fitted by   
\be
R\sim R_{\rm A} \left({\vrot \over 150 \ekms}\right) \times (2/5)^{N} \ {\rm pc},
\ee
($N=$I, II, ..., VII)  within an error of factor $\sim 2$, where $R_{\rm A}=630$ pc.  
This shows that the existence of the arms is discrete, and  that the ratio between the radii of two neighboring arms/rings is $\sim 2.5$, suggesting a Bode's law-like arm structure.  
Alternatively, it may be attributed to a logarithmic spiral with pitch angle of $p\sim 22\deg$.

\begin{table*}
\caption{Parameters of the GC arms inferred from $dv/dl$ method$^{\natural}$.} 
\begin{tabular}{lllllllll}  
\hline\hline  
Arm & $dv/dl$  & $\Delta l$ & $\vlsr^*$  &$R^{\dagger\dagger}$&$P$&$z=2h_z$&$\hat{i}$ &Arm volume ratio $^{\natural\natural}$ \\
 & (\kms deg$^{-1}$) &(deg)&  (\kms)$^{\#}$&(pc) & (Myr)&(pc)&(deg) &Vol$_N$/Vol$_{\rm Arm \ I + II}$\\ 
\hline
Arm I & 150.& $+0.25$& $-40$ &141.  &5.9 &26&4&For I+II = 120-pc ring\\
Arm II & 212.&$-0.1$&$+10$&101. &4.2 &20&&=1.0\\
Arm III &507.& $-0.1$&$+50$& 42.&1.7 &4.7&&$\sim 2\times 10^{-2}$ \\ 
Arm IV &1020.&$+0.08$  &$-80$  &21. &0.86 &3.6&&$\sim 3\times 10^{-3}$\\ 
Arm V & 2600. &$+0.025$ &$-50$ & 8.2 &0.34 &3.1& 21 &$\sim 4\times 10^{-4}$\\    
Arm VI (CND)   &9300. &0  &0  & 2.3$^{\dagger}$     &0.1&1.4&29&$\sim  10^{-5}$\\  
Arm VII (Minispirals){$^{\ddagger}$} &$\sim 10^4$ &--&--&$\sim 1.2$ &$\sim 0.05$ &$\sim 1.4$ & $\sim 36$ &$\sim 10^{-5}$\\ 
\hline\\   
\end{tabular} \\
$^{\natural}$ Inclination angle is assumed to be $i\simeq 90\deg$ (tilt ange ${\hat i}=0\deg$), except for Arm VII. \\ 
$^{\dagger\dagger}$ $R=\vrot R_0 (dv/dl)^{-1}$ for $\vrot=150$ \kms and $R_0=8.2$ kpc.\\
$^{\#}$ $\vlsr^*$ is $\vlsr$ at $l=-0\degd.056$ (\sgrastar). 
The reference center is taken at 
Sgr A$^*$ with $(l,b)= (359\degd.944227,	-0\degd.046157)$. \\ 
$^{\dagger}$ Consistent with the current measurement  $R\simeq 2.5$ pc \citep{2018PASJ...70...85T}.
\\ 
$^\ddagger$ Approximate values read on the maps in the literature \citep{2009ApJ...699..186Z,2017ApJ...842...94T}. \\
$^{\natural\natural}$ The ratio of the volume shared by Arm $N$ to that of the 120-pc ring composed of Arms I and II as calculated by Vol$_N=\pi R^2\times z$.
\label{tab1} 
\end{table*}

\ss{Face-on view of GC Arms}
\label{faceon}

\begin{figure}   
\begin{center}  
\includegraphics[width=7cm]{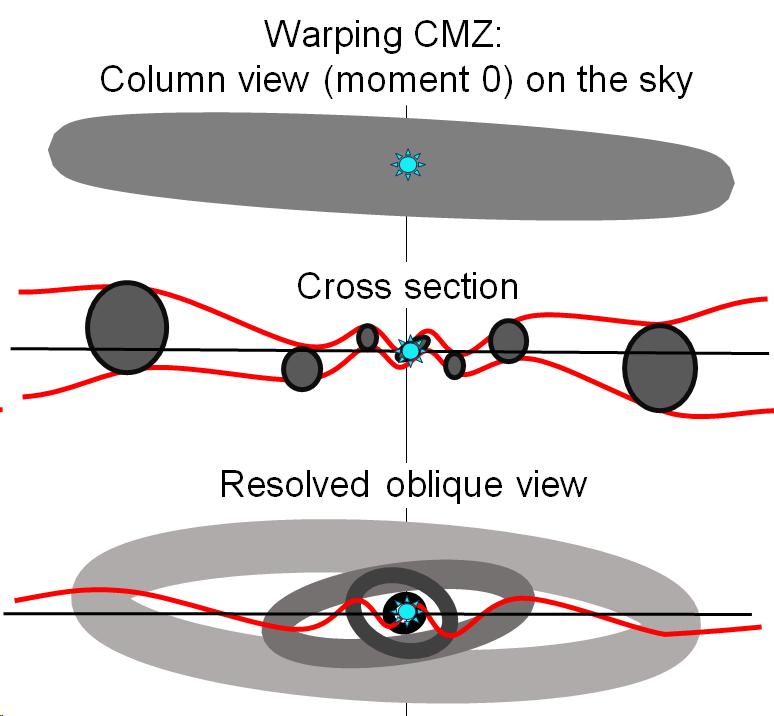} \\   
\vskip 3mm
\includegraphics[width=7.5cm]{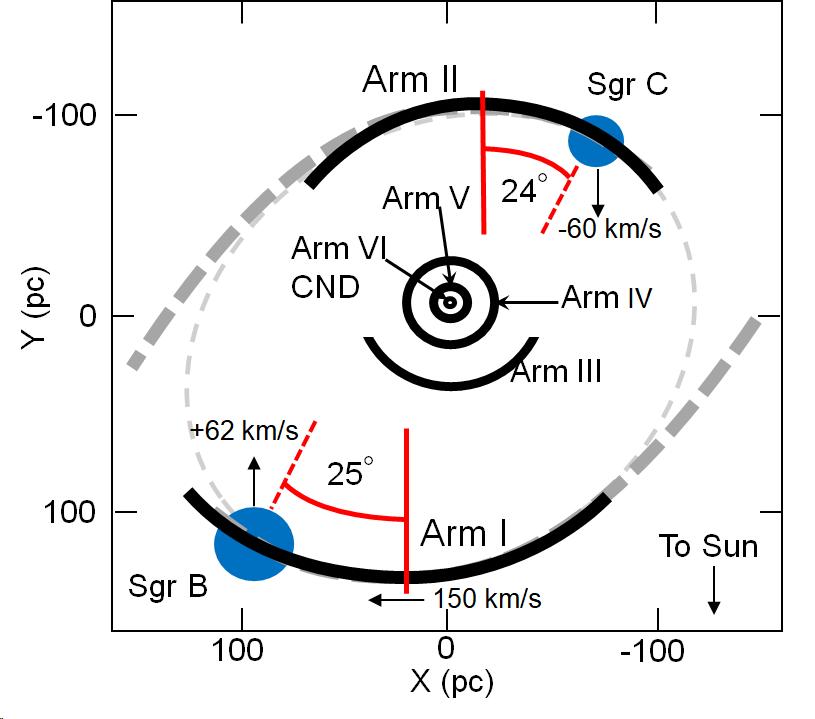}      \\ 
\includegraphics[width=7cm]{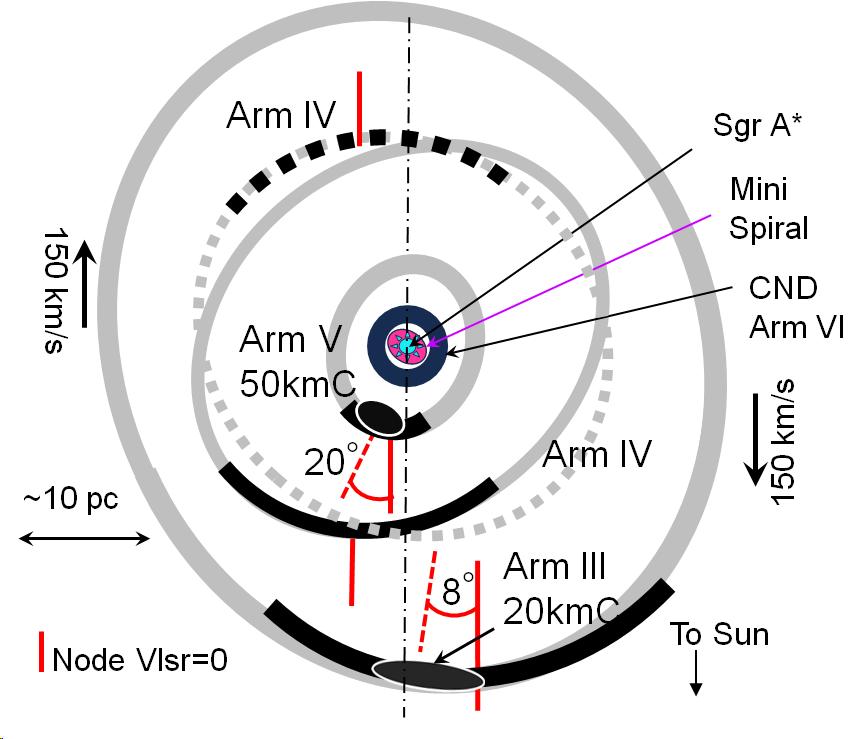}
\end{center}    
\caption{
[Top] Schematic view of CMZ projected on the sky, cross section of warping arms, and an oblique view of the resolved arms/rings. 
{[Middle] Proposed face-on distribution of the GC Arms. Thick lines show the Arms with determined radii. 
Vertical red bars indicate the nodes of $\vlsr^*=0$ \kms, and dashed red lines show the position angles $\theta$ of Sgr B and C with respect to the curvature centre. Grey dashed lines illustrate suggested connections to the outer disc \citep{sawada+2004}.
[Bottom] Same, but more inner arms. The thick curves represent the radii measured with the center of curvature shifted to the node of $\vlsr=0$ \kms. The grey ellipses show the possible orbits of the arms that satisfy the derived radii. Arms III and V are on the near side of Sgr A due to absorption of the continuum emission of Sgr A. {Near/far answer for arm IV was not given in this study.}}
 {Alt text: Schematic illustrations of the GC Arms.}
} 
\label{fot-arms}	 
\end{figure}

Using the estimated radii of the arms, their orientations in the CMZ are illustrated in figure \ref{fot-arms}.
The top panel shows a schematic sketch of the CMZ projected on the sky, cross section of warping arms, and an oblique view of the resolved arms and rings.

{The 2nd panel shows face-on views of the Arms using their calculated radii from the $dv/dl$ method, where 
the thick lines show face-on view of the Arms with determined radii. 
The longitudinal offset of the nodes at $\vlsr=0$ \kms are marked by red lines, and dashed red lines represent position angles $\theta$ of Sgr B and C with respect to the curvature centre. 
Grey dashed lines illustrate suggested connections to the outer disc \citep{sawada+2004}.
Position angles of the associated molecular clouds with respect \sgrastar\ are calculated by $\theta=\sin^{-1}\vlsr/\vrot$ ($\vlsr=150$ \kms).}

The bottom panel shows the inner arms, where the thick curves represent the measured radii with the curvature centers shifted to the $\vlsr=0$ \kms nodes. Grey ellipses illustrate possible orbits of the arms satisfying the derived radii.  
The result is globally consistent with the face-on views obtained using the intensity ratio of the CO line emission to OH absorption line 
\citep{sawada+2004,2017MNRAS.471.2523Y}.

We have so far assumed that the orbits of the arms are circular around a center at the longitude at which the LV ridge crosses the $\vlsr=0$ \kms
line.
Although this assumption leads to a reasonable galactocentric distance $R$ using the $dv/dl$ method,
in order to map the arms and clouds more precisely, we need a realistic gravitational potential and a flow line model around Sgr A (e.g. \citet{kruijssen+2015,kruijssen+2019}),
which is, however, beyond the scope of this paper.

\ss{Vertical {extent} decreasing toward nucleus}

In section \ref{bprofile} we have shown that the vertical {extent} of the Arms increases with the radius $R$.
In figure \ref{arm-plot} (middle panel) we plot the full {extent} $z$ of the Arms against the arm radius $R$.
Here, we also plotted the supposed vertical extent of the minispiral (Arm VII) of $\sim 1$ pc from the literature.
In this log-log plot, the {extent} is approximately expressed by a straight line given by 
\be 
z\sim 0.7 (R/1 \epc)^{0.7}  \epc.
\ee 
{Since the measured vertical extents of the arms may be regarded to represent upper limits to the thickness of the disc, this plot indicates that the thickness of the CMZ disc decreases toward the nucleus.}
It means that the disc inside Arms I and II becomes thinner from a few pc to $\sim 1$ pc near Sgr A.  
 
Note, however, that these latitude profiles do not include such clouds as the Sgr B complex in Arm I and 50kmC in Arm V, which extend more vertically than expected from the $R$-$z$ relation.
This raises the question of whether these clouds are physically associated with the arms and how such vertical protrusions could have formed from the thin arms.
Alternatively, such "high-$z$" clouds may manifest large bents (oscillations) of the trajectories or more deviated inclinations of orbits from $i \sim 90\deg$.  

We also point out that the vertical {extent} of the 200-pc expanding molecular ring (EMR) ($z\sim 150$ pc) at radius $\sim 200$ pc \citep{Sofue2017} deviates significantly from the fit to the Arms, as shown by a large cross in figure \ref{arm-plot}.
This will be touched upon in later subsection (\ref{EMR}).

\ss{Tilt angle of disc increasing toward nucleus: Warping CMZ}

As shown in the moment-0 maps (figures \ref{moment_45m} and \ref{moment_aces}), the 120-pc ring consisting of Arms I + II, Arm V, and Arm VI (CND) exhibits elliptical structures on the sky, indicating that the rings deviate significantly from the edge-on orientation.
Arm III does not appear as a clear ellipse, but its ridge is tilted a few degrees from the horizontal, which may represent a tilt of the ring.
Arm IV is too divergent in the moment-0 map to define a corresponding ellipse.
The minispiral (arm VII) is well known for its highly tilted orientation with a minor to major axis ratio of $b/a\sim 0.5$.

We then calculate the "tilt angle" ${\hat i} = 90\deg-i$ of the ring's rotation axis from the Galaxy's rotation axis. The inclination is measured using the ratio of the minor and major axes of an ellipse as well as the tilte angle of the major axis of the arms (ellipses) on the sky in the moment-0 maps.
Plotting the results in the last panel of figure \ref{arm-plot}, we see that the tilt angle increases rapidly with decreasing radius towards the nucleus.

This behavior can be explained in terms of the gas accretion caused by the galactic shock wave as follows:
The angular momentum $A_z$ of the gas about the rotation axis is effectively transferred by the oval motion in the barred potential, while the perpendicular component of the angular momentum $A_x$ is conserved.
This causes rapider loss of $A_z$ than $A_x$ until the radius becomes comparable to the $z$ {extent}, when $A_z$ becomes comparable to $A_x$, as observed in the central region with Arms V to VII. 
The magnetic twisting mechanism \citep{1986PASJ...38..631S} may work similarly, which acts to transfer the angular momentum of a rotating gas disc penetrated by strong vertical magnetic field \citep{2022ApJ...925..165H}. 

\ss{Relationship of the Arms to the general CMZ structure} 
\label{volumeratio}

We next estimate the relative luminosity ($\sim$mass) of the GC Arms to that of Arms I + II, or the 120-pc ring, using the arm radii $R$ in table \ref{tab1} and their vertical {extent} $z$.
Assuming about the same molecular-line brightness (within an order of magnitude), the relative luminosity of Arm $N$ to that of Arm I + II is estimated by 
\be
L_N/L_{\rm I+II} \sim (R_N/R_{\rm I+II}) 
 (z_N/z_{\rm I+II}),
\ee
where $N$ stands for the arm number, I to VII, and the subscript, I+II, stands for averages of the quantities of Arms I and II.
We obtain the ratios to be  
$L_i/L_{\rm I}\sim 1$ (for Arm I, II); 
$\sim 2\times 10^{-2}$ (III); 
$\sim 3\times 10^{-3}$ (IV);
$\sim 4\times 10^{-4}$ (V); and
$\sim 1.4\times 10^{-5}$ (VI). 

Arms I and II share most of the mass of the CMZ, composing the 120-pc molecular ring (or the "great ring").
Inside the ring the disc shares only a small portion of the CMZ's volume due to the decrease both in radius and {vertical extent}, or Arms III to VI share 
two orders of magnitudes smaller portions of the entire volume, hence the mass. 
This is consistent with the low infrared extinction in the nuclear stellar disc \citep{2022A&A...668L...8N}. 

The small volume (mass) of the disc inside Arm I and II means that the inflow from the CMZ ring into the nucleus is extremely slow, in other words, the ring is large enough to supply the inner arms, albeit with very low efficiency.

\section{Discussion}
 \label{sec_discussion} 

\ss{General remarks}

\citet{henshaw+2023} published a thorough review of recent progress in the study of 3D molecular gas distribution of the CMZ in the $(l,b,\vlsr)$ space (see the literature therein).
There seems to be a consensus that Arms I and II constitute the main structure of the CMZ, composing a large ring-like structure (the 120-pc ring) of radius $\sim 100$-120 pc rotating clockwise around \sgrastar\ as seen from the North Galactic Pole.
However, apart from some mentions of the existence of Arms III and IV, there has been no detailed study of the spiral arm or ring structure inside the 120-pc ring.
In this paper, we showed that there are many more inner arms including Arm III and IV of radii $R\sim 40$ and $\sim 20$ pc, respectively, and Arm V of radius $\sim 8.2$ pc which was identified here for the first time.
The innermost structures such as the CND and minispirals can be understood as the systems organized with a unified rule as Arm VI and VII, respectively.
The thus identified arms are summarized in figure \ref{fot-arms} and table \ref{tab1}.

{Since the present analysis focuses on the kinematical and geometrical structure of the CMZ, we have not calculated the density and mass.
Therefore, the derived arms/rings appear rather symmetric to \sgrastar.
The highly asymmetric distribution of the mass both in space and velocity known since 1980's \citep{bally+87,bally+88} is 
somehow suppressed in the present analysis. 
Converting $\Tb(l,b,\vlsr)$ cube to $\rho(x,y,z)$ cube for more quantitative modeling of CMZ would be a subject for the future.}
  
Theories and simulations to understand the distribution and motion of molecular gas in the GC have been extensively developed over decades, predicting different types of flow models in the CMZ and circum-nuclear regions
\citep{2008A&A...489..115R,2011ApJ...735....1W,2012ApJ...751..124K,2015MNRAS.453..739K,2017MNRAS.466.1213K,2017MNRAS.469.2251R,sorma+2019,sorma+2020,tress+2020}:
The simulations suggest that symmetric spirals of grand design mimicking Arms I and II are generated as a result of galactic shock waves in a bar potential, carrying the gas to the innermost regions and giving rise to CND and mini-spiral-like structures.
However, even if such mechanism works, the efficiency of the inward flow must be extremely low, because the mass ratio between the innermost arms and the entire CMZ is very low as shown in table \ref{tab1}. 
The proposed new view of the innermost spiral structure, including the CND and minispirals (associated with arms VI and VII, respectively), would provide further observational constraints on the model and adds information to a more precise understanding of the CMZ. 

Below we discuss some specific topics related to the individual arm structures.

\ss{Comparison with far-infrared dust map}

\sss{Arm I and II in dust emission}

\begin{figure} 
\begin{center}    
\hskip -5mm\includegraphics[width=9cm]{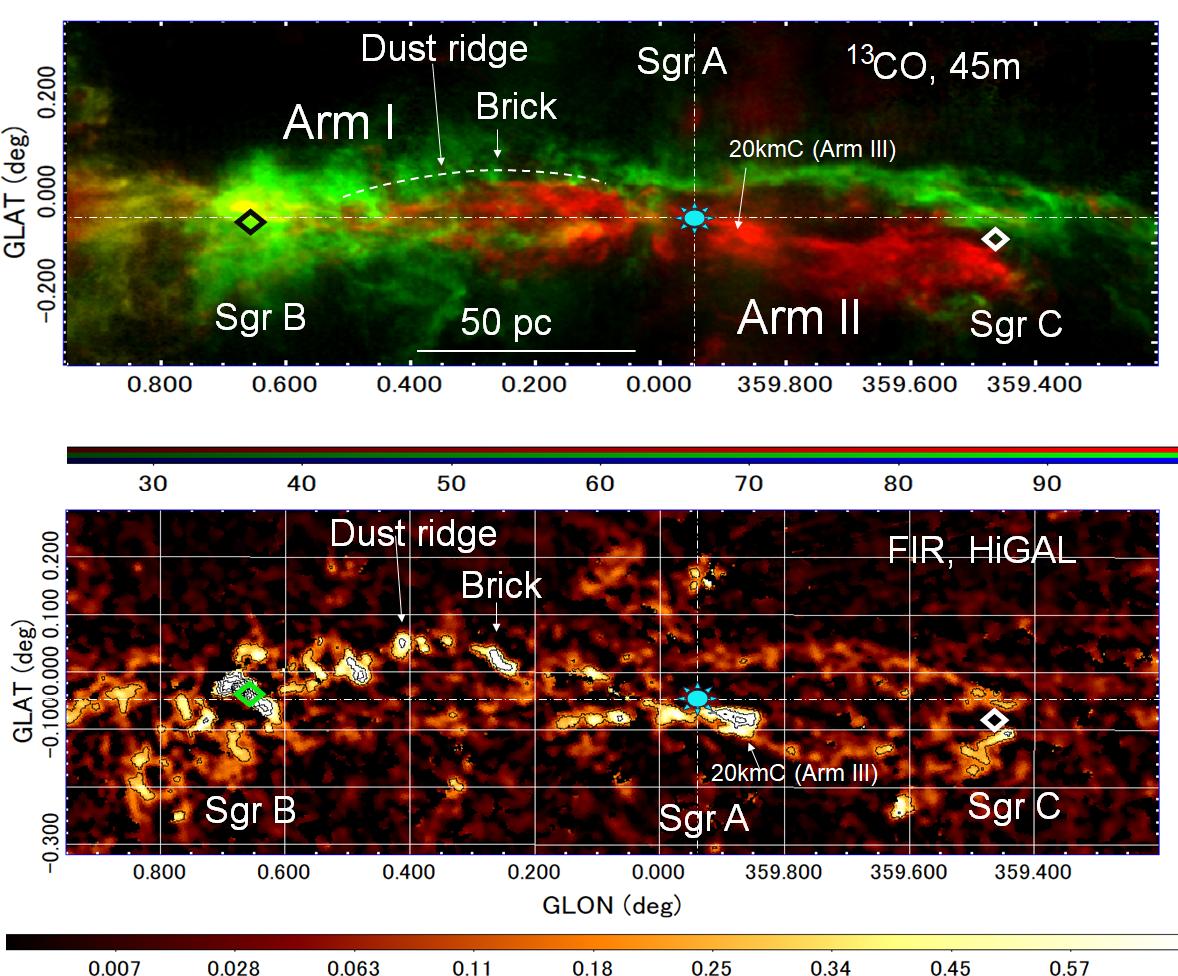}  
\end{center}
\caption{[Top] Masked moment 0 map in \coth\ Arm I (green) and II (red) (same as the top panel of figure \ref{moment_45m}) compared with [bottom] a Herschel FIR-derived dust column-density map \citep{2015ApJ...815..130G} with the intensity indicated by the bar in units of [grams cm$^{-2}$] and contours every 0.7 starting at 0.2.
Note that the 20kmC at $\sim$G-0.0.025-0.07, which is located on the near side of Sgr A, is contaminated in the moment 0 map of Arm II (far side) in the top panel.
 {Alt text: Moment 0 map of Arm I+II compared with far-infrared dust map.}} 
\label{FIRvs13CO}	 
\end{figure}

We show in figure \ref{FIRvs13CO} a preliminary comparison of the arm structure in the \coth\ line with the image of the CMZ in a dust column density map computed from the Herschel HiGAL survey at 70, 160, 250 350 and 500 $\mu$m. The dataset was reprocessed with an algorithm designed to obtain more accurate measurements of cold, dense filamentary structures, densities and temperatures, by improved subtraction of diffuse foreground and background emission. The five Herschel bands were convolved to the resolution Jo of the 500 $\mu$m SPIRE image and a process based on the CUPID-findback algorithm was repeatedly applied until consecutive iterations differed by less than 5\% in all pixels. (See also \citet{2011AJ....142..134E}).
Taking an estimated value for the dust beta of 1.6, we derived a dust temperature map which, combined with the flux map, yielded the dust column density map.  Additional details about the method and checks on its reliability are given in \citet{2015ApJ...815..130G}.  

It is clear that the warped dust ring coincides closely with the molecular-line ring composed of Arm I and II, where Arm I and II are located in the near and far sides of \sgrastar, associated with Sgr B and C, respectively. 
Note, however, that Arm II near Sgr A in this moment 0 map is contaminated by Arm III associated with the 20kmC which is visible in the FIR map as a short horizontal belt near \sgrastar.
The infinity shape suggested in the integrated-intensity maps in molecular lines and dust emission \citep{2011ApJ...735L..33M} can be traced in the FIR dust map, whereas it does not show up so clearly here in the \coth\ map resolved in the $(l,b,\vlsr$) space.

\sss{Dust Ridge and the Brick}
 
The positive longitude side of Arm I in the CO-line is well correlated with the Dust Ridge composed of a chain of the massive dark cloud "Brick" G+0.0253+0.016 and Clouds b, c, d, e/f in the far infrared emission and absorption
\citep{2019MNRAS.485.2457H,
2020ApJS..249...35B,
2023MNRAS.525..962P,
2023ApJ...959...36G}.
However, the Brick in CO in this map shows up only partially, because the radial velocity is about 20 \kms displaced from the main ridge of Arm I as shown in figure \ref{ArmI} so that the CO-line moment 0 map misses the major part of the cloud.

\sss{Other Dust clouds}

The dust clouds called "Stone, Straw, and Sticks" draw a horizontal belt around $(l,b)\sim (+0\deg.08, -0\deg.08)$ \citep{2020ApJS..249...35B} appear in the moment 0 map of Arm II in figure \ref{moment_45m} (panel D).
They appear as an LV ridge in figures \ref{ArmII}, \ref{ArmIII} and \ref{aces_lv_max} (top panel), running from $(l,\vlsr)=(0\deg.05,50\ekms)$ to $(0\deg.15,60\ekms)$.
Coincidence in the LV plane indicates that the clouds are associated with Arm II.
We point out that mid-infrared (4--8 $\mu$m) images of the GC \citep{2006JPhCS..54..176S} reveal no prominent absorption toward these clouds, indicating that they are in the far side of the central bulge, supporting the association with Arm II (far side). 
On the other hand the Brick and Dust Ridge exhibit heavy extinction, locating them in the near side, which is consistent with the association with Arm I (near side).

\ss{20- and 50-\kms\ Clouds} 
\label{ss20vs50} 

The inner region $|l|\lesssim 0\degd.2 \ (\sim 30\ \epc)$ around Sgr A$^*$ contains two well-known molecular clouds. 20- and 50kmC \citep{2005ApJ...620..287H,Takekawa17a,2009PASJ...61...29T,2019ApJ...872..121U}. 
We comment on these clouds, which are visible and resolved in our data. 

\sss{Line-of-sight locations of 20- and 50kmC}

\begin{figure}   
\begin{center}  
{\large 20kmC \hskip 2.5cm 50kmC}\\
\vskip 3mm
{\hcnaces\ ACES} \\
\vskip 2mm
A \hskip 4cm B\\
\includegraphics[width=8.5cm]{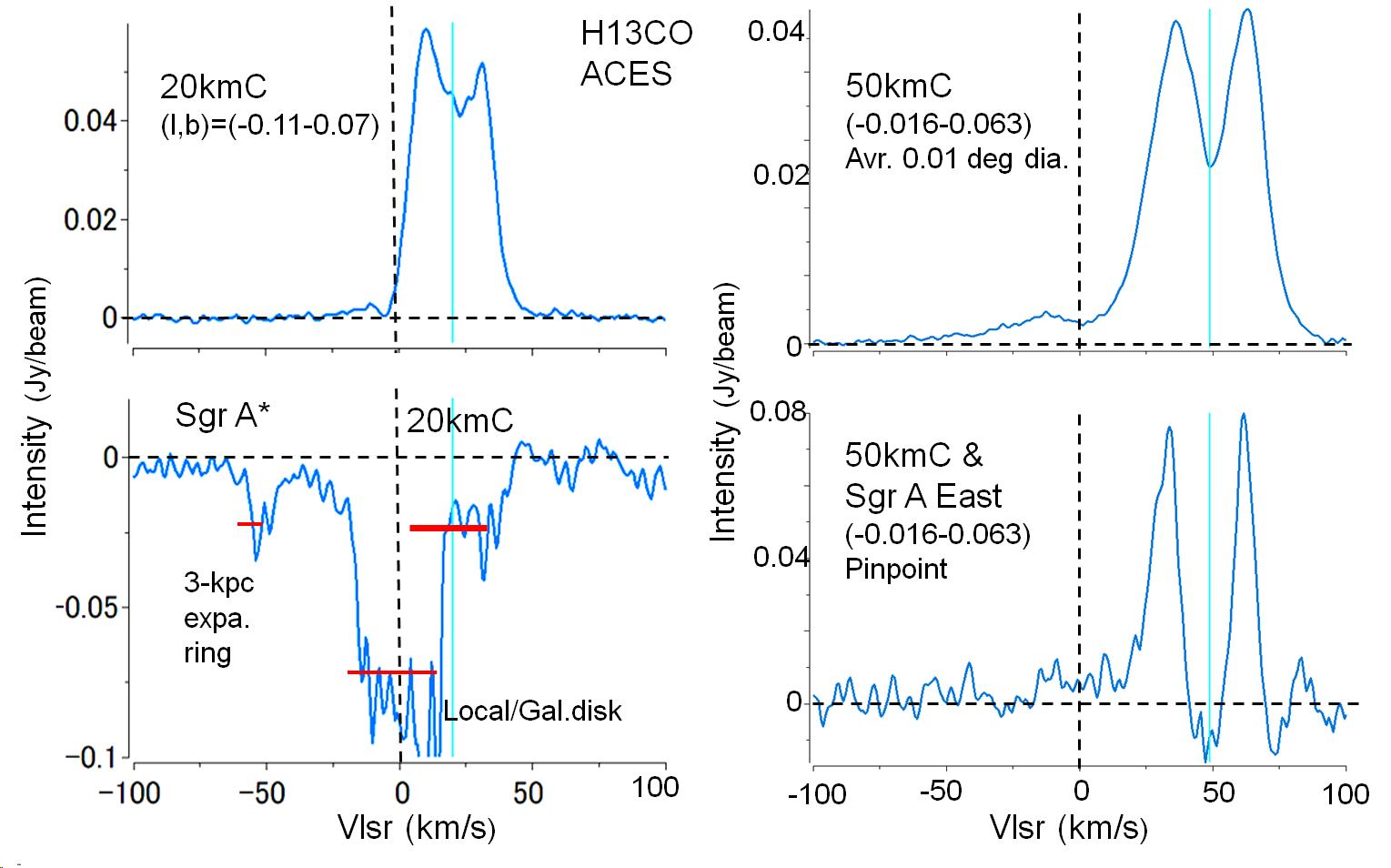} \\
\vskip -3cm C \hskip 4cm D\\
\vskip 3cm
{HI ATCA}\\
E \hskip 4cm F\\
\includegraphics[width=4.2cm]{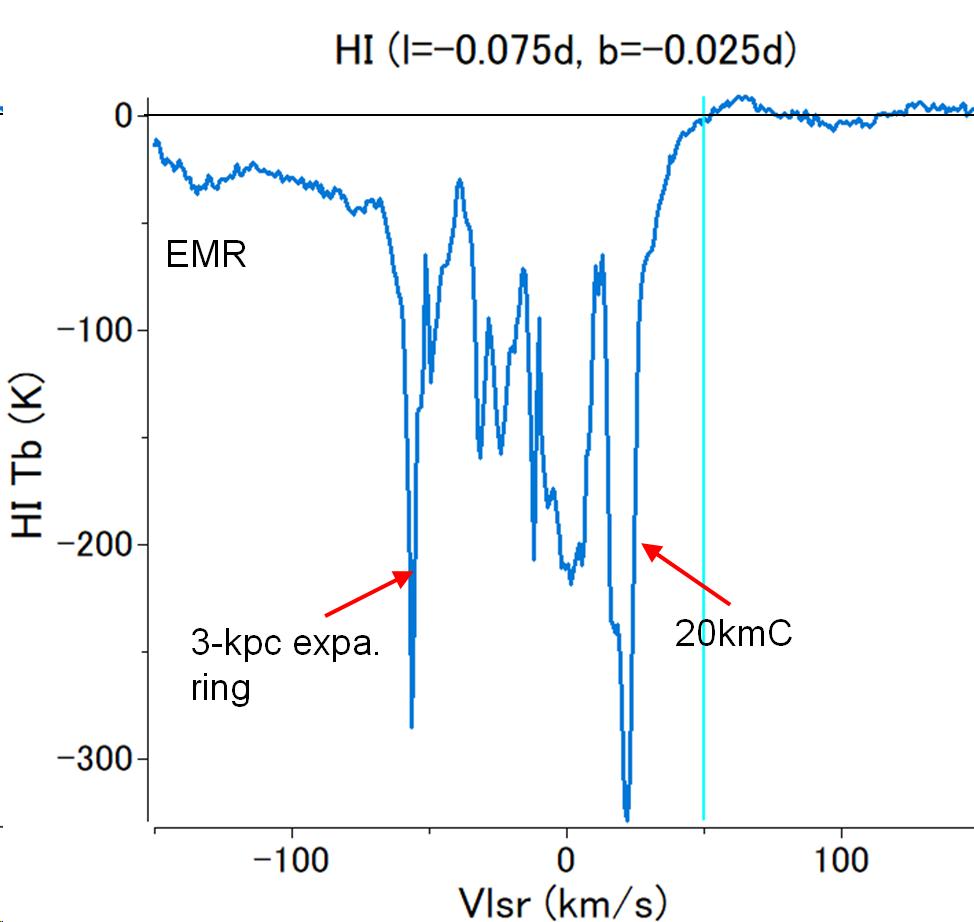}    
\includegraphics[width=4.2cm]{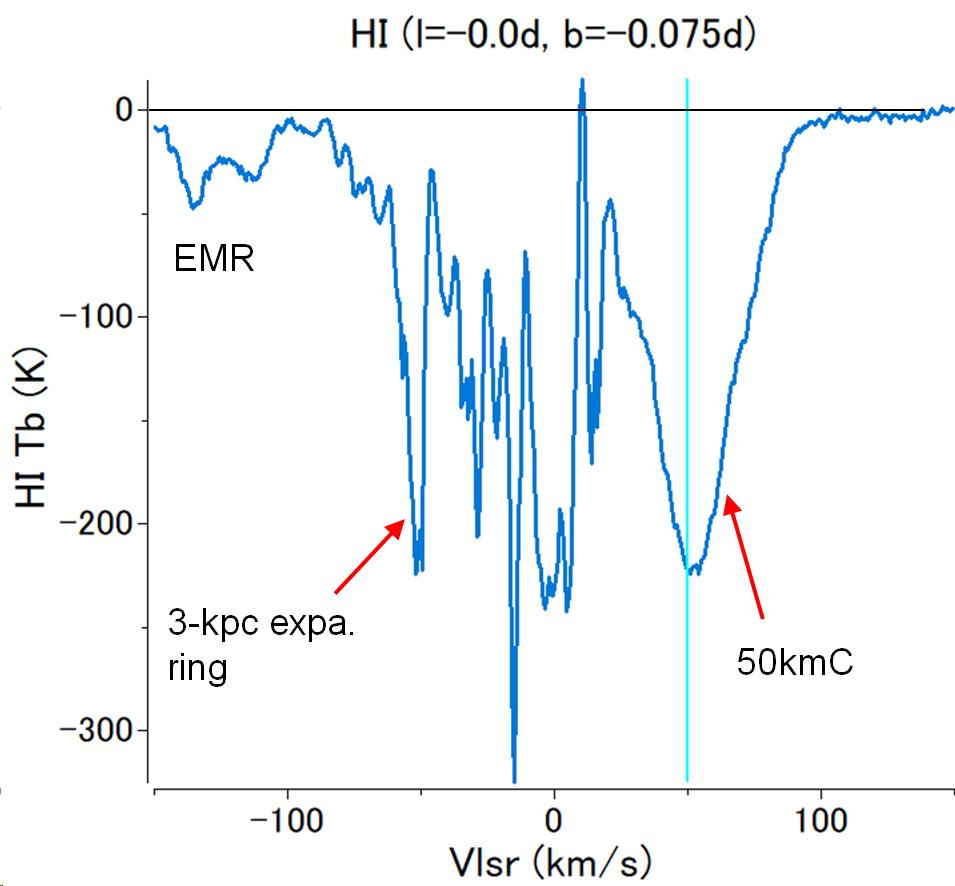}  \\
E  \includegraphics[width=2.3cm]{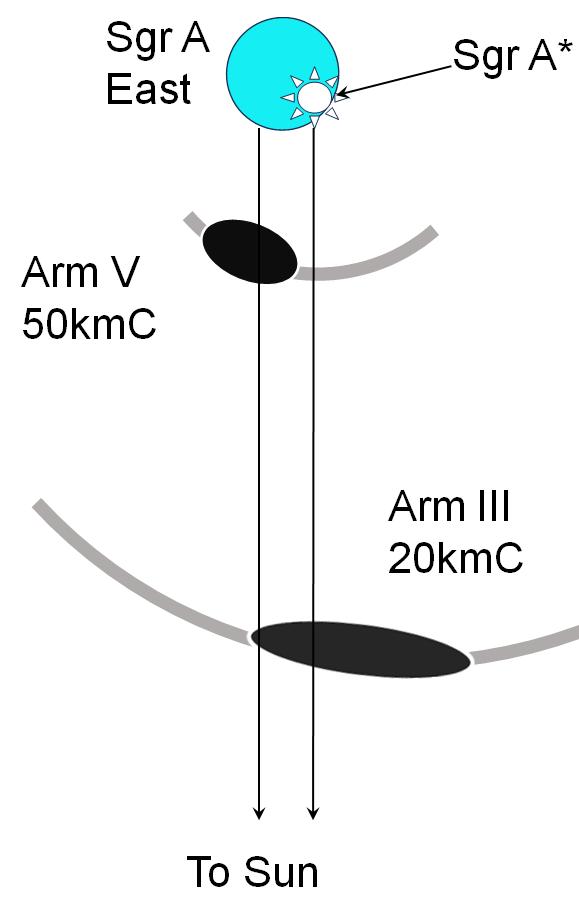}  
\end{center}    
\caption{[A]: \hcnaces\-line profile of the 20kmC at $(l,b)=(-0\deg.11,-0\deg.07)$ by ACES, showing self absorption. 
[B] Same of 50kmC at $(-0\deg.016,-0\deg.063)$ averaged in $0\deg.01$ diameter, showing  absorption. 
Note the larger line width than 20kmC about twice.
[C] \hcnaces-profile toward \sgrastar, showing absorption of the continuum by 20kmC along with local and disc clouds on the near-side.
[D] Pinpoint profile of 50kmC toward the radio bright edge of Sgr A East, showing negative absorption, hence, near-side location.
[E] HI line absorption profile of 20kmC at G-0.075-0.025 averaged in a field of diameter $0\deg.05$ using ATCA HI survey \citep{mcclure+2012}.
Note the sharp absorption line of 20kmC.
[F] Same for 50kmC at G+0.0,-0.075. Note the broad absorption line.
[E] Schematic illustration of the near-far solutions for 20- and 50kmC using absorption of the radio continuum emission.
Note that 50kmC has much wider line widths than 20kmC in both the lines, indicating more dynamic condition due to closer location to the GC.
 {Alt text: \hcnaces-line profiles of 20kmC and 50kmC, showing self absorption  and absorption against continuum of \sgrastar\ and Sgr A East., compared with the HI line profiles.}
 } 
\label{abs2050kmC}	 
\end{figure}

We first consider the line-of-sight locations of the two clouds using absorption profiles of the molecular-line spectra shown in figure \ref{abs2050kmC}. 
The 20kmC shows a clear double-horn profile indicative of self absorption. 
The pinpoint spectrum on \sgrastar\ exhibits negative-intensity absorption, indicating that the continuum emission is absorbed by 20kmC, so that the cloud is on the near side of \sgrastar. 
50kmC also shows double-horns toward the edge of Sgr A East, and a pinpoint spectrum exhibits negative intensity absorption.
This evidences that the cloud is on the near side of Sgr A East, consistent with the earlier result \citep{2009PASJ...61...29T}. 

We applied the same method to the HI line using the Australia-Telescope Compact Array (ATCA) survey of the GC \citep{mcclure+2012}. 
The bottom panels of figure \ref{abs2050kmC} show the HI-line spectra tward 20kmC. 
Both spectra show deep absorption lines. 
Note also that 50kmC has a much wider line width.

We have thus proved that radio continuum emission from Sgr A is absorbed by the 20- and 50kmC in the molecular and HI lines, and conclude that the two clouds are on the near side of \sgrastar.
This is consistent with the near side location of the two clouds inferred from the analyses of dust extinction by infrared observations \citep{2021A&A...647L...6N,2024arXiv241017321L,2024arXiv241017320W}.

\sss{20kmC} 

\begin{figure*}   
\begin{center}     
A \hskip 4.5cm B \hskip 5cm C\\
\includegraphics[height=5cm]{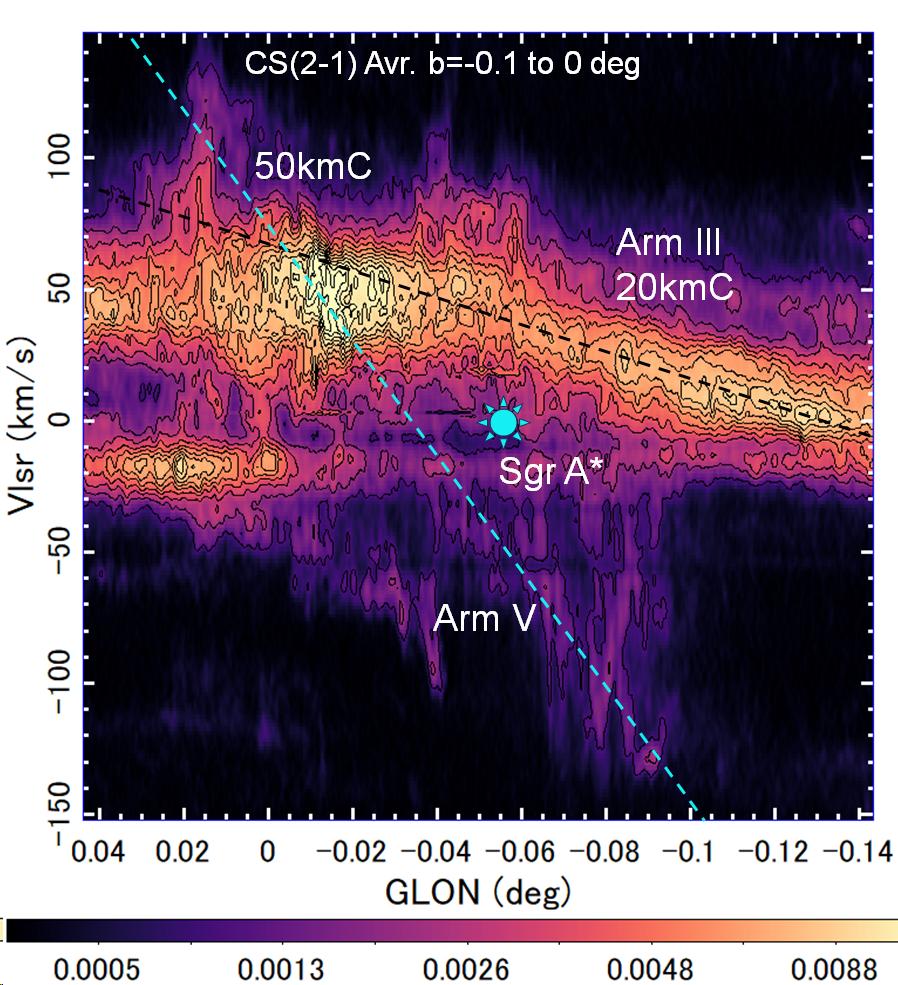} 
\includegraphics[height=5cm]{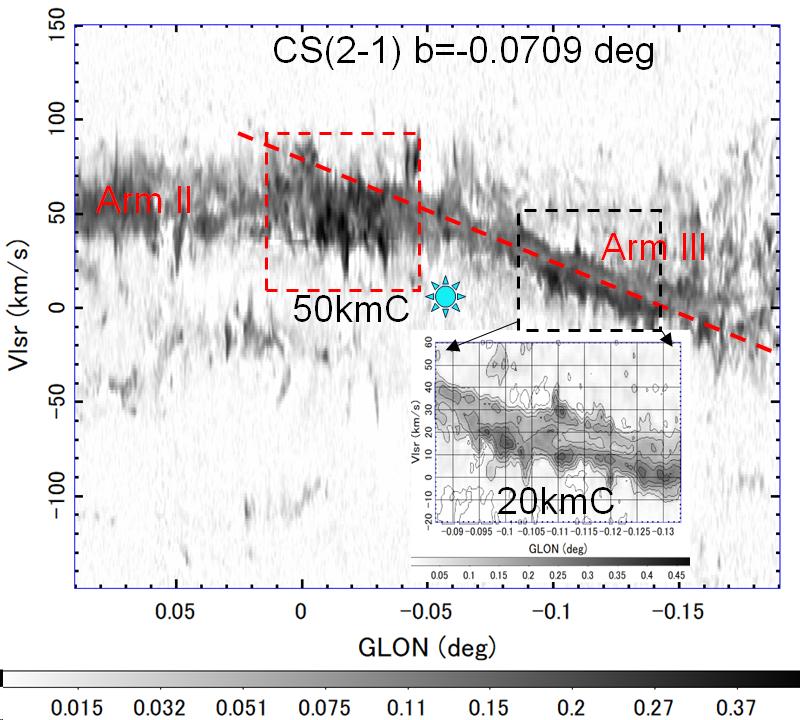}
\includegraphics[height=5cm]{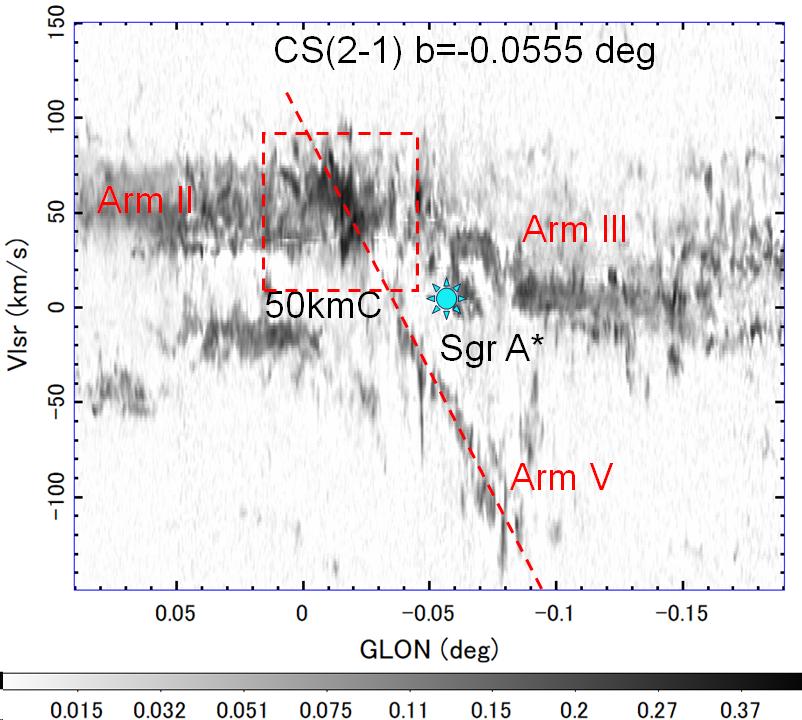}
\end{center}
\caption{
[A] Averaged LVD from $b=-0\deg.1$ to $0\deg$ in \cs\  around 20- and 50kmC.
Intensity scale is in Jy beam$^{-1}$ and contours are every 1 mJy beam$^{-1}$.
Dashed lines traces Arm III and V.
[B] LVD at $b=-0\deg.0709$ across 20- and 50kmC. 
Note the self absorption belt along Arm III. 
50kmC is enclosed by the dashed box, and is enlarged in figure \ref{2050ABC}.
[C] LVD at $b=-0\deg.055$ showing 50kmC exactly on the extension of Arm V. 
Arm V is clearly visible in this diagram.
50kmC is enlarged in figure \ref{2050ABC}.
Note that Arm II and 3-kpc expanding ring also intersect 50kmC.
{Alt text: LVD of 50kmC and Arm III with 20kmC averaged from $b=-0\deg.1$ to $0\deg$, and close up of LVDs of 20kmc at $b=-0\deg.0709$ and 50kmC at $b=-0\deg.0555$. }
} 
\label{2050new}	 
\end{figure*}

In figure \ref{2050new} we show the LVD around 20- and 50kmC averaged in all the latitude range in the present analysis from $0\deg$ to $-0\deg.1$.
The middle panel shows an LVD at $-0\deg.071$ across the centers of 20- and 50kmC and along Arm III, and the right panel is an LVD at $b=-0\deg.055$ including Arm V. 
 
The 20kmC is aligned along the straight ridge of Arm III, constituting the main structure of the arm.
There is no clear-cut boundary on either side of the cloud along the LV ridge, and the arm extends continuously toward both sides.
This indicates that 20kmC is not an isolated cloud, but is the major part of Arm III.

The LV ridge of Arm III (20kmC) is separated by an absorption belt, which is a 2D view of the double-horn spectra. 
The double-horn absorption may be attributed to 
(i) an expanding cylinder (not a shell), 
(ii) absorption of the background continuum emission, or
(iii) self absorption. 
The possibility of an expanding cylinder (i) is unlikely, considering the formation mechanism.
We may also rule out an expanding spherical shell, which postulates an LV ellipse rather than absorption belt.
Absorption of background light (ii) is also unlikely because the region is $\sim 7$ pc away from Sgr A on the sky where the radio brightness is too low to cause absorption in the molecular line.
We may therefore conclude that the feature is due to (iii) self-absorption along the arm.

We may thus conclude that the 20kmC is the densest part of Arm III with its 3D centre position at $(x^*,y^*,z^*)\sim (-7.7, 42, -3.6)$ (pc).
Here, $(x^*,y^*,z^*)$ are the cartesian coordinates with respect to \sgrastar\ with the three axes in the directions toward positive-galactic longitude, toward the Sun, and toward the North Galactic Pole, respectively.

\sss{50kmC}

The location of 50kmC is more complicated, because its LV ridge overlaps with Arm II, III, V and the far-side 3-kpc expanding ring \citep{2008ApJ...683L.143D}, and the proximity on the sky suggests association with Sgr A.
Here, we may rule out Arm II and far 3-kpc ring because they are on the opposite side of Sgr A.

Adopting the kinetic energy ($\sim 7\times 10^{49}$ erg), velocity dispersion ($\sigma_v\sim 28$ \kms) and mean gas density ($\sim 10^4$ \htwocc) derived by \citet{2009PASJ...61...29T}, we estimate the molecular mass to be $M_{\rm mol}\sim 10^4\Msun$.
This mass is two orders of magnitudes smaller than the dynamical mass of $M_{\rm dyn}\sim 3\sigma_v^2 r/G \sim 10^6\Msun$ for $r\sim 1$ pc, and therefore 50kmC is not an isolated bound "cloud", but is an expanding shell or an arm orbiting (streaming) in the Galactic potential. 

\begin{figure*}   
\begin{center}      
A \hskip 4.5cm B \hskip 4.5cm C\\
\includegraphics[height=5cm]{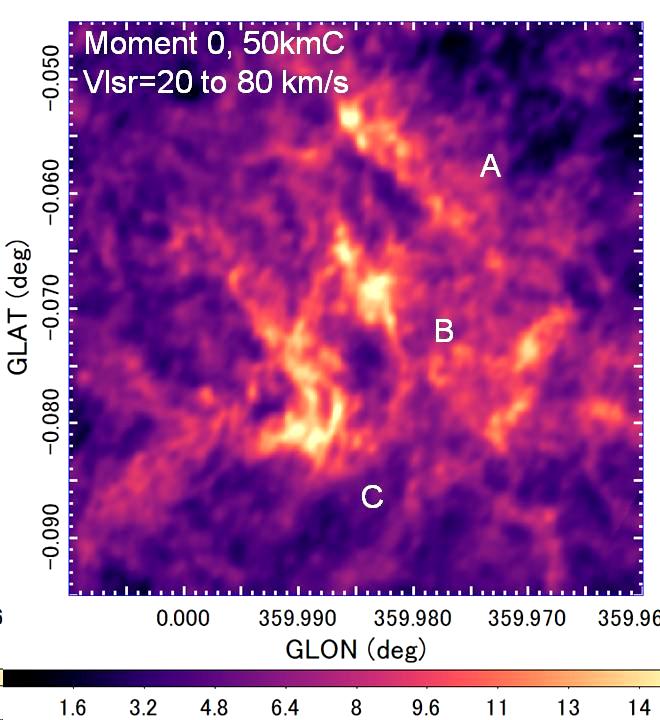} 
\includegraphics[height=5cm]{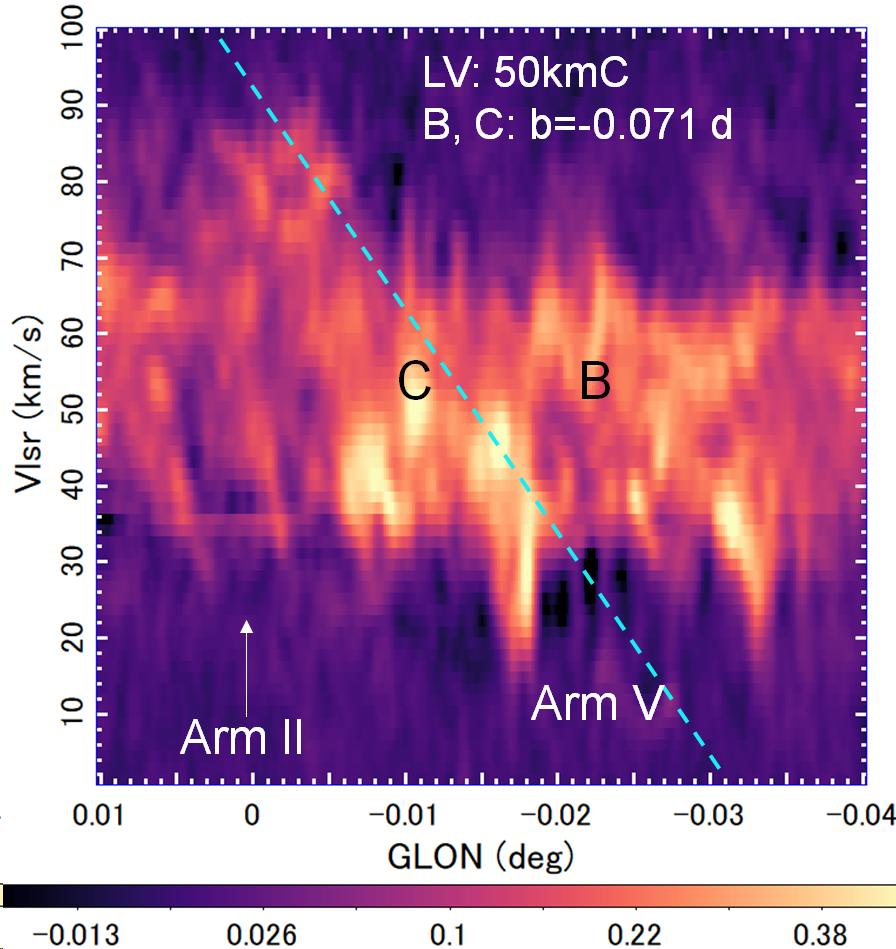}
\includegraphics[height=5cm]{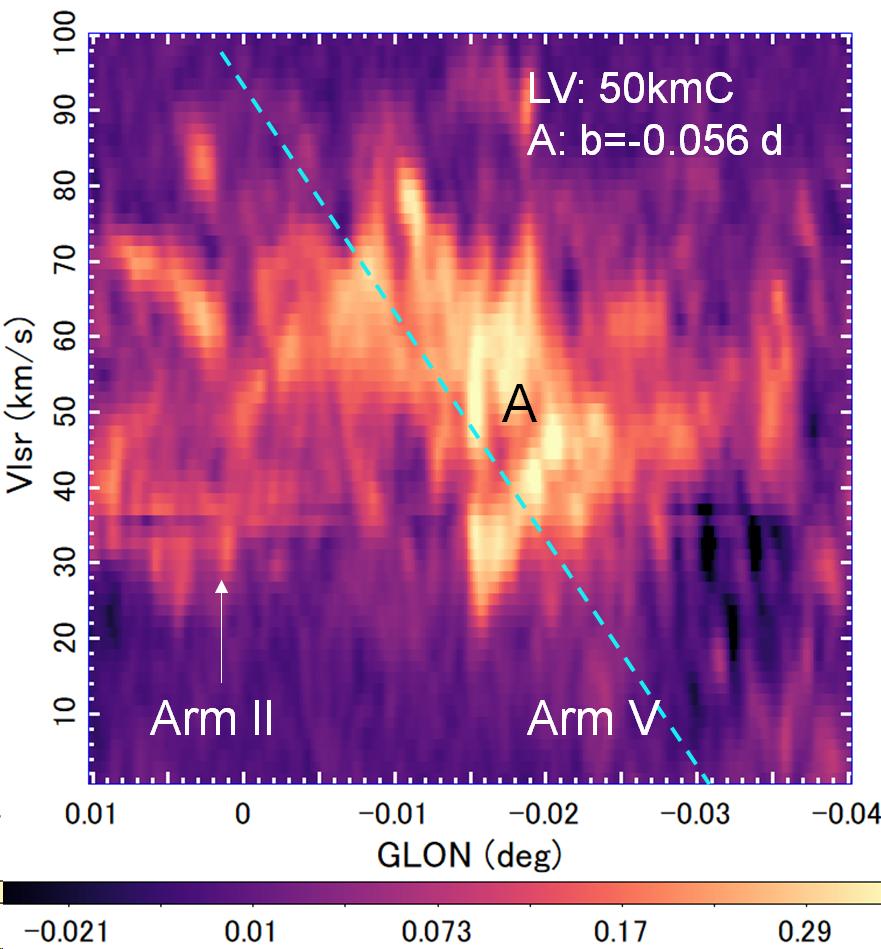}
\end{center}
\caption{
[A] Moment 0 map of 50kmC in \cs\ integrated from $\vlsr=20$ to 80 \kms\  (intensity scale in \kms\ Jy beam$^{-1}$). The cloud is composed of three sub clouds A, B and C, which do not appear in the masked moment maps in figure \ref{moment_aces}. 
[B] LVD at $b=-0\deg.071$ across 50kmC-B and C (intensity scale in Jy beam$^{-1}$) . 
The dashed line is the extension of Arm V.
[C] Same, but at $b=-0\deg.056$ across 50kmC-A. 
{Alt text: Moment 0 map of 50kmC, LVDs of 50kmC-A, B, and C. }
} 
\label{2050ABC}	 
\end{figure*}

For detailed inspection of kinematics we present \cs\ LVDs and moment 0 map in figures \ref{2050new} and \ref{2050ABC}. 
The three components of 50kmC, clump A (northwest), B (center) and C (southeast) \citep{2009PASJ...61...29T}, are resolved, and reveal LV ridges coherently inclined with velocity gradients equal to that of Arm V.
It is also stressed that the clumps have large velocity widths of $2\sigma_v\sim 50$--60 \kms.  

As to the location of 50kmC, we consider the following three possibilities: \\
\noindent (i) 50kmC is located near Sgr A because of their proximity on the sky \citep{2009PASJ...61...29T}.
The large velocity width is attributed to expansion of a supernova remnant. 
However, the velocity gradient has not been discussed.
The 3D position of the cloud centre with respect to \sgrastar\ is roughly $(x^*,y^*,z^*)\sim(5.6, 0, -3.4)$ (pc)\\
(ii) 50kmC is part of Arm V, because the LVD exactly overlaps with Arm V and the velocity gradient is equal to that of Arm V.
Also, the large velocity width is explained by the Galactic rotation. 
This model puts the cloud centre at $(x^*,y^*,z^*)\sim(5.6, 7.7, -3.4)$ (pc).\\
(iii) 50kmC is part of Arm III, because the LVDs approximately overlap with each other.
This assumes $(x^*,y^*,z^*)\sim(5.6, 42, -3.4)$ (pc).
However, the velocity gradient and large velocity width are not explained.

We here consider that the second scenario is plausible, but possibilities of (i) and (iii) cannot be ruled out at this moment.

\ss{200-pc EMR}
\label{EMR}

Besides the main structures of the CMZ like the GC Arms, figures \ref{ArmI} to \ref{aces_lv_max} also exhibit more complicated features surrounding these.
One of such features is the expanding molecular ring (EMR) commonly appearing in these LVDs, which are marked in figure \ref{aces_lv_max} by the nearly horizontal dashed lines EMR+ and EMR-.
It has long been debated whether the EMR is due to an expanding ring caused by an explosion at the center 
\citep{Kaifu+1972,Scoville1972,Sofue2017}, 
or a "parallelogram" due to non-circular gas flow in a bar potential\citep{binney+1991,2015MNRAS.449.2421S,tress+2020}. 

Discussion of these features is beyond the scope of this paper, but we point out that the EMR is fainter than the main Arms I and II by a factor of $\sim 10^{-2}$ in surface brightness of the molecular lines, and the total mass is an order of magnitude smaller than that of the CMZ \citep{Sofue2017}.
Furthermore, the EMR has a large vertical extension above and below the CMZ of $\sim \pm 100$ pc, which deviates significantly from the {vertical extent}-radius relation in figure \ref{arm-plot}.
Therefore, in order for the EMR (parallelogram) to act as a mass supplier to the CMZ, the gas transported from the outer Galactic disc with a thickness of $\sim 20$--30 pc must first be lifted to that height and then quickly compressed to the CMZ of thickness $\sim 10$--20 pc.
In this context, it has been recently argued that the western wing of the EMR may be a high-velocity ($\sim \pm 100$--200 \kms) and high-altitude ($z/2\sim 20$--60 pc) molecular inflow with a length of $\sim 200$ pc, acting to transport the gas into the CMZ \citep{2024A&A...689A.121V}.

\section{Summary}

Analyzing the molecular-line cubes of the Galactic Centre taken with the ALMA (\csaces\ and \hcnaces), Nobeyama 45-m telescope (\coth), and ASTE 10-m telescope (\hcnaste), we studied the kinematic behavior of GC Arms I to VI identified in the longitude-velocity diagrams (LVDs).
The galactocentric radii of the Arms are determined by the $dv/dl$ method assuming a flat rotation curve.
Applying the LV-masking method, we also obtained moment 0 maps integrated in the velocity range within $\pm \sim 7.5$--10 \kms from the LV ridges of the Arms.
We find that the radius of the $N$th arm is approximately given by $R\sim 630\times 0.40^{N}$ pc, suggesting a logarithmic spiral or Bode's law-like discreteness of the orbits.
If we consider the minispirals to constitute Arm VII, the relation holds from $N=1$ (I) to 7 (VII).
The vertical full {extent} of the arms is approximated by $z\sim 0.7 (R/1 \epc)^{0.7}$ pc.

Unifying the derived parameters of the arms and rings, we summarize the results in figure \ref{fot-arms} as a schematic view of the warping CMZ:
Arms I and II share most of the mass (volume) of CMZ; the inner arms share a few percent of the CMZ mass (volume); hence the accretion is slow; the {vertical extent} of the disc decreases toward the centre; and the warping amplitude or the arm's tilt from the galactic plane increases toward the centre. 


\begin{ack}
This paper makes use of the following ALMA data: ADS/JAO.ALMA$\#$2021.1.00172. 
 
ALMA is a partnership of ESO (representing its member states), NSF (USA) and NINS (Japan), together with NRC (Canada), NSTC and ASIAA (Taiwan), and KASI (Republic of Korea), in cooperation with the Republic of Chile. 

The Joint ALMA Observatory is operated by ESO, AUI/NRAO and NAOJ. 
  
The data analysis in this paper was partially performed at the Astronomical Data Center of the National Astronomical Observatories of Japan.

KK acknowledges the support by JSPS KAKENHI Grant Numbers JP23K20035 and JP24H00004. 

C.\ Battersby  gratefully  acknowledges  funding  from  National  Science  Foundation  under  Award  Nos. 2108938, 2206510, and CAREER 2145689, as well as from the National Aeronautics and Space Administration through the Astrophysics Data Analysis Program under Award ``3-D MC: Mapping Circumnuclear Molecular Clouds from X-ray to Radio,” Grant No. 80NSSC22K1125.

COOL Research DAO \citep{cool_whitepaper} is a Decentralised Autonomous Organisation supporting research in astrophysics aimed at uncovering our cosmic origins.

J.M.D.K. gratefully acknowledges funding from the European Research Council (ERC) under the European Union's Horizon 2020 research and innovation programme via the ERC Starting Grant MUSTANG (grant agreement number 714907). 
 
KMD acknowledges support from the European Research Council (ERC) Advanced Grant MOPPEX 833460.vii.
 
 L.C., V.M.R. and I.J.-S. acknowledge support from the grant PID2022-136814NB-I00 by the Spanish Ministry of Science, Innovation and Universities/State Agency of Research MICIU/AEI/10.13039/501100011033 and by ERDF, UE. 
 
V.M.R. also acknowledges support from the grant RYC2020-029387-I funded by MICIU/AEI/10.13039/501100011033 and by "ESF, Investing in your future", from the Consejo Superior de Investigaciones Cient{\'i}ficas (CSIC) and the Centro de Astrobiolog{\'i}a (CAB) through the project 20225AT015 (Proyectos intramurales especiales del CSIC); and from the grant CNS2023-144464 funded by MICIU/AEI/10.13039/501100011033 and by “European Union NextGenerationEU/PRTR”.

I.J.-S. acknowledges support from ERC grant OPENS, GA No. 101125858, funded by the European Union. Views and opinions expressed are however those of the author(s) only and do not necessarily reflect those of the European Union or the European Research Council Executive Agency. Neither the European Union nor the granting authority can be held responsible for them.

P. Garc\'ia is sponsored by the Chinese Academy of Sciences (CAS), through a grant to the CAS South America Center for Astronomy (CASSACA).
 
R.S.K. thanks the 2024/25 Class of Radcliffe Fellows for highly interesting and stimulating discussions, financial support from the European Research Council via  ERC Synergy Grant ``ECOGAL'' (project ID 855130), the German Excellence Strategy via 
(EXC 2181 - 390900948) ``STRUCTURES'', the German Ministry for Economic Affairs and Climate Action in project ``MAINN'' (funding ID 50OO2206), Ministry of Science, Research and the Arts of the State of Baden-W\"{u}rttemberg through bwHPC and the German Science Foundation through grants 
INST 35/1134-1 FUGG and 35/1597-1 FUGG, and also for data storage at SDS@hd funded through grants INST 35/1314-1 FUGG and INST 35/1503-1 FUGG. 

D. Riquelme-V\'asquez acknowledges the financial support of DIDULS/ULS, through the project PAAI 2023.

J.Wallace gratefully acknowledges funding from National Science Foundation under Award Nos. 2108938 and 2206510.


\end{ack}
 

\section*{Data availability} 
The single-dish data underlying this article are available at
https:// www.nro.nao.ac.jp/ $\sim$nro45mrt/html/ results/data.html. 
The interferometer data were taken from the internal release version of the 12m+7m+TP (Total Power)-mode data from the ALMA cycle 8 Large Program "ALMA Central Molecular Zone Exploration Survey" (ACES, 2021.1.00172.L).
 
\section*{Conflict of interest}
The authors declare that there is no conflict of interest.


\begin{appendix}   
\section{IMSHIFT-relieving method} 
\label{aprelief} 

\begin{figure}   
\begin{center}    
45-m \coth, Relieved LVD, Avr. $|b|\lesssim 0\degd.3$\\ 
 \includegraphics[width=8cm]{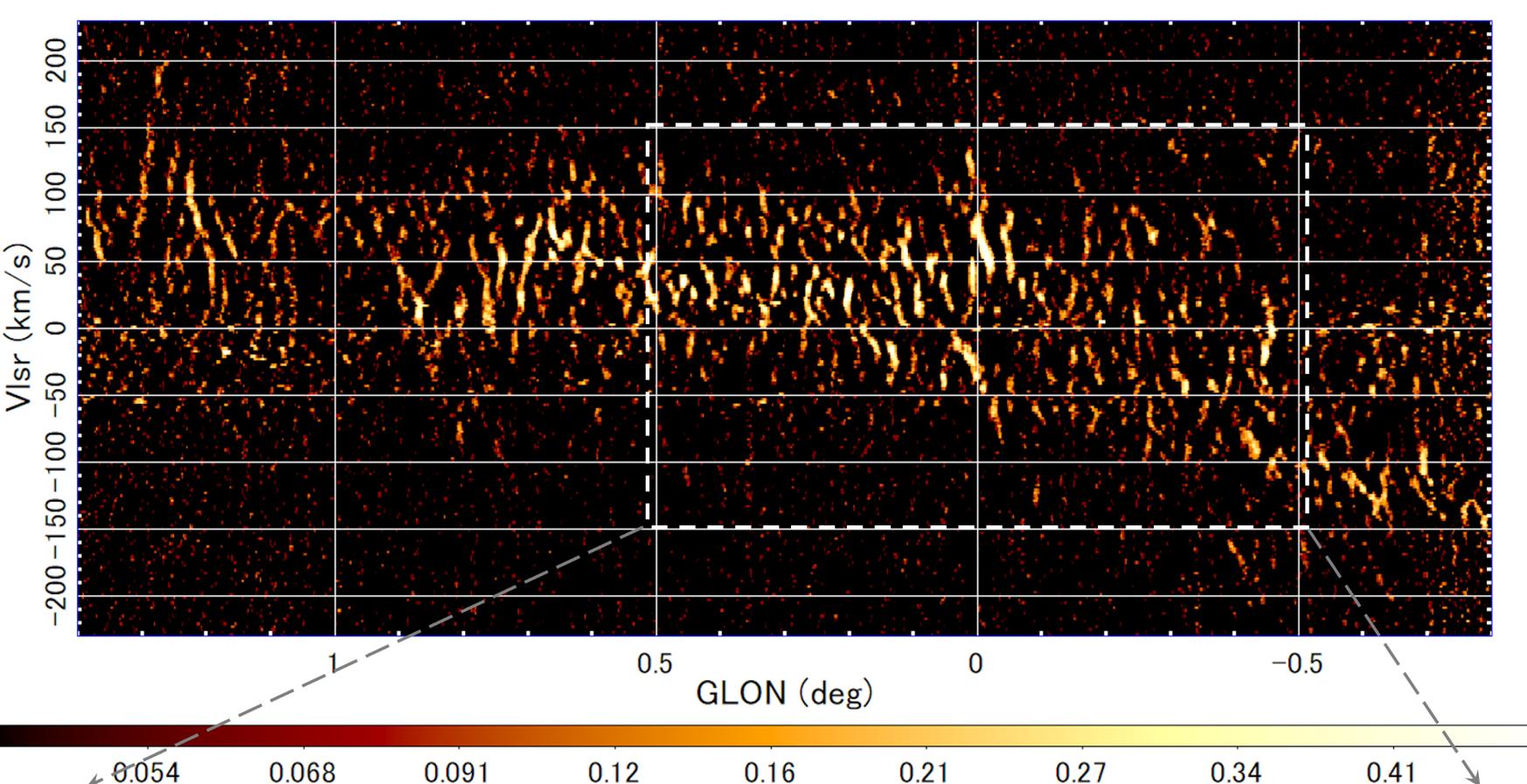}\\
45-m \coth\ LVD chan. Relief  \\
 \includegraphics[width=8cm]{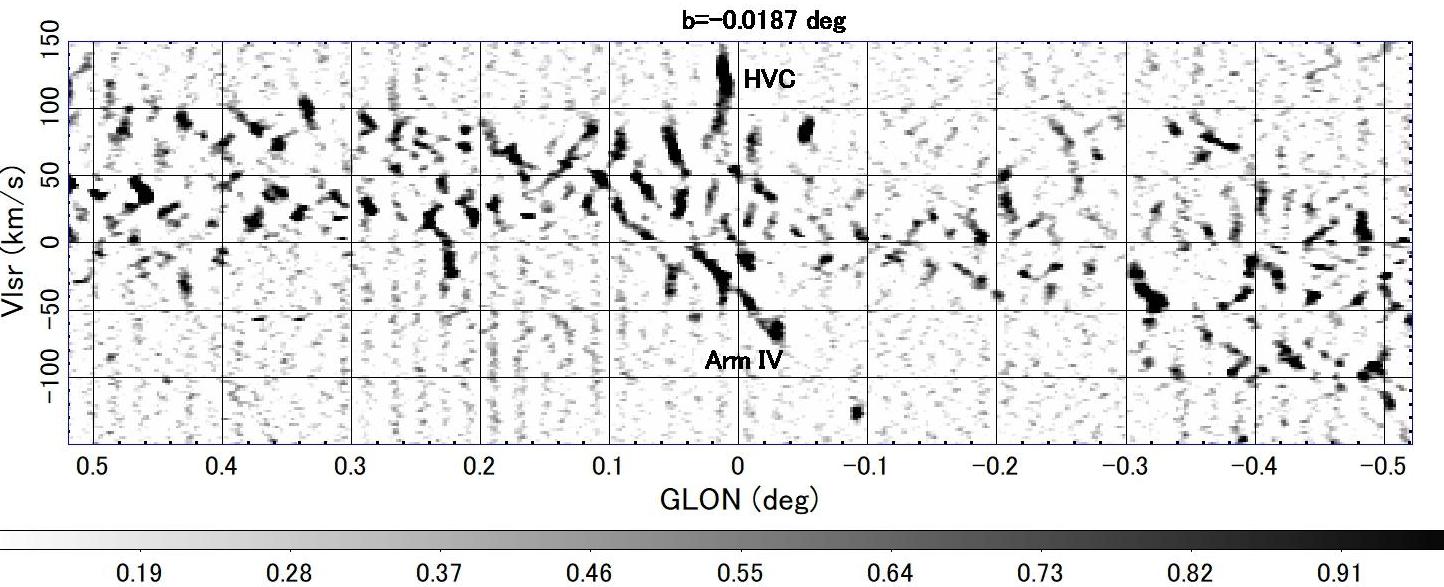}   
\end{center}
\caption{\coth\ IMSHIFT-relieved LVDs (longitude-velocity diagrams) in the CMZ in \coth\ line observed with the 45-m telescope in the whole area (top) and the central region at $b=-0\degd.02$  (bottom).  {Alt text: 2 LVDs by applying the IMSHIFT-relieving method. }}  
\label{45m13lvx5}	 
\end{figure}

In order to abstract tilted LV stripes representing rotating arms in the CMZ using single dish observations, we apply the IMSHIFT relieving technique, which is a modification of the "background-filtering" (BGF) (pressing) method  
\citep{sof1993}.
This method subtracts extended components with scale sizes greater than a threshold value (here 5 pixels) in one direction (here in galactic longitude), so that it enhances oblique and vertical LV stripes.
This method, therefore, suppresses the horizontal LV stripes (contamination) due to the fore- and background Galactic disc.  
We confirmed that there 
are no significant differences in the results when the relieving size is from $\delta x\sim 3$ to 10 pix. 
Figure \ref{45m13lvx5} shows an example of relieved LVD averaged in $|b|\lesssim 0\degd.3$ in the whole CMZ in \coth\ line, and the bottom panel is enlargement in the central region at a fixed latitude.

The method consists of the following procedure.
Let the original map represents intensity distribution $T(x,y)$. 
The relieved intensity is defined by
\be
\Delta T(x,y)=(\Delta T^+ + \Delta T^-)/2 
\ee 
where
$\Delta T^+=T(x,y)-T(x+\delta x,y)$ and 
$\Delta T^- =T(x,y)-T(x-\delta x,y)$. 
We then replace the pixel values to zero, if $\Delta T<0$. 
In the present analysis, we adopt a relieving size of $\delta x = 5 {\rm pix}\sim 37\asec.5=1.5$ pc in the longitude direction.  
However, the following point may be kept in mind when using it:
The method suppresses structures wider than the threshold width, the obtained LVDs are not useful to discuss large-scale arms and rings, particularly in the outer CMZ. 

\section{Channel LVDs}
\label{apchannel}

\begin{figure}  
\begin{center}       
\vskip -5mm
ASTE 10m \hcnaste \\ 
\includegraphics[width=7cm]{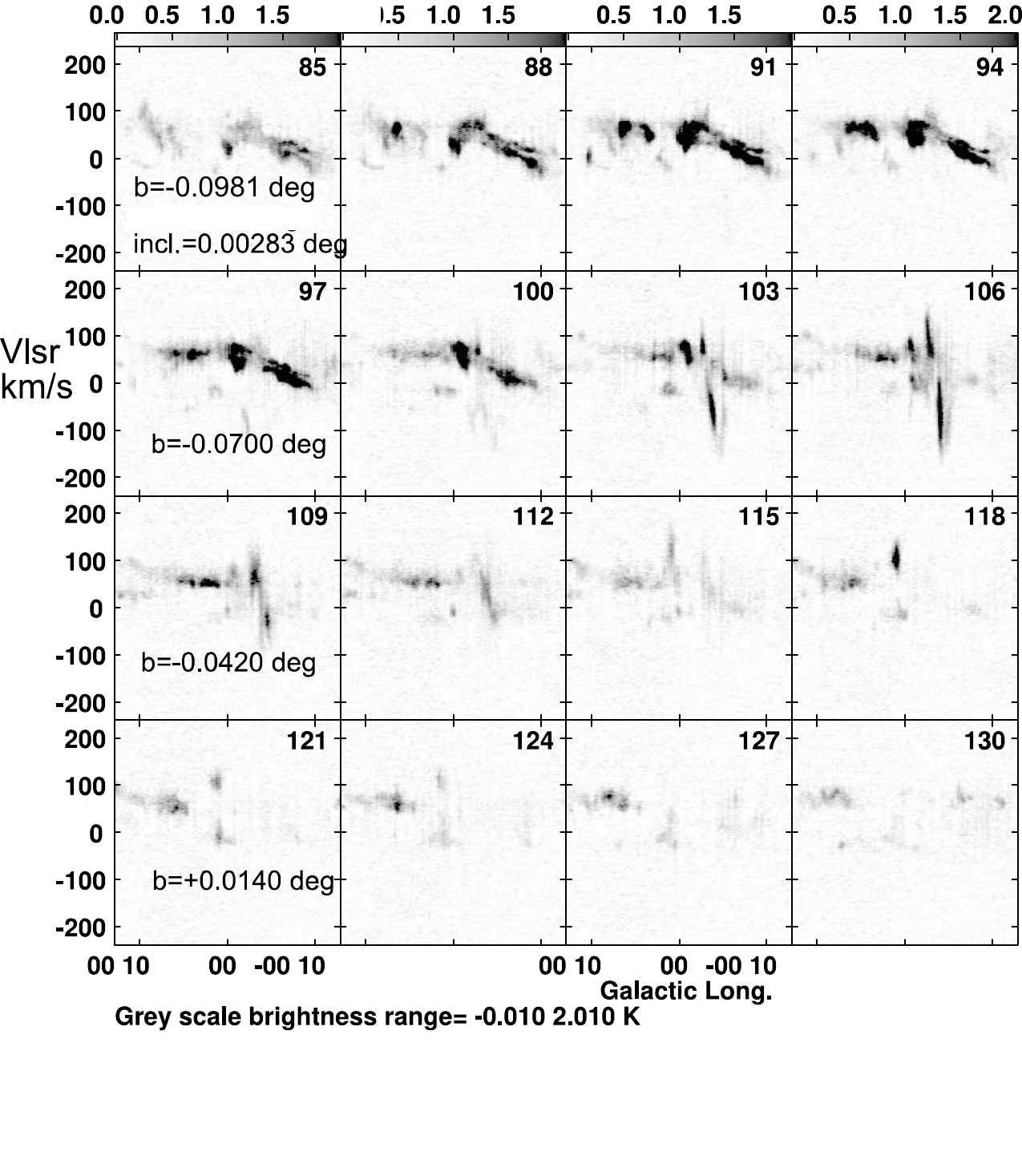}  \\
\vskip -7mm
ALMA \csaces \\  
\includegraphics[width=7cm]{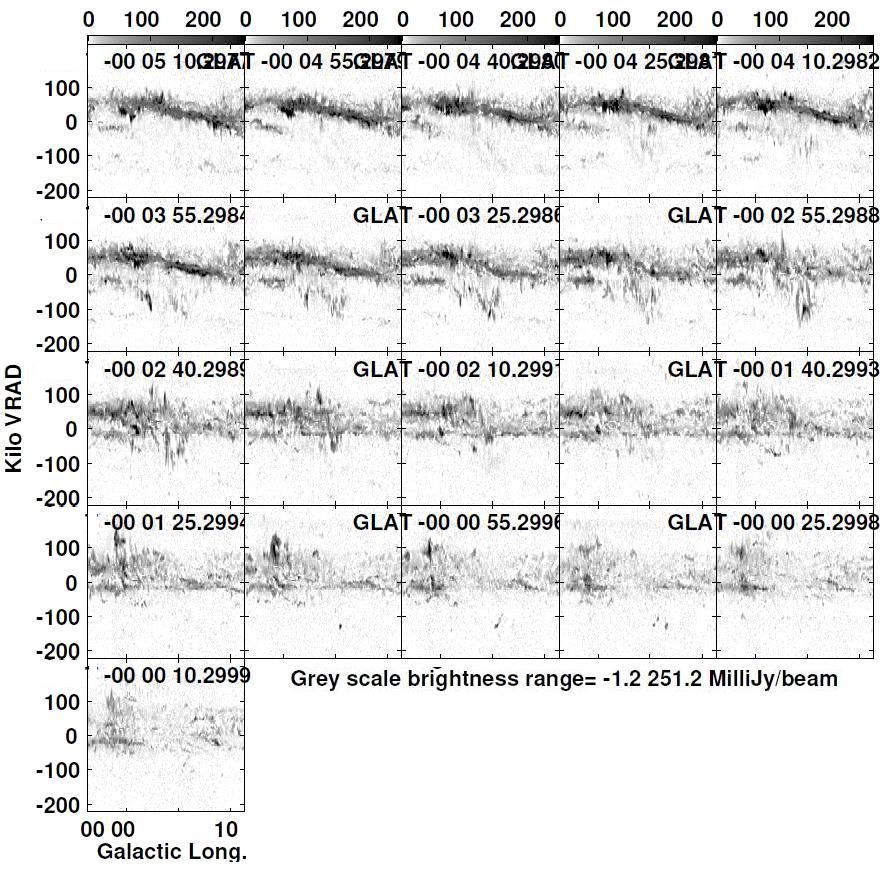}   \\   
ALMA \hcnaces \\
\includegraphics[width=7cm]{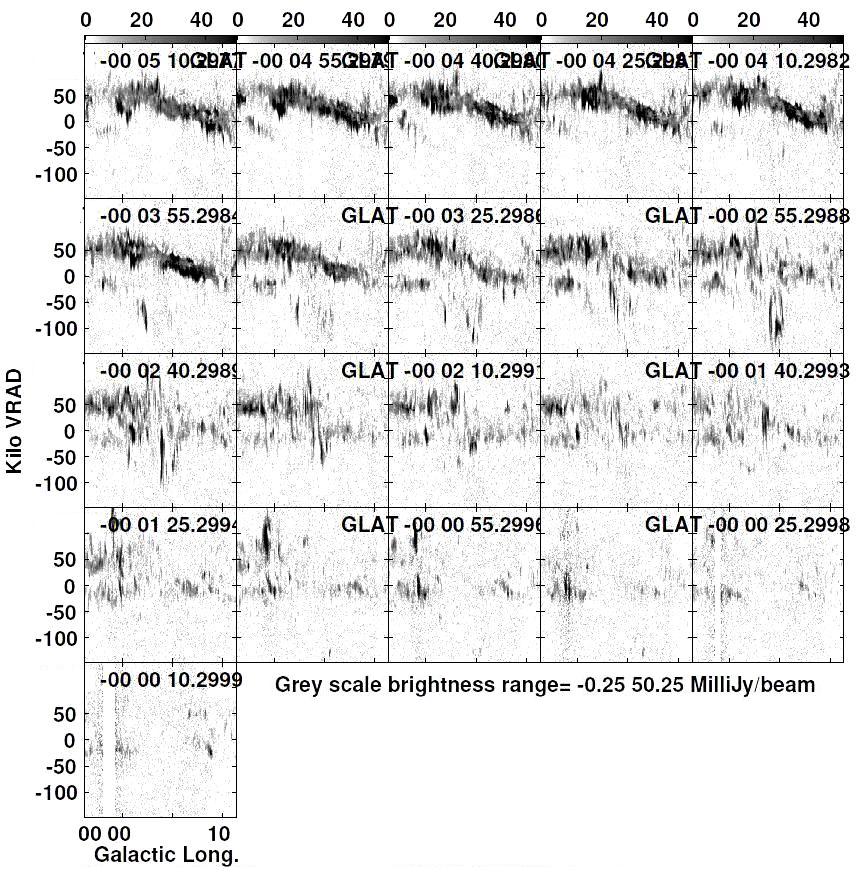}    
\end{center}   
\caption{[Top] Channel LVDs in \hcnaste\ by ASTE 10-m telescope.
[Middle] ACES latitude-channel maps of LVDs in \csaces\ (top) and [bottom] \hcnaces\ (bottom) of the central $\pm 0\degd.12$ about \sgrastar. 
 {Alt text:   Latitudinal channel LVDs in \hcnaste\ form ASTE-10 m, channel LVDs in \csaces\ and \hcnaces\ by ACES. }
}
\label{lvd-channels}	   
\end{figure}

We present latitudinal channel maps of LVDs of the central $l\sim \pm 0\degd.2$ region in the \hcn\ line from ASTE 10-m in the top panel of figure \ref{lvd-channels} in order to present high-density LV arms. 
The 2nd and bottom panels show \csaces\ and \hcnaces-line channel LVDs from ACES of the central $\sim \pm 0\degd.1$ region at higher resolutions, respectively.
These figures along with the original cubes were used to find and identify an arm as a straight LV ridge extending over $\sim 100$ \kms, and to confirm that the arm is not artifact specific to a certain channel, but is a real object by comparing the feature with those continuously appearing in the neighboring multiple channels.

\end{appendix}

\end{document}